\documentclass{article}

\usepackage{graphicx} % Required for inserting images
\usepackage[margin=1in]{geometry}%
\usepackage{authblk}  % 用于作者和机构信息处理
\usepackage{amsmath}
\usepackage{amsmath,empheq}
\usepackage{booktabs}  % 三线表
\usepackage{multirow}  % 合并单元格
\numberwithin{table}{section}  % 将表格编号绑定到章节
\usepackage{tabularx}
\usepackage{caption}  % 优化标题格式
\usepackage{float} % 在导言区添加
\usepackage{adjustbox}
\usepackage{appendix}
\usepackage{booktabs,siunitx}
\usepackage{longtable} %
\numberwithin{figure}{section}  % 将图编号关联到章节
\usepackage[colorlinks=true, linkcolor=blue, citecolor=blue, urlcolor=blue]
{hyperref}

\title{\textbf{Stochastic Price Dynamics in Response to Order Flow Imbalance: Evidence from CSI 300 Index Futures}}
\author{
Chen Hu$^{1}$\thanks{Email: huchen@glqh.com; khuaug@gmail.com}, 
Kouxiao Zhang$^{2}$\thanks{Email: zhangkouxiao@glqh.com;}
}

\affil{Guolian Futures.,Ltd, Shanghai, China}

\date{May 2025}
\begin{document}
  
\maketitle

\begin{abstract}
We conduct modeling of the price dynamics following order flow imbalance in market microstructure and apply the model to the analysis of Chinese CSI 300 Index Futures. There are three findings. The first is that the order flow imbalance is analogous to a shock to the market. Unlike the common practice of using Hawkes processes, we model the impact of order flow imbalance as an Ornstein-Uhlenbeck process with memory and mean-reverting characteristics driven by a jump-type Lévy process. Motivated by the empirically stable correlation between order flow imbalance and contemporaneous price changes, we propose a modified asset price model where the drift term of canonical geometric Brownian motion is replaced by an Ornstein-Uhlenbeck process. We establish stochastic differential equations and derive the logarithmic return process along with its mean and variance processes under initial boundary conditions, and evolution of cost-effectiveness ratio with order flow imbalance as the trading trigger point, termed as the quasi-Sharpe ratio or response ratio. Secondly, our results demonstrate horizon-dependent heterogeneity in how conventional metrics interact with order flow imbalance. This underscores the critical role of forecast horizon selection for strategies. Thirdly, we identify regime-dependent dynamics in the memory and forecasting power of order flow imbalance. This taxonomy provides both a screening protocol for existing indicators and an ex-ante evaluation paradigm for novel metrics. 
\\ 
\\
% Keywords Section
\textbf{Keywords:} 
Order flow imbalance, Ornstein-Uhlenbeck process, SDE,  Market microstructure
\end{abstract}
  
\tableofcontents

\section{Introduction}
   
High-frequency trading, or HFT is operationally defined as algorithmic systems exploiting latent microstructure invariants to generate sub-second trading signals. The extant literature has witnessed exponential growth in HFT studies since the 2000s, with scholarly attention partitioned across strategy classes such as market-making, latency arbitrage, and short-term statistical arbitrage.

Extant research has extensively focused on market microstructure to characterize price dynamics. This study specifically investigates order flow imbalance and trade flow imbalance within market microstructure, with emphasis on examining and uncovering the temporal adaptability of existing indicators. Cont et al. \cite{cont1} introduced a novel order flow imbalance factor within the market microstructure framework and systematically investigated various valuable micro-level microstructure metrics including both order flow imbalance and trade flow imbalance. A growing body of subsequent research has applied this methodology across diverse financial instruments, consistently revealing its robust adaptability. Cont et al. \cite{cont2} analyzed the application of the aforementioned metrics in deeper level market structures and their mutual influences among stocks, and investigated the out-of-sample predictive performance of these indicators. Furthermore, they proposed that the prediction horizon for these indicators should be sufficiently extended. In \cite{shen1}, Shen conducted extensive empirical research on China's CSI 300 index futures based on the metric analysis from \cite{cont1}, thoroughly testing the statistical characteristics and validity of the foundational indicators. However, the study did not sufficiently explore the variations in order flow imbalance across different historical windows and forecasting horizons. In the field of order book dynamics and market depth analysis, Cont et al.\cite{cont3, cont4} conducted a comprehensive research. In terms of price response and incentive mechanisms, Anantha et al. \cite{nittur1} and Rosenbaum et al. \cite{rosenbaum1} leveraged Hawkes \cite{hawkes1} process to theoretically model the price impact of indicators. The Hawkes process characterizes temporal dependencies through event-driven intensity function for self-exciting process, and has become an important methodological framework for studying indicator-based price responses in financial markets. 

Another fundamentally important process is the Ornstein-Uhlenbeck process, commonly abbreviated as the O-U process. This process was originally developed by Ornstein and Uhlenbeck \cite{Uhlenback1} to model the velocity of particles in standard Brownian motion. It has been widely applied in financial domains, particularly in volatility research, pairs trading, spread trading, interest rate modeling, option pricing, equilibrium price oscillation analysis, and indicators and signals analysis. For example, Azze et al. \cite{azze1} developed a detailed model for spread-based option portfolios. Guerreiro et al. \cite{guerreiro1} applied the fractional Ornstein-Uhlenbeck process to VIX analysis. Bormetti et al. \cite{Bormetti1, Bormetti2} utilized the O-U process for volatility modeling, specifically for volatility surface fitting and asset price dynamics research. Holý et al. \cite{Holy1}, Leung and Li \cite{Li1} conducted research on optimization and entry/exit strategies in pairs trading. Perelló et al. \cite{Perello1} applied the O-U process to investigate and characterize long-term interest rates in multiple countries. Petroni et al. \cite{Petroni1} utilized a composite model combining the O-U process with an independent jump process to describe Asian options. Bergault et al. \cite{Bergault1} constructed a modeling framework for multi-asset optimal execution in statistical arbitrage with underlying spread exhibiting O-U process properties. Lipton and Lopez de Prado \cite{Prado1} applied an O-U process to model mean-reverting price behavior in equilibrium markets and optimized market making execution strategies. Furthermore, many financial processes have been modeled as Ornstein-Uhlenbeck  processes driven by finite jump processes, or specifically, O-U processes driven by finite Lévy processes. For instance, Larsson et al. \cite{Larsson1} studied optimization problems in entry-exit strategies for pairs trading. Alfonsi et al. \cite{Vadillo1} employed a non-homogeneous O-U process driven by both Brownian motion and normal inverse Gaussian Lévy processes in their risk assessment study of temperature and electricity derivatives.

Our research principally examines the price forecasting capability of order flow imbalance metrics through analytical modeling approaches. Regarding the modeling of financial indicators, Lehalle and Neuman \cite{Lehalle1} employed an O-U process to characterize market indicators. The framework models a class of limit order book imbalances through an O-U process. Building upon this approach, we empirically validate and emphasize the investigation of an alternative order flow imbalance metric derived from the indicator formulation in \cite{cont1}. In their work \cite{Ayed1, Ayed2}, Bel Hadj Ayed et al. modeled price dynamics through an implicitly assumed O-U process trend and derived optimal execution strategies within this framework. In the research, this O-U process assumes being driven by a Wiener process. Through empirical analysis, we find that when modeling order flow imbalance as an Ornstein-Uhlenbeck process, the process may be driven by a heavy-tailed jump component rather than a Wiener process. We embed this driving process within a generalized Lévy process formulation. This heavy-tailed process assumption may better align with empirical observations of financial price changes, which frequently exhibit fat-tailed distributions rather than normal ones. Our analysis derives the log-return process characterizing the post-order flow imbalance evolution, yielding an initial model formulation that underwent empirical validation through data testing.
       
\subsection{Main contributions}
First, many previous studies based on microstructure metrics, including those provided by \cite{cont1} typically examine the metrics after accumulating data over a historical window, and consider subsequent changes over symmetric time intervals to evaluate the effectiveness of the metric. For example, after calculating a 5-second historical window Order Flow Imbalance ($OFI$) metric, the analysis would focus only on the corresponding future 5-second market changes. In our study, we treat the accumulated Order Flow Imbalance over a certain period as a step shock and examine its response over varying subsequent time intervals. We find that the effectiveness of the metric is a function of its response time. Unlike numerous studies that model this process as a Hawkes process, we follow the analysis in \cite{Lehalle1} and characterize the response to order flow imbalance shocks (as defined in \cite{cont1}) as an Ornstein-Uhlenbeck process with memory and mean-reverting properties. Following the analytical approach of \cite{Ayed1, Ayed2}, we replace the drift term in the conventional geometric Brownian motion with this process, based on the stability of the correlation between order flow imbalance and contemporaneous mid-price change. Through the solution of the coupled stochastic differential equations, we obtain the logarithmic return process for the price dynamics under our modeling assumptions, including explicit expressions for both the mean and variance processes. Furthermore, we derive a time-varying Sharpe-like ratio under order flow imbalance shocks to characterize the evolution of trading efficiency. This ratio, which we term the "response ratio", quantifies the trade-off and competition between deterministic drift induced by order flow imbalance and stochastic diffusion movements. These results establish a foundation for future research into trading optimization.

For the aforementioned hypothesized model for logarithmic return process, we conduct empirical tests using one year of tick-level data from CSI 300 index futures. To statistically examine the impulse response of order flow imbalance metrics, we exhaustively test various combinations of historical windows and forecasting horizons. For each set of parameters, we perform linear regressions of the order flow imbalance against subsequent average mid-price changes, and obtain results that support our model.

Prior research focused mainly on the examination of individual microstructure indicators in isolation. Through our investigation, we introduce the concept of indicator matching. We systematically examine the pairing effects between order flow imbalance and various other indicators, revealing that optimal indicator selection depends critically on the chosen forecasting horizon.

In our study of the market, we not only analyze the overall effects over one year but also compute the monthly metrics and statistical properties separately. Based on this, we classify market operations into distinct modalities. In some months, the market is considered to be in an efficient modality, making it difficult to trade. However, in other months, metrics and statistical data reveal that the market is in an inefficient modality. It is precisely during these periods of insufficient pricing that high-frequency trading is well-suited to participate, thereby assisting the market in improving pricing efficiency.

Meanwhile, a meaningful microstructure metric should exhibit consistent statistical effects across different market modalities, demonstrating only quantitative variations rather than qualitative changes. Conversely, metrics that show significant differences in statistical properties across modalities can be classified as weak metrics. This also provides a reference criterion for subsequent research to evaluate the validity of microstructure metrics. We propose several screening conditions to identify robust metrics.

\subsection{Outline}
In section 2, we analyze the dynamics of the mid-price changes to order flow imbalance metrics and propose a theoretical model combining geometric Brownian motion and the U-O process. Section 3 conducts empirical tests based on the analysis in Section 2. Section 4 examines the combined effects between multiple metrics and order flow imbalance, and provides an integrated model recommendation. Section 5 investigates changes in the market regime and proposes criteria for evaluating the effectiveness of microstructure metrics. Sections 6 and 7 outline future research directions and present conclusions, respectively.

\section{Model specification}
\subsection{Data and several major metrics}
In \cite{cont1}, the $OFI$ (Order Flow Imbalance) and $TI$ (Trade Imbalance) metrics were highlighted as key subjects of investigation. We now summarize their definitions as follows. Additionally, we analyze other metrics for comparative purposes alongside $OFI$ and $TI$. Based on the multi-metric analysis presented in \cite{sahalia1}, we further select several additional matrics along with their first-order lagged terms. These metrics are formally defined in this section, and their empirical performance will be tested in subsequent analyses.

\subsubsection{Data}
To align with subsequent metric definitions, we first provide a brief explanation of the CSI 300 index futures data. The exchange disseminates the finest-grained market data for CSI 300 index futures in the form of snapshot updates every 500 milliseconds. A market tick is generated and transmitted whenever a trade occurs or there are changes in price or quantity of the limit order book within this 500-millisecond window. Each tick contains the latest transaction price, trading volume within the 500ms interval, and the current state of the limit order book. Since these updates are not triggered by every individual market change, all references to 'tick data' hereafter specifically denote these snapshot updates. Each market snapshot constitutes a trading event.

\subsubsection{OFI}
For the reader's convenience, we reproduce the definition from \cite{cont1} below as follows.
Events affecting the order book occur randomly at times $\tau_n$. The number of trading events occurring in the interval $\left[0,t\right]$ is defined as $N\left(t\right)=\max\left\{n\mid\tau_n\le t\right\}$. The order flow imbalance metric over the time interval $\left[t_{k-1},t_k\right]$ is defined as the sum of contributions $e_n$ from all order flow events in this interval:
\begin{equation}
    OFI_k=\sum_{n=N\left(t_{k-1}\right)+1}^{N\left(t_k\right)}e_n
\end{equation}
The order flow event contribution $e_n$ reflects the state of supply-demand imbalance in the market. A positive value generally favors long positions, while a negative value benefits short positions. Next, we describe the definition. The best bid price and quantity represent demand, whereas the best ask price and quantity represent supply. Let $n$ denote the timestamp of the current market snapshot and $n-1$ the previous one. We compare four variables between these two timestamps:
\begin{itemize}
    \item $p_{n-1}^B$, $q_{n-1}^B$, $p_{n-1}^A$, $q_{n-1}^A$ represent the best bid price, bid quantity, best ask price, and ask quantity at time $n-1$.
    \item $p_n^B$, $q_n^B$, $p_n^A$, $q_n^A$ represent the corresponding variables at time $n$.
\end{itemize}
The following relationships hold:
\begin{itemize}
    \item If $p_n^B > p_{n-1}^B$ or $q_n^B > q_{n-1}^B$, it indicates \textbf{increasing demand};
    \item If $p_n^B < p_{n-1}^B$ or $q_n^B < q_{n-1}^B$, it indicates \textbf{decreasing demand};
    \item If $p_n^A < p_{n-1}^A$ or $q_n^A > q_{n-1}^A$, it indicates \textbf{increasing supply};
    \item If $p_n^A > p_{n-1}^A$ or $q_n^A < q_{n-1}^A$, it indicates \textbf{decreasing supply};
\end{itemize}
The order flow event contribution $e_n$ is defined as a function of the current and previous market snapshots. Here, $I_{\{T/F\}}$ denotes an indicator function that equals 1 when the subscript condition in the brackets is true and 0 otherwise:
\begin{equation}
e_n=I_{\{p_n^B\geq p_{n-1}^B\}}q_n^B-I_{\{p_n^B\le p_{n-1}^B\}}q_{n-1}^B-I_{\{p_n^A\le p_{n-1}^A\}}q_{n-1}^A+I_{\{p_n^A\geq p_{n-1}^A\}}q_{n-1}^A
\end{equation}

\subsubsection{TI}
Similarly, for the reader's convenience, we reproduce the definition of trade imbalance from \cite{cont1} below as follows. The trade imbalance metric over the time interval $\left[t_{k-1},t_k\right]$ is defined as the sum of all trade event contributions $\omega_n$ within this interval:
\begin{equation}
    TI_k=\sum_{n=N\left(t_{k-1}\right)+1}^{N\left(t_k\right)}\omega_n
\end{equation}
The trade event contribution $\omega_n$ represents the state of trade imbalance in the market. We summarize its formulation from \cite{cont1} as follows:
The net trading volume within the latest 500-millisecond interval is obtained from the difference between the current ($V_n$) and previous ($V_{n-1}$) snapshot volumes. This volume can be categorized as either \textit{aggressive buys} or \textit{aggressive sells}:
\begin{itemize}
\item An \textit{aggressive buy} occurs when an eager buyer submits an order at a price $\geq$ the best ask price ($p^A_{n-1}$)
\item An \textit{aggressive sell} occurs when an eager seller submits an order at a price $\leq$ the best bid price ($p^B_{n-1}$)
\end{itemize}
While the actual execution mechanics involve more complexity, a commonly adopted approach is the Lee-Ready algorithm as follows. Let $M_p$ denote the current snapshot's bid-ask midpoint price:
\begin{equation}
    M_p=\frac{\left(P_n^A+P_n^B\right)}{2}
\end{equation}
Let $p_n^{last}$ and $p_{n-1}^{last}$ denote the transaction prices of the current and previous snapshots respectively, and $V_n$, $V_{n-1}$ their corresponding trading volumes. We define the incremental volume as $v = V_n - V_{n-1}$, and let $I_{\{T/F\}}$ be an indicator function that equals 1 when the subscript condition in brackets is true and 0 otherwise. Then the trade event contribution $\omega_n$ is given by:
\begin{equation}
    \omega_n=\left(I_{\{p_n^{last}>M_p\}}-I_{\{p_n^{last}<M_p\}}+I_{\{p_n^{last}=M_p \& p_n^{last}\geq p_{n-1}^{last}\}}-I_{\{p_n^{last}=M_p \& p_n^{last}<p_{n-1}^{last}\}}\right)v
\end{equation}

\subsubsection{Lambda}
For the reader's convenience, we reproduce the definition of Lambda metric from \cite{sahalia1} below. The Lambda metric over time interval $\left[t_{k-1},t_k\right]$ measures the price impact per unit traded volume. Let $p_k^{\max}$ and $p_k^{\min}$ denote the highest and lowest prices observed within $\left[t_{k-1},t_k\right]$, respectively. The Lambda metric is then defined as follows:
\begin{equation}
    Lambda_k=\frac{p_k^{max}-p_k^{min}}{N\left(t_k\right)-N\left(t_{k-1}\right)}
\end{equation}
where
\begin{align*}
p_k^{\max} = \max_{\substack{n \in \{N(t_{k-1})+1, \ldots, N(t_k)\}}} p_n \\
p_k^{\min} = \min_{\substack{n \in \{N(t_{k-1})+1, \ldots, N(t_k)\}}} p_n
\end{align*}

\subsubsection{AvgEn}
The $AvgEn$ metric is defined based on the $OFI^{avg}$ metric, where $OFI^{avg}$ represents the average value of the aforementioned $e_n$ over a continuous trading interval. $AvgEn$ is then defined as the differential of the $OFI_{avg}$ metric, specifically:
\begin{equation}
    AvgEn_k=OFI_{N\left(t_k\right)}^{avg}-OFI_{N\left(t_{k-1}\right)}^{avg}
\end{equation}
where
\begin{equation*}
    OFI_{N\left(t_k\right)}^{avg}=\frac{\sum_{n=1}^{N\left(t_k\right)}e_n}{N\left(t_k\right)}
\end{equation*}
Please note that in our analysis of market microstructure, market resting periods must be excluded. Additionally, the beginning and ending segments of each trading period should be omitted to ensure that microstructure metrics are computed using only adjacent active trading times. Trading events influenced by resting periods cannot be used in market microstructure indicator calculations. For example, in a two-hour morning trading session, both the initial and final periods must be discarded. Consequently, when a new trading period begins, $N(t_k)$ must be reset to count from 1.

\subsection{Price correlation and autocorrelation of the metrics}
\begin{figure}[htbp]
  \centering
  \includegraphics[width=0.8\textwidth]{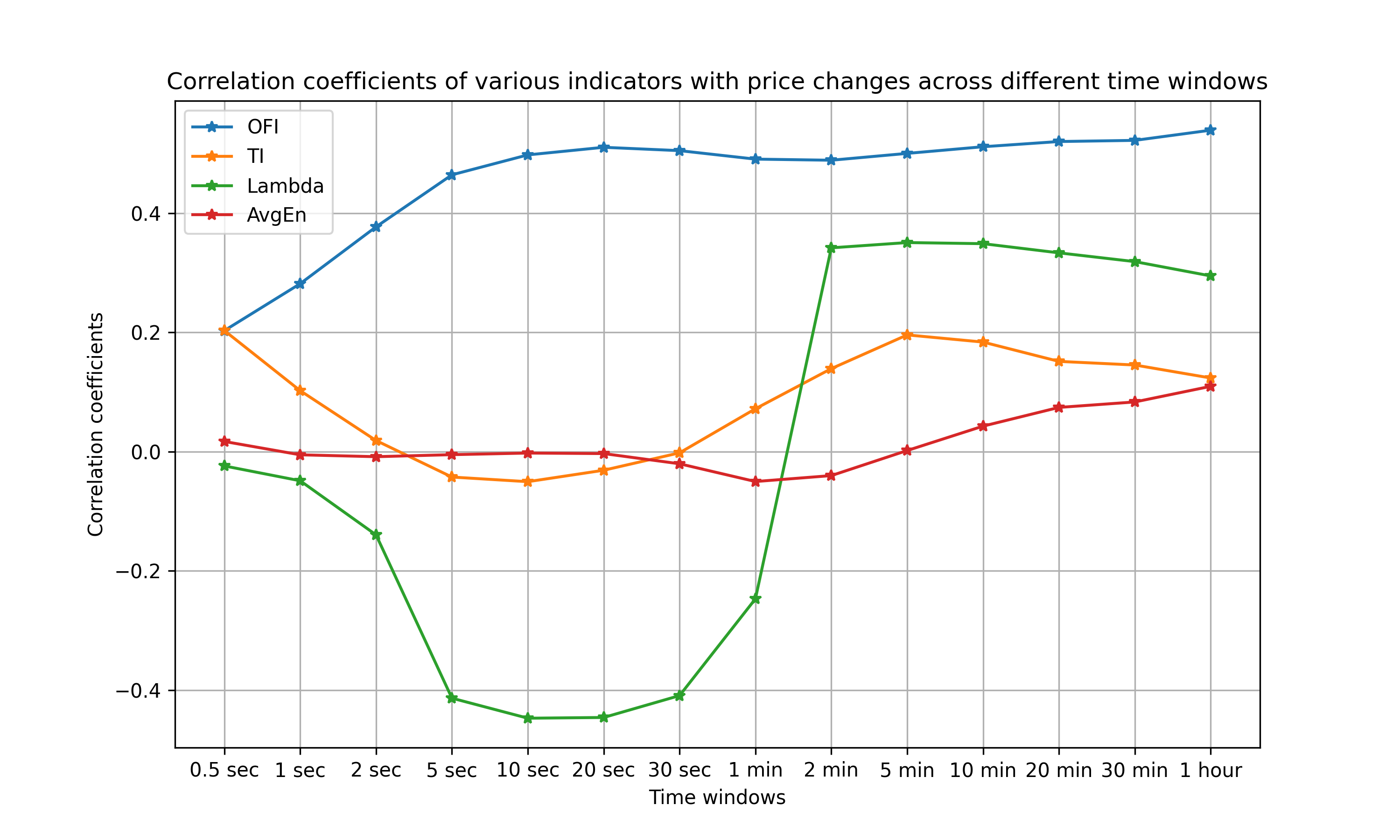} % 调整width控制大小
  \caption{Correlation coefficients between metrics and corresponding price changes within the windows} % 自动编号
  \label{fig:example} % 标签需放在caption之后
\end{figure}

All the following analysis is based on one year CSI 300 futures tick data, with total of 6 million ticks, except otherwise specified. We compute correlation coefficients between the values of the aforementioned four metrics across different time windows and their corresponding mid-price changes within those windows. The results are in Table ~\ref{tab:correlation} and visualized in Figure~\ref{fig:example} as follows:

\begin{table}[htbp]
  \centering
  {\footnotesize
  \begin{tabularx}{\textwidth}{|*{15}{X|}}  % 15列自动调整宽度
    \hline
     &0.5s&	1s&2s&	5s&	10s&	20s&	30s&	1m	&2m&	5m&	10m&	20m&	30m&	1h\\
    \hline
    OFI&	0.20 &	0.28 &	0.38 &	0.46 &	0.50 &	0.51 &	0.50 &	0.49 &	0.49& 	0.34& 	0.51 &	0.52 &	0.52 &	0.54 \\
    %\hline
    TI&	0.20 &	0.10 &	0.02 &	-0.04 &	-0.05 &	-0.03 &	0.00 &	0.07 &	0.14 &	0.20 &	0.18 &	0.15 &	0.15& 	0.12 \\
    Lmda &	-0.02& 	-0.05& 	-0.14& 	-0.41& 	-0.45& 	-0.45& 	-0.41& 	-0.25& 	0.34 &	0.35& 	0.35& 	0.33& 	0.32& 	0.29 \\
    AgEn&	0.02& 	-0.01& 	-0.01& 	-0.01& 	0.00& 	0.00& 	-0.02& 	-0.05& 	-0.04& 	0.00& 	0.04 &	0.07& 	0.08& 	0.11 \\
    \hline
    % 更多行...
  \end{tabularx}
  }
  \caption{Correlation coefficients between metrics and corresponding price changes within the windows}  % 标题在表格后
  \label{tab:correlation}  % 标签用于引用
\end{table}

Correlation analysis with mid-price changes reveals that the $OFI$ metric maintains stable coefficients throughout all time windows, persistently exceeding 0.5 for intervals beyond 5 seconds, demonstrating superior price representation efficacy. 

The $TI$ metric also exhibits some correlation coefficients, though it demonstrates weak negative correlations in shorter time windows (5-10 seconds). For intervals exceeding 2 minutes, however, it transitions to positive correlations. The $TI$ metric measures aggressive trading volume. One interpretation is that $TI$ captures executed aggressive trades, which in the short term exhaust all pending orders in one direction, resulting in negative price correlations. However, over extended time windows, aggressive trades begin to exhibit directional predictive power.

The $Lambda$ metric quantifies price change or volatility per unit traded volume. For shorter time windows (5-30 seconds), it shows negative correlation with midpoint price changes, while demonstrating strong positive correlation for intervals beyond 2 minutes. This indicates an asymmetric price pattern: gradual rises followed by sharp declines in short-term periods, but rapid rises with gradual declines in longer cycles - a finding empirically validated in CSI 300 index futures markets.

In contrast, the $AvgEn$ metric does not exhibit significant correlation characteristics. To summarize, since our primary objective is price change prediction, we conduct further analysis on the autocorrelation properties of both $OFI$ and $TI$ metrics, which demonstrate meaningful price correlations.

\begin{figure}[H]
  \centering
  \begin{minipage}[b]{0.48\textwidth}
    \centering
    \includegraphics[width=\linewidth]{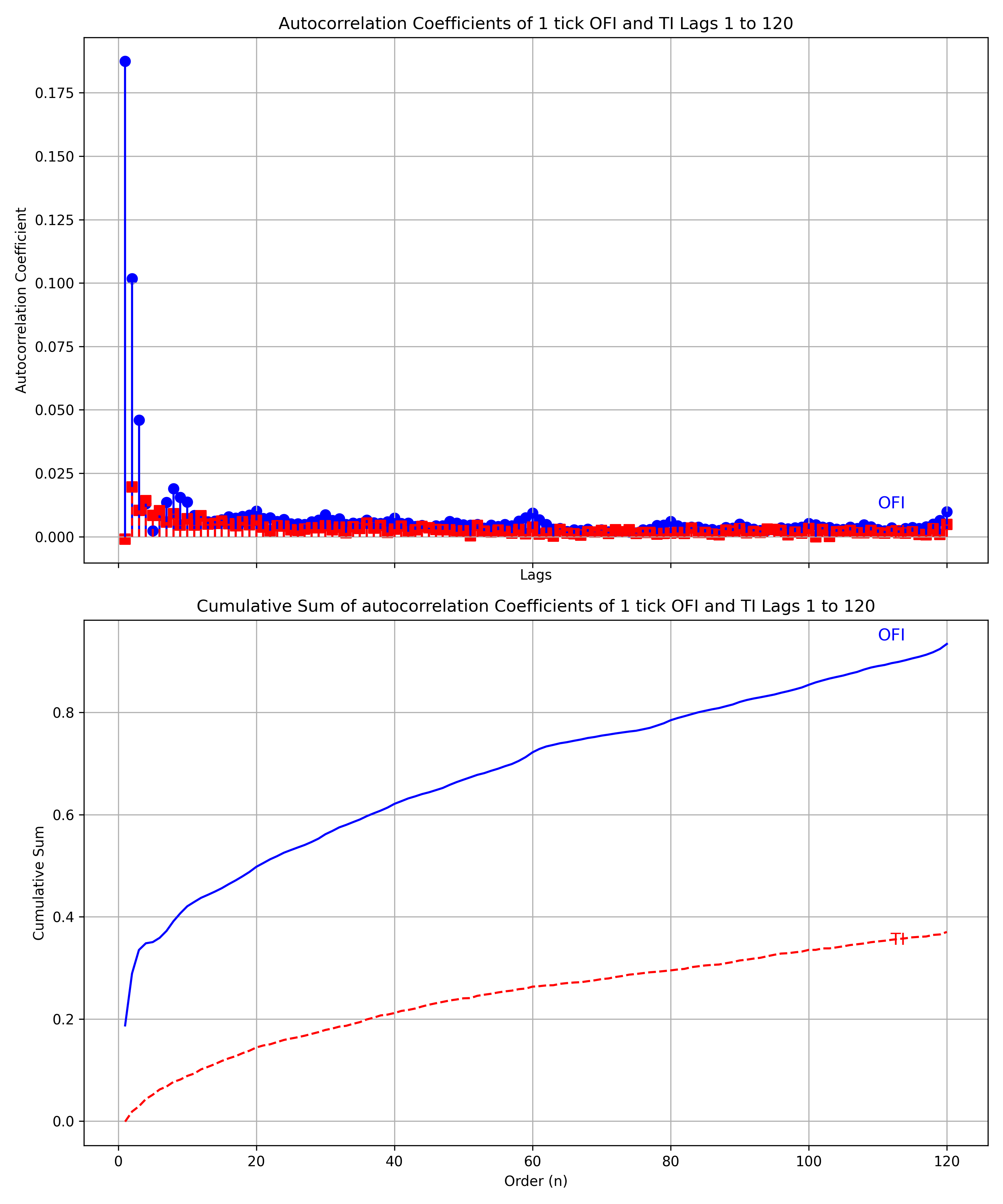}
    \caption*{(a) 1-120 tick lag}  % 使用caption*取消编号，手动添加(a)标签
  \end{minipage}
  \hfill
  \begin{minipage}[b]{0.48\textwidth}
    \centering
    \includegraphics[width=\linewidth]{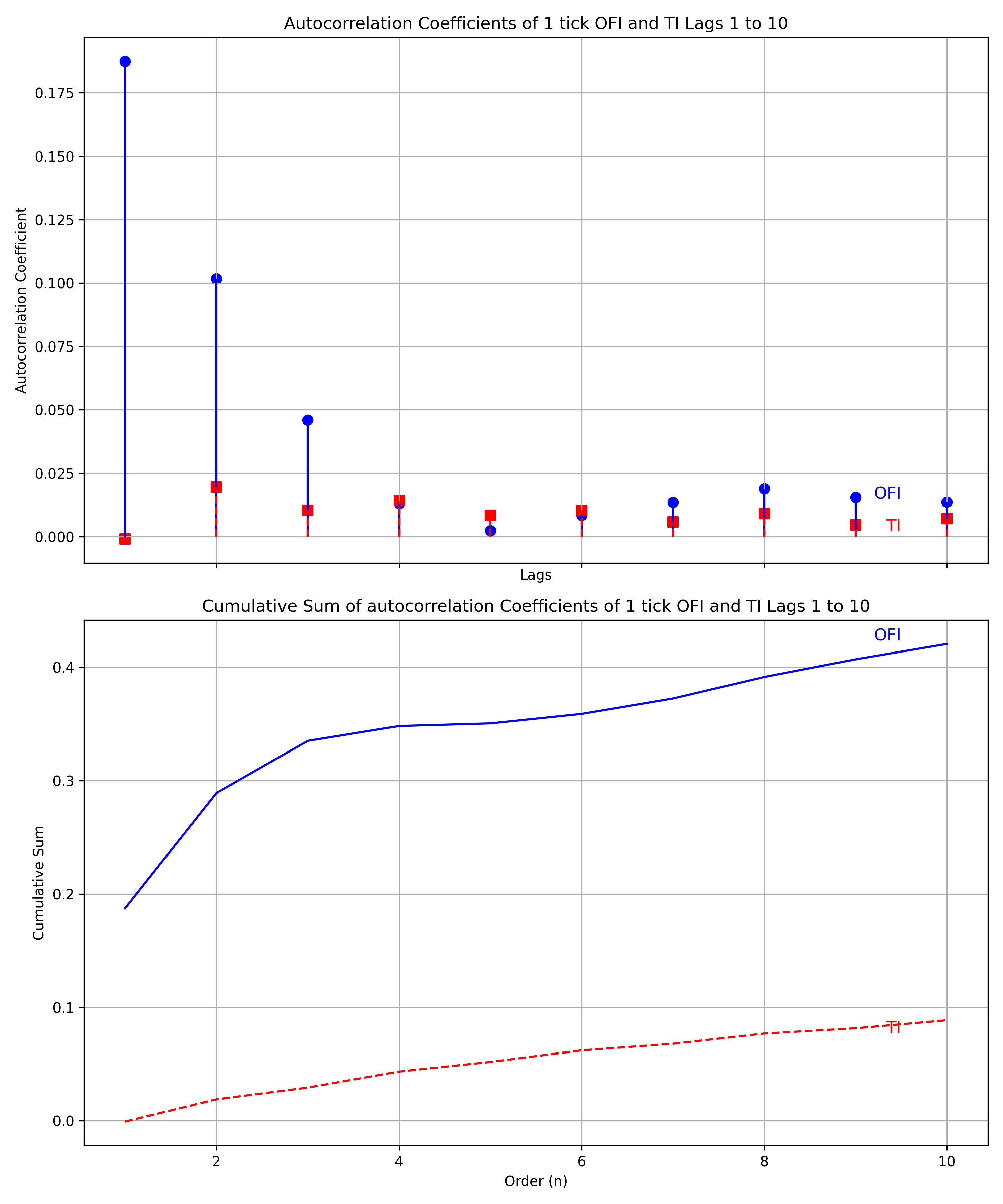}
    \caption*{(b) 1-10 tick lag}  % 使用caption*取消编号，手动添加(b)标签
  \end{minipage}
  \caption{One-year autocorrelation of $e_n$ and $\omega_n$ metrics, along with their cumulative autocorrelation profiles. (a) full-range delays 1-120 ticks lag (b) short-term detail view 1-10 ticks lag }
  \label{fig:autocorr_composite}
\end{figure}

We note that the finest granularity of the $OFI$ metric is calculated per order flow event as the $e_n$ measure (i.e., 1-tick $OFI$), while the $TI$ metric's minimum granularity is derived from the $\omega_n$ measure per trade event (i.e., 1-tick $TI$). We compute autocorrelation for these finest-granularity metrics. Furthermore, we cumulatively sum the autocorrelation coefficients starting from lag $1$ to preliminarily assess the duration of an indicator's memory effect. Figure~\ref{fig:autocorr_composite} presents both metrics' autocorrelations from 1-tick to 120-tick delays (where 1 tick = 500ms, corresponding to autocorrelations from 500ms to 1-minute intervals). The right panel of the figure provides an enlarged view of both the first 10 autocorrelations and their cumulative sums for detailed examination. 

% original begin{minipage}[b]{0.48\textwidth}
% original \includegraphics[width=\linewidth]{

% adjust \begin{minipage}[b]{0.45\textwidth}
% adjust \includegraphics[width=0.95\linewidth]{

%\begin{figure}[htbp] 

We perform analysis on one year of $e_n$ metric data and find that, at short time scales, the $e_n$ metric exhibits strong autocorrelation. This autocorrelation decays proportionally over time. When we cumulatively sum the autocorrelation, the resulting series displays gradual decay. This leads us to infer that the $e_n$ metric—and by extension, the $OFI$ constructed from it—may retain memory effects far beyond the initial tick interval. When a large $OFI$ is observed, the resulting price drift may persist for an extended period.

\begin{figure}[H]
  \centering
  % 上方的子图
  \begin{minipage}[t]{\textwidth}
    \centering
    \includegraphics[width=0.8\linewidth]{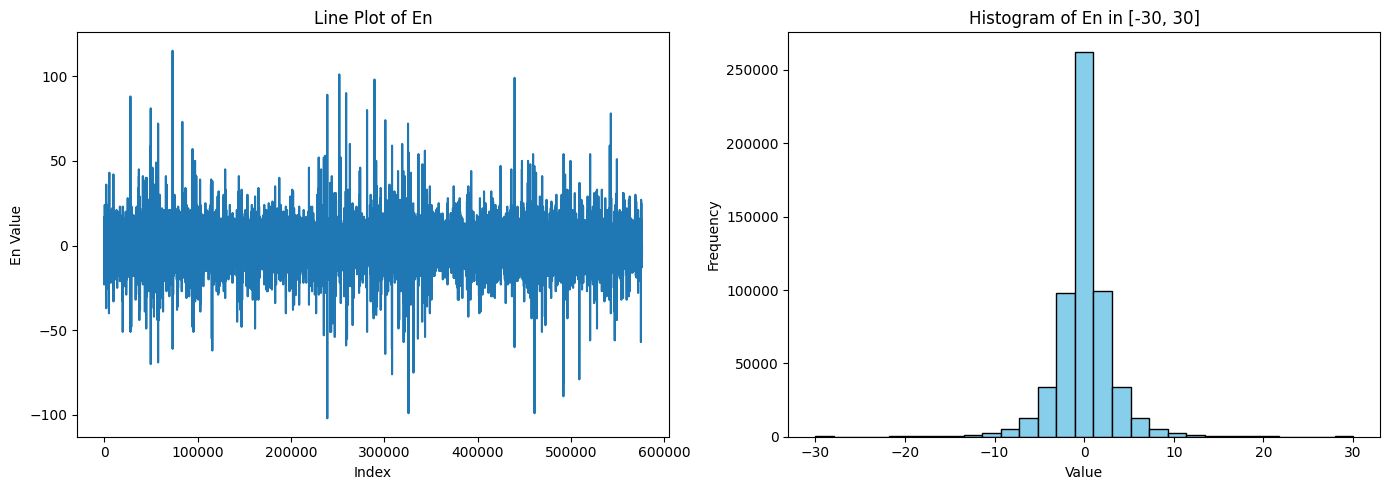}
    \caption*{(a) time series and histogram}  % 使用caption*取消编号，手动添加(a)标签
  \end{minipage}
  % 添加垂直间距
  \vspace{0.6cm} 
  % 下方的子图
  \begin{minipage}[t]{\textwidth}
    \centering
    \includegraphics[width=0.35\linewidth]{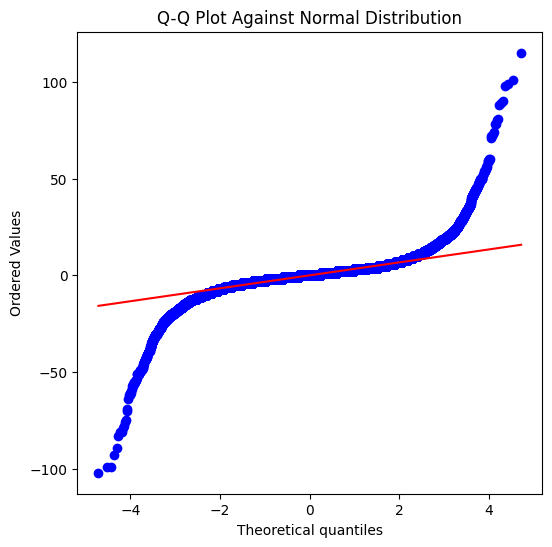}
    \caption*{(b) QQ plot}  % 使用caption*取消编号，手动添加(b)标签
  \end{minipage}
  \caption{$e_n$ metric of CSI 300 futures Nov 2024 contract  (a) Time series and histogram (b) QQ-plot}
  \label{fig:qqplot}
\end{figure}

Below, we present a tick-by-tick statistical analysis of one month of $e_n$ data. Figure~\ref{fig:qqplot} shows (left) the sequential $e_n$ values for the CSI 300 November 2024 contract and (right) its histogram, with Q-Q plots in the lower panel. Table~\ref{tab:en_statistics} provides summary statistics.
%\begin{figure}[htbp]

\begin{table}[htbp]
  \centering
  {\footnotesize
  \begin{tabularx}{\textwidth}{|*{4}{X|}}  % 15列自动调整宽度
    \hline
     Mean:&	-0.00506&	Std:&	3.59765\\
    \hline
    Skewness:&	-0.00807&	Kurtosis:&	34.14931 \\
    \hline
    % 更多行...
  \end{tabularx}
  }
  \caption{$e_n$ metric of CSI 300 futures Nov 2024 contract statistics}  % 标题在表格后
  \label{tab:en_statistics}  % 标签用于引用
\end{table}
The metric exhibits strong mean-reversion characteristics centered around zero, with symmetric but fat-tailed distributions and possible frequent jump-like extreme values. In the following section, we develop model to analyze the price response to this metric.

\subsection{Modeling analysis of the aggregate price drift impact from metric}
Based on the preceding analysis, we hypothesize that the $e_n$ sequence - and, by extension, the $OFI$ sequence formed by the summation of non-overlapping $e_n$ terms - probably follows an Ornstein-Uhlenbeck (O-U) process driven by a symmetric jump Lévy process. We assume the $OFI$ metric quantifies persistent market impacts, exhibiting stationary fluctuations around a long-run equilibrium. Four empirical regularities motivate this hypothesis:
\begin{enumerate}
  \item The impact of the $e_n$ metric sequence on prices may resemble a step-shock response. We observe that its second-order autocorrelation is approximately half of the first-order autocorrelation, while the third-order autocorrelation is again halved. The $e_n$ metric exhibits a proportionally decaying autocorrelation, a characteristic consistent with the dynamics of the O-U process.
  \item The $e_n$ metric process exhibits mean-reversion characteristics with a long-term equilibrium of zero. Both its time-series plot and histogram confirm stationarity around this mean—a behavior consistent with O-U process dynamics.
  \item The process exhibits near-zero skewness and excess kurtosis ($34\gg 3$), confirming symmetric leptokurtic characteristics. This is visually and quantitatively supported by Figure~\ref{fig:qqplot} and Table~\ref{tab:en_statistics}.
  \item In terms of market microstructure, the $e_n$ metric captures directional market momentum, comprising both limit order book provisions and aggressive market orders in that direction. We hypothesize that the aggregate external driving force has an upper bound over a given period. When partially realized, the residual momentum persists but decays gradually.
\end{enumerate}
We therefore propose that this process is an O-U process driven by a symmetric Lévy process. Based on this proposition, we now derive a series of theoretical results.

A discrete-time O-U process driven by a symmetric Lévy process can be expressed as:
\begin{equation}
    \Delta x_t=\theta\left(\mu-x_t\right)\Delta t+\sigma\sqrt{\Delta t}\epsilon_t+\eta\Delta J_t
\end{equation}
It can be shown that when the latter two terms constitute symmetric, zero-mean Wiener and Lévy processes, the $k-th$ order autocorrelation coefficient of the process—that is, the autocorrelation of $x_t$—satisfies:
\begin{equation*}
    \rho_k=e^{-\theta k\Delta t}
\end{equation*}
Here, $k$ denotes both the time-step difference and the order of autocorrelation. When we set $\Delta_t$ = 1, the $k-th$ order autocorrelation coefficient approximately follows:
\begin{equation}
    \rho_k=\left(1-\theta\right)^k
\end{equation}
Therefore, when we observe proportionally decaying autocorrelation coefficients, we may hypothesize that the metric follows the increments of an O-U process.

The preceding analysis is based on tick-level $e_n$ data, whereas the $OFI$ metric aggregates $e_n$ across multiple ticks. When $e_n$ exhibits proportionally decaying autocorrelation, it can be shown that the $OFI$—constructed as the sum of non-overlapping $e_n$ terms—also displays proportional autocorrelation decay. We define a new variable $y_t$ as the cumulative sum of $x_t$ increments over non-overlapping p-step intervals:
\begin{equation*}
    y_t=\sum_{i=1}^{p}x_{n\left(t-1\right)+i}
\end{equation*}
It can be shown that if the k-th order autocorrelation of $x_t$ follows $\rho_k=e^{-\theta k\Delta t}$, then the $k-th$ order autocorrelation of $y_t$ satisfies:
\begin{equation}
    \rho_k^{\left(y\right)}=e^{-\theta pk\Delta t}
\end{equation}
This implies that the scale-invariance of autocorrelation decay persists under $OFI$ aggregation, suggesting O-U dynamics hold for arbitrary summation windows.

Building on the previous section's findings—where the $OFI$ metric exhibits strong and stable correlations with contemporaneous price changes—we hypothesize that a significant $OFI$ variation can be modeled as a short-term shock, inducing a prolonged price drift response. We now formalize this price drift process based on the above assumptions.

\begin{figure}[htbp]
  \centering
  \includegraphics[width=0.8\textwidth]{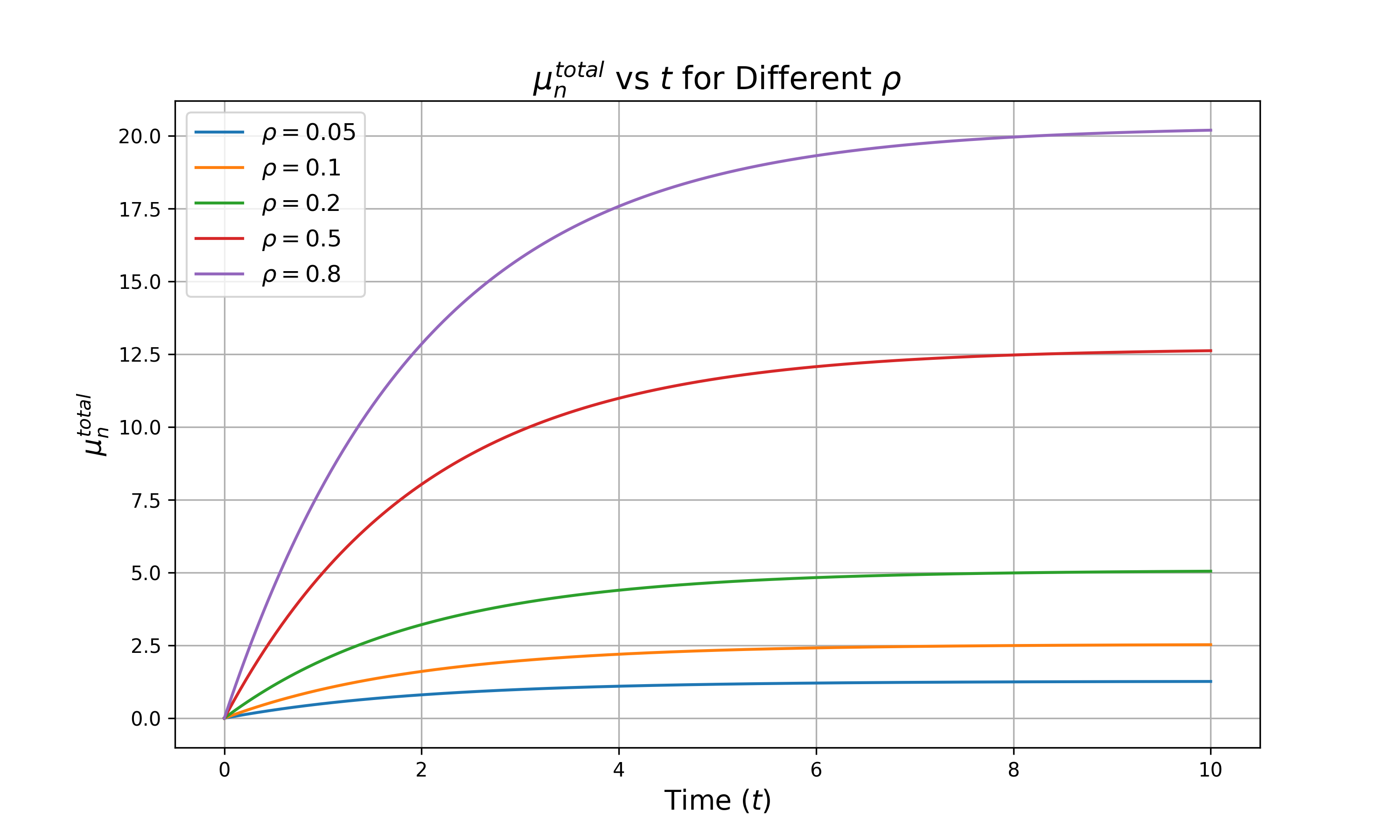} % 调整width控制大小
  \caption{Total impact of metrics on price drift under different correlation coefficients described by the O-U process in the model} % 自动编号
  \label{fig:mu_tot} % 标签需放在caption之后
\end{figure}

For the discrete-time O-U process driven by a Lévy process as in the equation (8), we seek to derive the formula for the discrete sum $S_n=\sum_{t=0}^{n}x_t$, in order to quantify the aggregate price impact of $OFI$ shocks. When focusing on the expected value of $S_n$, we may disregard the stochastic terms $\sigma\sqrt{\Delta t}\epsilon_t$ and $\eta\Delta J_t$, as established previously, these represent symmetric Wiener and Lévy processes with zero expectation. Thus, the expectation of $S_n$ can be approximated solely through the mean-reverting component. For this mean-reversion term, we obtain:
\begin{equation*}
\mathrm{E}\left[x_{t+1}\right]=\mathrm{E}\left[x_t\right]+\theta\left(\mu-\mathrm{E}\left[x_t\right]\right)\Delta t
\end{equation*}
Let $\mathrm{E}\left[x_t\right]=m_t$, then the above equation transforms to:
\begin{equation*}
    m_{t+1}=m_t+\theta\left(\mu-m_t\right)\Delta t
\end{equation*}
This constitutes a first-order linear difference equation, whose solution takes the form:
\begin{equation*}
    m_t=\mu+\left(m_0-\mu\right)e^{-\theta t}
\end{equation*}
Here, \( m_0 = \mathrm{E}\left[x_0\right] \) represents the initial expected value—precisely the magnitude of the $OFI$ step shock. We now compute the expectation of \( S_n \). Since \( m_t \) gives \( \mathrm{E}\left[x_t\right] \), the expected cumulative impact \( \mathrm{E}\left[S_n\right] \) is derived as:
\begin{equation*}
E\left[S_n\right]=\sum_{t=0}^{n}E\left[x_t\right]=\sum_{t=0}^{n}m_t
\end{equation*}
Substituting the expression for \( m_t \):
\begin{align*}
    \ E\left[S_n\right]=\sum_{t=0}^{n}\left[\mu+\left(m_0-\mu\right)e^{-\theta t}\right]\ \\
    E\left[S_n\right]=\sum_{t=0}^{n}\mu+\left(m_0-\mu\right)\sum_{t=0}^{n}e^{-\theta t}\ \\
    \ E\left[S_n\right]=\left(n+1\right)\mu+\left(m_0-\mu\right)\sum_{t=0}^{n}e^{-\theta t}
\end{align*}
Here, the summation \( \sum_{t=0}^{n} e^{-\theta t} \) represents a geometric series, which simplifies to:
\begin{equation*}
    \sum_{t=0}^{n}e^{-\theta t}=\frac{1-e^{-\theta\left(n+1\right)}}{1-e^{-\theta}}
\end{equation*}
Therefore:
\begin{equation*}
    E\left[S_n\right]=\left(n+1\right)\mu+\left(m_0-\mu\right)\frac{1-e^{-\theta\left(n+1\right)}}{1-e^{-\theta}}
\end{equation*}
Given the empirical results of zero long-run average of \(OFI_k\), we have \( \mu = 0 \) and initial shock \( m_0 = OFI \), the model reduces to:
\begin{equation}
    E\left[S_n\right]=OFI\frac{1-e^{-\theta\left(n+1\right)}}{1-e^{-\theta}}\ \ 
\end{equation}
To quantify the aggregate price impact of the \(OFI\) metric, the actual contribution to price drift requires an additional transformation. Given our observation of stable and significant correlation (\( \rho \)) between the metric and price changes, we posit that the total contribution to price drift scales proportionally with \( \rho \). Formally, defining \( \mu^{total} \) as $OFI$'s aggregate price drift contribution, we obtain:
\begin{equation*}
    \mu_n^{total}=\rho kE\left[S_n\right\}
\end{equation*}
After replacing \(n\) with \(t\), we have:
\begin{equation}
    \mu_t^{total}=OFI\rho k\frac{1-e^{-\theta t}}{1-e^{-\theta}}
\end{equation}
The above formulation captures the aggregate price drift—or equivalently, the maximum drift impact observable following a significant $OFI$ shock. With our observed empirical autocorrelation decay rate \( \theta = 0.5 \), Figure~\ref{fig:mu_tot} demonstrates the total price drift impact under varying correlation coefficients \( \rho \) in the O-U framework. Notably, Figure~\ref{fig:mu_tot} effectively mirrors the cumulative autocorrelation effects shown in the lower panel of Figure~\ref{fig:autocorr_composite}.

The influence of different metrics varies substantially due to their correlation disparities—for instance, $OFI$ exhibits significantly stronger long-term cumulative impact than $TI$, a finding robustly validated across multiple empirical tests. This aligns with both theoretical and empirical verifications in \cite{cont1}.

In summary, the expected value of the discrete sum of the discrete-time O-U process, \( E\left[S_n\right] \), conforms to the above equation. Note that this is only the expectation of \( S_n \) and does not include the effects of stochastic fluctuations. The actual \( S_n \) will fluctuate around this expected value, depending on the influence of the stochastic term \( \sigma\sqrt{\Delta t}\epsilon_t + \eta\Delta J_t \). When the time horizon is sufficiently long, i.e., \( n \rightarrow \infty \), we have:
\begin{equation*}
    E\left[S_\infty\right]=OFI\frac{1}{1-e^{-\theta}}
\end{equation*}
Therefore:
\begin{equation*}
    \mu_\infty^{total}=OFI\rho k\frac{1}{1-e^{-\theta}}
\end{equation*}
i.e., it converges to a constant value. The above equation summarizes the aggregate impact of $OFI$ as a jump shock on price. Thus, it can be observed that even the accumulation of short-term $OFI$ may exert a prolonged effect. Therefore, when we compute the $OFI$ over a given time period, the subsequent price drift—i.e., the price change—may persist across varying window lengths.

However, the above calculation only accounts for the contribution to price drift. Actual price changes are affected by volatility. If we model price changes as geometric Brownian motion, then physically, the second-order effect of random fluctuations in Brownian motion (i.e., volatility) reduces long-term average returns because noise diffusion causes the geometric mean of asset prices to be lower than the arithmetic mean. Through Itô's Lemma, when transforming the geometric Brownian motion of asset prices to log prices, an additional drift term \( -\frac{\sigma^2}{2} \) emerges. This term originates from the second-order variation of Brownian motion. In the next subsection, we analyze how incorporating volatility effects influences the metric's impact on both the logarithmic return average and standard deviation process of prices, and conduct modeling of an new indicator, the Quasi-Sharpe ratio.

\subsection{A model of logarithmic-return dynamics and risk-adjusted performance}
We proceed to derive the asset price log-return process, its mean and variance process when $OFI$ acts as a shock, thereby obtaining a quasi-Sharpe ratio process. We first assume the asset follows geometric Brownian motion. Below we briefly outline geometric Brownian motion (GBM). $GBM$ is a stochastic process widely used in financial modeling, whose dynamics are described by the following stochastic differential equation (SDE):
\begin{equation*}
    dS_t=\mu S_t dt+\sigma S_t dW_t
\end{equation*}
Where $S_t$ is the asset price, $\mu$ is the drift term, or the asset price average change rate, $\sigma$ is the volatility, calculated as the return standard deviation; $W_t$ is the standard Brownian Motion or the Wiener process. Consider the mean and standard deviation of the logarithmic return of the asset is \( R_t = \ln{\left(\frac{S_t}{S_0}\right)} \), then:
\begin{equation}
    R_t=\left(\mu-\frac{\sigma^2}{2}\right)t+\sigma W_t
\end{equation}
Then, the expectation is:
\begin{equation*}
    E\left[R_t\right]=E\left[\left(\mu-\frac{\sigma^2}{2}\right)t+\sigma W_t\right]=\left(\mu-\frac{\sigma^2}{2}\right)t
\end{equation*}
To compute the standard deviation of the return rate, since \( \sigma W_t \) follows a normal distribution and \( W_t \sim \mathcal{N}\left(0, t \right) \), its variance is \( \sigma^2 t \). Therefore, the standard deviation is given by \( \mathrm{Std}\left(R_t\right) = \sigma \sqrt{t} \).

We now proceed to model the impact of the $OFI$ metric. Consider a shock induced by changes in the $OFI$ and its effect on the geometric Brownian motion described above. 
\begin{enumerate}
\item Based on prior analysis of the shock's characteristics, we infer that $OFI$ process follows an O-U process driven by a symmetric Lévy process. 
\item From empirical results, the $OFI$ maintains a relatively stable correlation coefficient with the contemporaneous mid-price changes across any time window. Therefore, we model the metric's influence on price by linearly transforming the metric process into the drift rate \( \mu \) of the geometric Brownian motion.
\end{enumerate}
 As a result, when the metric follows the aforementioned O-U process, we assume the resulting drift term also adheres to this O-U process. That is, the drift rate \( \mu_t \) follows an O-U process driven by a Lévy process. By definition of the process, we have:
\begin{equation*}
    d\mu_t=\theta\left(\mu_l-\mu_t\right) dt+dL_t
\end{equation*}
Therefore, Our model is described by the following system of stochastic differential equations:
\begin{subequations}
\begin{empheq}[left=\empheqlbrace]{align}
d S_t &= \mu_t S_t dt + \sigma S_t dW_t \label{eq:price} \\
d\mu_t &= \theta\left(\mu_l-\mu_t\right) dt+dL_t \label{eq:drift}
\end{empheq}
\end{subequations}
Where:
\begin{itemize}
    \item $\mu_t$: The time-varying drift rate.
    \item $\mu_l$: The long-run average of the drift rate (mean reversion level).
    \item $\theta$: The mean reversion speed.
    \item \( L_t \): A zero-mean, finite second-moment Lévy process with symmetric distribution.
\end{itemize}
It should be noted that \( W_t \) and \( L_t \) may exhibit correlation, as the stochastic perturbations in prices could potentially represent the combined effects of $OFI$-related disturbances along with other perturbations.

Below, based on the characteristics of financial time series and empirical results, we assume the long-term mean of this O-U process (i.e., $OFI$) to be zero (\( \mu_l=0 \)). We then derive the mean and standard deviation of the logarithmic return process when subjected to an initial drift rate \( \mu_0 \). From these derived moments of returns, we can obtain the Quasi-Sharpe ratio and its temporal evolution. Given this initial drift effect \( \mu_0 \), and considering the stable long-term correlation between prices and $OFI$ established in previous analysis, we have:
\begin{equation*}
    \mu_0=OFI_0\rho k
\end{equation*}
That is, the initial shock to the drift is proportional to both the indicator's initial value and its correlation coefficient with price. Assuming the indicator's long-term mean is zero (i.e., when \( \mu_l=0 \)), the above drift rate dynamics formula becomes:
\begin{equation*}
    d\mu_t=-\theta\mu_t dt+dL_t
\end{equation*}
Given the initial condition \( \mu_t = \mu_0 \), an explicit solution exists for the above equation:
\begin{equation}
    \mu_t=\mu_0e^{-\theta t}+\int_{0}^{t}{e^{-\theta\left(t-s\right)}dL_s} \label{eq:mu_t_solution}
\end{equation}
Where:
\begin{itemize}
    \item \( \mu_0 e^{-\theta t} \): The mean-reverting portion of the drift rate dynamics.
    \item \( \int_{0}^{t}{e^{-\theta\left(t-s\right)}dL_s} \): The cumulative contribution of the Lévy process \( L_t \), exhibiting characteristics of a stationary process.
\end{itemize}

We first analyze the drift term mean and variance directly. The first term in \eqref{eq:mu_t_solution} is deterministic, while the second term represents the cumulative contribution of the Lévy process \( L_t \). Since \( L_t \) is a Lévy process with long-term zero mean, i.e.:
\begin{equation*}
    E\left[L_t\right]=0,\ \forall t\geq0
\end{equation*}
Moreover, since the increments of the Lévy process are independent, for any integrable function \( f\left(u\right) \), we have:
\begin{equation}
    E\left[\int_{0}^{t}{f\left(u\right)dL_u}\right]=\int_{0}^{t}f\left(u\right)E\left[dL_u\right]=0
\end{equation}
Thus, the mean of the second term in aforementioned \eqref{eq:mu_t_solution} equals zero for all \( t \). It follows that the expectation of the drift rate becomes:
\begin{equation}
    E\left[\mu_t\right]=\mu_0e^{-\theta t} \label{eq:drift_mean}
\end{equation}
The calculated variance of the drift rate is given by following expression, with proof in Appendix~\ref{app:mut_var}:
\begin{equation}
    \mathrm{Var}\left(\mu_t\right)=\frac{\sigma_L^2}{2\theta}\left(1-e^{-2\theta t}\right)
\end{equation}
Here, \( \sigma_L^2 = E\left[L_t^2\right]/t \) represents the per-unit-time variance of the Lévy process. Since we know this Lévy process has finite second moments, the above assumption is justified.

\begin{figure}[htbp]
  \centering
  \includegraphics[width=0.8\textwidth]{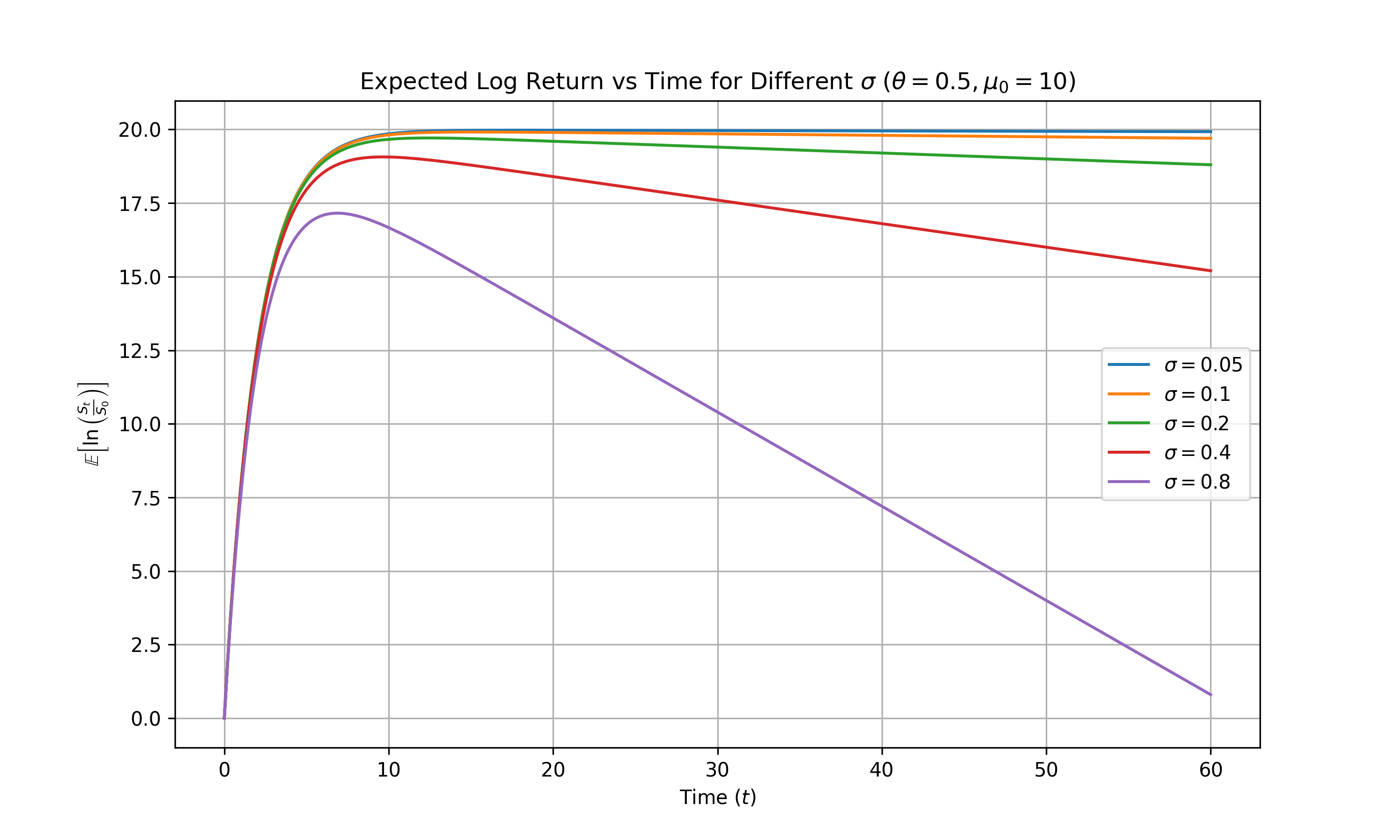} % 调整width控制大小
  \caption{Model-predicted dynamics of expected log-returns across varying price volatility levels} % 自动编号
  \label{fig:rt_mean} % 标签需放在caption之后
\end{figure}

We proceed to compute the expectation and variance processes of the logarithmic returns after coupling the two equations. The asset return is defined as \( R_t = \ln\left(\frac{S_t}{S_0}\right) \). Integrating the geometric Brownian motion yields:
\begin{equation}
    R_t=\int_{0}^{t}\mu_s ds-\frac{\sigma^2}{2}t+\sigma W_t \label{eq:rt_solution}
\end{equation}
Therefore, the expected logarithmic return evolves over time as:
\begin{equation*}
    E\left[R_t\right]=E\left[\int_{0}^{t}\mu_s d s\right]-\frac{\sigma^2}{2}t
\end{equation*}
Beginning with the left-hand integral term, we substitute the expectation of the drift rate obtained earlier in \eqref{eq:drift_mean}:
\begin{equation*}
    E\left[\int_{0}^{t}\mu_s d\ s\right]=\int_{0}^{t}E\left[\mu_s\right] ds=\int_{0}^{t}{\mu_0e^{-\theta s}} ds=\mu_0\frac{1-e^{-\theta t}}{\theta}
\end{equation*}
Therefore, the time-dependent expectation of the logarithmic return is given by:
\begin{equation}
    E\left[R_t\right]=\mu_0\frac{1-e^{-\theta t}}{\theta}-\frac{\sigma^2}{2}t \label{eq:logret_exp}
\end{equation}
The equation is visualized in Figure \ref{fig:rt_mean}, illustrating the time-dependent behavior of logarithmic return expectations across varying volatility parameters. The initial value \( \mu_0 = 10 \) is chosen to emphasize its influence.
The return relationship shown in the above figure is empirically validated in \cite{shen1} (pp. 15, Figure 2.7: 'Mean daily profit for various forecast windows').

We further derive the expression for logarithmic returns. Based on the log-return formula \( R_t = \int_{0}^{t}\mu_s\,ds - \frac{\sigma^2}{2}t + \sigma W_t \), we substitute the closed-form solution of \( \mu_t \) in \eqref{eq:mu_t_solution} into it. The integral component evaluates to:
\begin{equation*}
    \int_{0}^{t}{\mu_sds}=\frac{\mu_0}{\theta}\left(1-e^{-\theta t}\right)+\frac{1}{\theta}\int_{0}^{t}{\left(1-e^{-\theta\left(t-u\right)}\right)dL_u}
\end{equation*}
Therefore, the expression for the logarithmic return is given by:
\begin{equation}
    \ln{\left(\frac{S_t}{S_0}\right)}=\frac{\mu_0}{\theta}\left(1-e^{-\theta t}\right)+\frac{1}{\theta}\int_{0}^{t}{\left(1-e^{-\theta\left(t-u\right)}\right)dL_u}-\frac{1}{2}\sigma^2t+\sigma\ W_t  \label{eq:rt_result}
\end{equation}

\begin{figure}[htbp]
  \centering
  \includegraphics[width=0.8\textwidth]{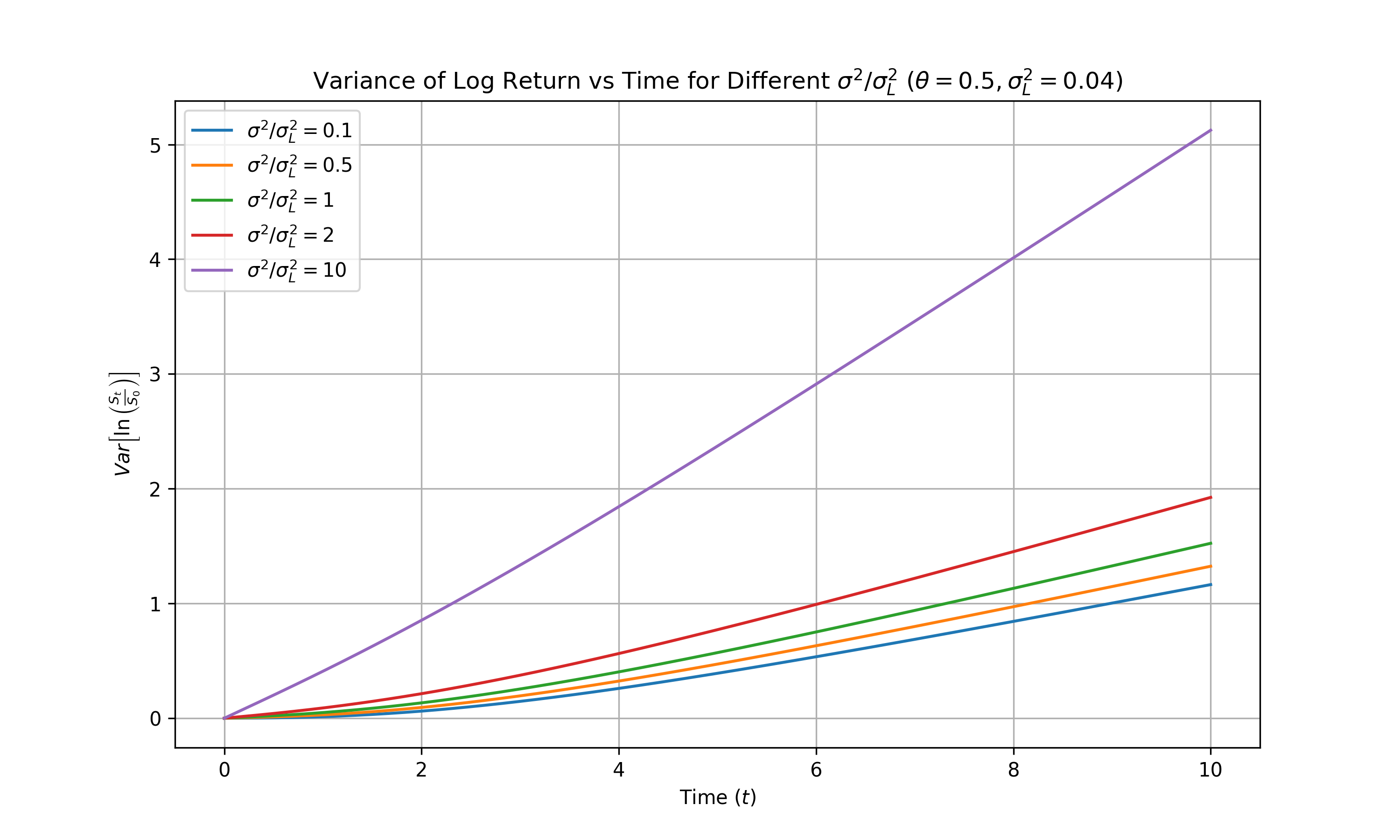} % 调整width控制大小
  \caption{Temporal evolution of total variance under different Brownian-to-Lévy variance ratios in the model} % 自动编号
  \label{fig:logrt_var} % 标签需放在caption之后
\end{figure}

We now proceed to compute the variance of the returns. Let us restate the following assumptions:
\begin{enumerate}
    \item We assume \( \mu_s \) is a Lévy-driven O-U process with mean 0, symmetric distribution, finite variance, and increments \( \sigma_L^2 \).
    \item The drift's Lévy process \( L_t \) and GBM's Wiener process \( W_t \) are independent.
\end{enumerate}
Under the above assumptions, the time-dependent variance of the logarithmic return is derived as follows (see Appendix~\ref{app:logarithmic_ret_var} for detailed derivation):
\begin{equation}
    \mathrm{Var}\left[\ln{\left(\frac{S_t}{S_0}\right)}\right]=\sigma^2t+\frac{\sigma_L^2}{\theta^2}\left[t-\frac{2}{\theta}\left(1-e^{-\theta t}\right)+\frac{1}{2\theta}\left(1-e^{-2\theta t}\right)\right]  \label{eq:logret_var}
\end{equation}
Figure \ref{fig:logrt_var} illustrates the temporal evolution of the logarithmic return variance. The magnitude of total variance variation depends on the ratio between the Brownian motion variance and Lévy process variance. Holding the Lévy variance constant, we plot curves for different ratios.

The standard deviation of logarithmic returns is the square root of the variance, yielding:
\begin{equation}
    \mathrm{Std}\left(R_t\right)=\sqrt{\sigma^2t+\frac{\sigma_L^2}{\theta^2}\left(t-\frac{2}{\theta}\left(1-e^{-\theta t}\right)+\frac{1}{2\theta}\left(1-e^{-2\theta t}\right)\right)}
\end{equation}
In summary, we obtain the following key results:
\begin{itemize}
    \item Expected logarithmic return over time:
    \begin{equation*}
        E\left[R_t\right]=\mu_0\frac{1-e^{-\theta t}}{\theta}-\frac{\sigma^2}{2}t
    \end{equation*}
    \item Standard deviation of logarithmic returns over time:
    \begin{equation*}
    \mathrm{Std}\left(R_t\right)=\sqrt{\sigma^2t+\frac{\sigma_L^2}{\theta^2}\left(t-\frac{2}{\theta}\left(1-e^{-\theta t}\right)+\frac{1}{2\theta}\left(1-e^{-2\theta t}\right)\right)}
    \end{equation*}
\end{itemize}

Referencing the Sharpe ratio definition, we introduce a quasi-Sharpe ratio by neglecting the risk-free rate and using logarithmic returns. This quasi-Sharpe ratio (or response ratio) is defined as the ratio of expected log-return to log-return standard deviation. Its time evolution is given by:
\begin{equation}
\mathrm{QuasiSharpe}\left(t\right)=\mathrm{\mathrm{QS}}\left(t\right)=\frac{\mu_0\cdot\frac{1-e^{-\theta t}}{\theta}-\frac{1}{2}\sigma^2t}{\sqrt{\sigma^2t+\frac{\sigma_L^2}{\theta^2}\left(t-\frac{2}{\theta}\left(1-e^{-\theta t}\right)+\frac{1}{2\theta}\left(1-e^{-2\theta t}\right)\right)}} \label{eq:logret_qs}
\end{equation}
The following Figure \ref{fig:quasiQQ} displays the temporal evolution of the quasi-Sharpe ratio (response ratio) under 
\begin{itemize}
\item Varying \( \sigma^2/\sigma_L^2 \) configurations
\item Two initial conditions: \( \mu_0=10 \) (strong drift) vs \( \mu_0=0.1 \) (weak drift)
\item Fixed Lévy variance \( \sigma_L^2 = \text{0.04} \)
\end{itemize}

%\begin{figure}[htbp]
\begin{figure}[H] 
  \centering
  \begin{minipage}[b]{0.48\textwidth}
    \centering
    \includegraphics[width=\linewidth]{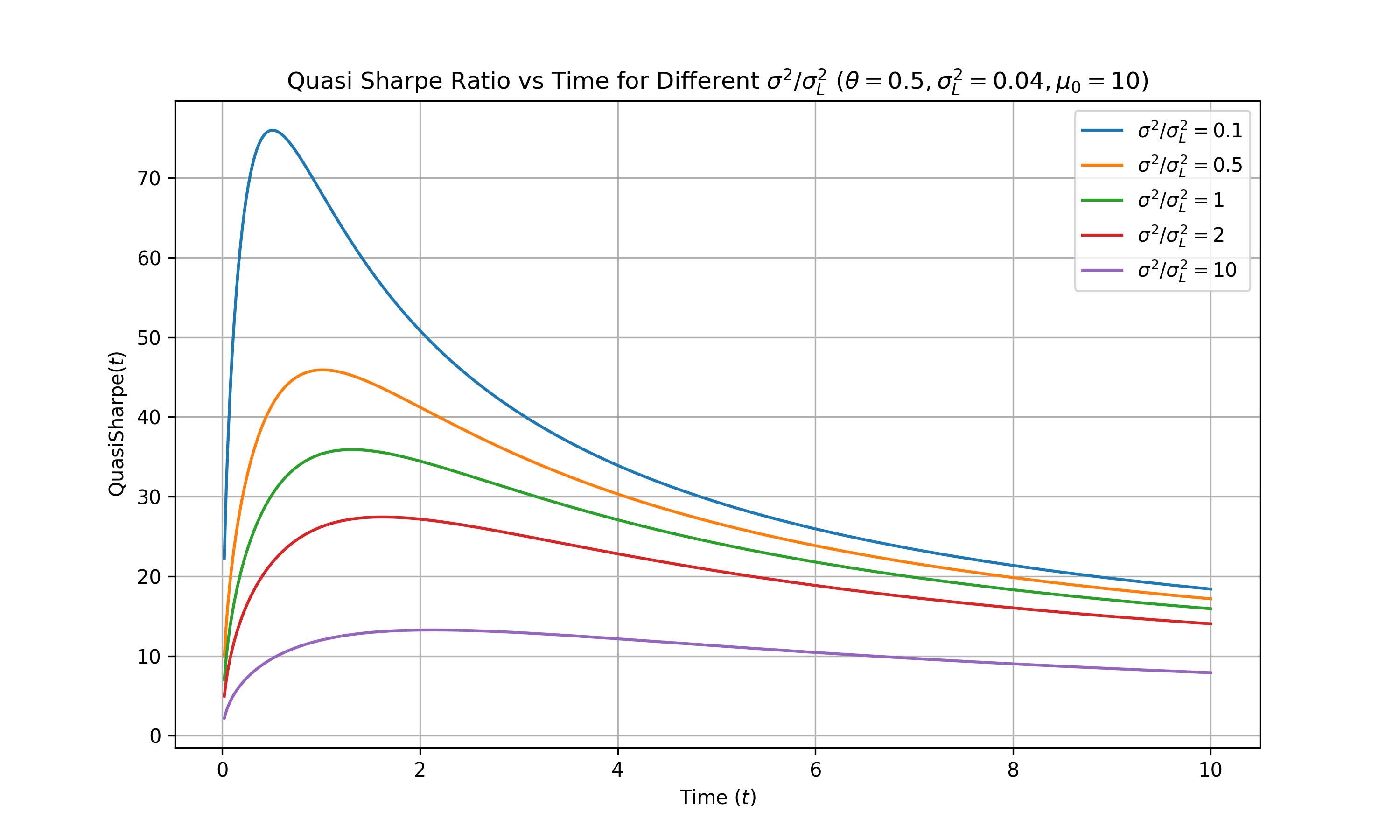}
    \caption*{(a) $\mu_0 = 10 $}  % 使用caption*取消编号，手动添加(a)标签
  \end{minipage}
  \hfill
  \begin{minipage}[b]{0.48\textwidth}
    \centering
    \includegraphics[width=\linewidth]{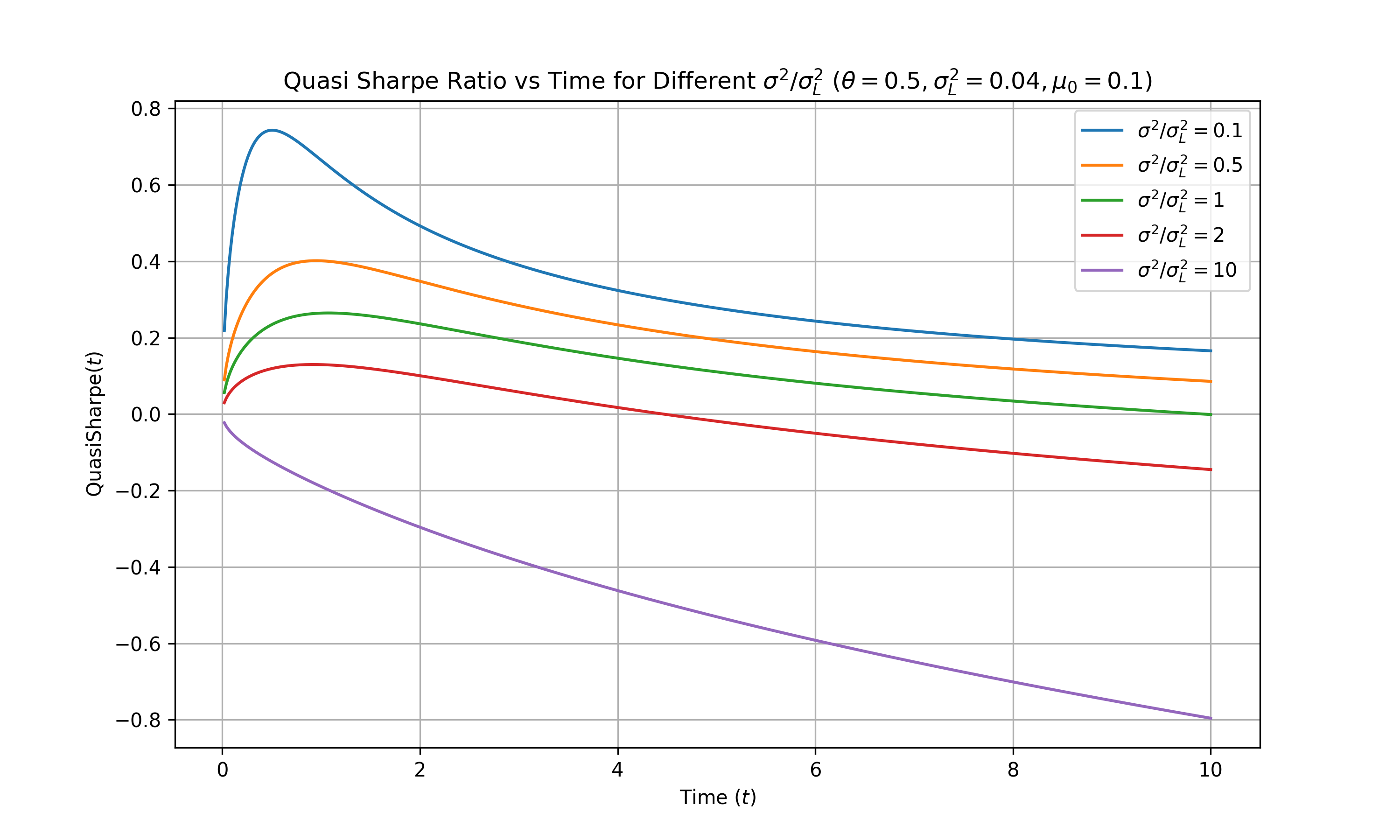}
    \caption*{(b) $\mu_0 = 0.1 $}  % 使用caption*取消编号，手动添加(b)标签
  \end{minipage}
  \caption{Model-theoretic time evolution of Quasi-Sharpe ratio under varied brownian-lévy variance ratios and initial drift conditions (a) strong initial drift $\mu_0 = 10 $ (b) weak initial drift $\mu_0 = 0.1 $ }
  \label{fig:quasiQQ}
\end{figure}

Systematic analysis of the quasi-Sharpe ratio characteristics reveals:
\begin{itemize}
\item \textbf{Volatility dominance effect}: Under the condition where Brownian motion variance is dominated by Lévy variance (\( \sigma^2 \ll \sigma_L^2 \)), the response ratio attains elevated magnitudes. This aligns with financial intuition, as stronger signal-to-noise ratio emerges when indicator-driven perturbations (governed by the Lévy process) prevail over random market noise (captured by the Brownian component).
\item \textbf{Asymptotic zero initialization}: In the limit \( t \rightarrow 0 \), while \( \text{E}[R_t] \rightarrow 0 \) and \( \sigma(R_t) \rightarrow 0 \), the expectation decays at a faster rate (\( \mathcal{O}(t) \) vs \( \mathcal{O}(t^{1/2}) \)), resulting in \( \lim_{t\to 0} QS(t) = 0 \). This boundary condition will be rigorously examined later.
\item \textbf{Existence of extremal values}: Under the condition \( \sigma^2 \leq c\sigma_L^2 \) for some constant \( c = \mathcal{O}(1) \), or given substantial initial drift \( |\mu_0| \geq M_0 \), the response ratio \( QS(t) \) possesses a local maximum, corresponding to the optimal holding time \( t^* \) that satisfies:\[t^* = \arg max_{t>0} QS(t)\]
\item \textbf{Long-run equilibrium}: When \( t \rightarrow \infty \): With \( \text{E}[R_t] = \mathcal{O}(t) \) and \( \text{Var}(R_t) = \mathcal{O}(t) \), the quasi-Sharpe ratio \( QS(t) \) tends to a limiting constant;
\end{itemize}

Our derivation above assumes the drift follows an Ornstein-Uhlenbeck (O-U) process. In the following Figure \ref{fig:logrt_gbm}, we simultaneously plot the quasi-Sharpe ratios for both the constant-drift geometric Brownian motion and the O-U drift case for comparison. The dashed lines represent constant drift scenarios, while solid lines correspond to O-U drift. Curves with identical colors share the same Brownian-to-Lévy variance ratio \( \sigma^2/\sigma_L^2 \).
\begin{figure}[H]
  \centering
  \includegraphics[width=0.8\textwidth]{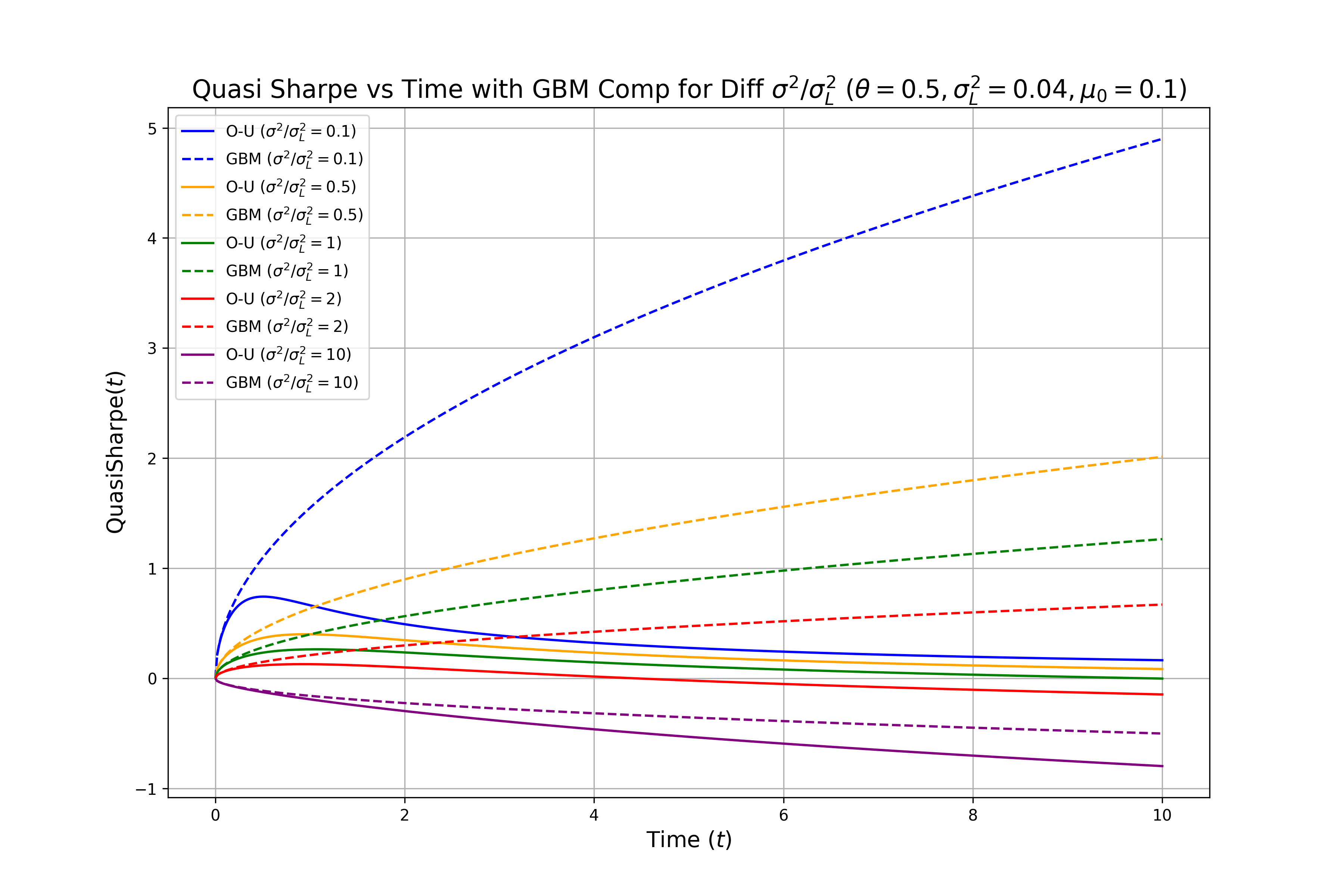} % 调整width控制大小
  \caption{Comparison of response ratio dynamics: Ornstein-Uhlenbeck drift vs. constant-drift geometric brownian motion} % 自动编号
  \label{fig:logrt_gbm} % 标签需放在caption之后
\end{figure}
The figure demonstrates that near \( 0 \), the indicator-driven process asymptotically approaches a standard Brownian motion with constant drift. We now analyze the response ratio behavior in the neighborhood of \( 0 \) to verify this claim. Through Taylor series expansion and related methods, it can be shown that in the right limit of \( 0 \), the quasi-Sharpe ratio converges to that of a standard Brownian motion with fixed drift. The asymptotic relationship is given by:
\begin{gather*}
    \lim_{t\rightarrow0^+}{\mathrm{QS}}\left(t\right)=0 \\
    \mathrm{QS}\left(t\right)\sim\left(\frac{\mu_0-\frac{1}{2}\sigma^2}{\sigma}\right)\cdot\sqrt t
\end{gather*}
The detailed derivation is provided in Appendix~\ref{app:quasi_sharpe}. As evident from the above analysis, this expression asymptotically coincides with the quasi-Sharpe ratio formula for standard geometric Brownian motion, demonstrating that the process reduces to a fixed-drift GBM when \( t \) is near \( 0 \). This consistency indirectly validates the derivation of the quasi-Sharpe ratio.

Let us proceed to a more detailed analysis of the mean logarithmic-return:
\begin{equation*}
    E\left[R_t\right]=\mu_0\frac{1-e^{-\theta t}}{\theta}-\frac{\sigma^2}{2}t
\end{equation*}
The mean log-return achieves its maximum when \( t \)  satisfies:
\begin{equation*}
    t=-\frac{1}{\theta}\ln{\left(\frac{\sigma^2}{2\mu_0}\right)}
\end{equation*}
with the maximum value being:
\begin{equation}
    E\left[R_t\right]_{max}=\frac{\mu_0}{\theta}+\frac{\sigma^2}{2\theta}\ln{\left(\frac{\sigma^2}{2e\mu_0}\right)}
\end{equation}
The detailed proof is provided in Appendix~\ref{app:max_main}. The equation can be interpreted as follows: The first term \( \frac{\mu_0}{\theta} \) constitutes the dominant contribution to the drift rate, where \( \frac{1}{\theta} \) acts as an amplification factor for the mean-reverting OU process. The second term, typically negative, represents the damping effect of volatility on returns. From the properties of the O-U process, we know that the \( k \)-th order autocorrelation coefficient is given by \( \rho_k=\left(1-\theta\right)^k \). Therefore, when the autocorrelation coefficient decays more slowly (i.e., when \( \theta \) is smaller), this corresponds to longer memory in the indicator, resulting in a larger first drift term. This demonstrates that autocorrelation coefficients can be employed for qualitative analysis of a metric's memory effects. The dominant drift term for the long-term contribution of the metric is proportional to \( 1/\left(1-\rho\right) \), where \( \rho \) represents the ratio of the autocorrelation coefficients of adjacent order.

Based on the above analysis and modeling, we draw the following conclusions:
\begin{itemize}
    \item When modeling order flow imbalance as a price shock, the subsequent price dynamics can be represented as a specialized geometric Brownian motion, where the drift term follows an O-U process driven by a symmetric Lévy process. Both processes are zero-mean, symmetric about the origin, and exhibit finite second moments.
    \item The logarithmic return process derived from this modeling framework satisfies \eqref{eq:rt_result}.
    \item The mean log-return modeled by this process follows \eqref{eq:logret_exp}. It initially increases and then decreases with time \( t \), exhibiting an extrema. This behavior resembles an exponential saturation process. The variance of log-returns follows \eqref{eq:logret_var}. It increases monotonically with \( t \), showing nonlinear variation in the short term and linear variation in the long term.
    \item We define a quasi-Sharpe ratio metric as the mean log-return divided by the standard deviation. This indicator initially increases and subsequently decreases over time when the initial shock is sufficiently large, exhibiting a well-defined maximum, see equation \eqref{eq:logret_qs}.
    \item When transforming the aforementioned price dynamics into a trading strategy, the strategy's essence represents an equilibrium between: (i) the cumulative price impact from order flow imbalance shocks, and (ii) the uncertainty introduced by price random walks. For different markets and tradable instruments, specific parameter estimation is required to execute trades near the optimal quasi-Sharpe ratio region.
\end{itemize}
The preceding section established the theoretical modeling framework for price dynamics. We now proceed to empirical validation of the return process.

\section{Empirical data analysis }
\subsection{Data analysis method}
To validate the above derivation - specifically that the average mid-price changes following order flow imbalance $OFI$ converges to an exponentially saturating process combining O-U processes as the holding period increases - we conduct empirical testing on mean price changes after single $OFI$ events. This study currently focuses on verifying only the expected price change(i.e., the time-dependent expected logarithmic return process); variance validation will be addressed in subsequent research.

We perform multiple regressions across various historical time windows and forecast horizons. The analysis examines the relationship between the factors $OFI$ and future mid-price changes. The regression equation is given below, where for each historical window \( \left[t_{h-1},t_h\right] \), we compute the aggregate $OFI$ metric \( OFI_h \) as the sum of all marked \( e_n \) values between observations \( N_{t_{h-1}+1} \) and \( N_{t_h} \). For each window length \( h \), we then regress the subsequent mid-price changes in different forecast horizons \( t \) against $OFI$ .
\begin{gather*}
\Delta P_{k,t}=\widehat{\alpha_t}+\widehat{\beta_t}OFI_{k,h}+\widehat{\epsilon_k} \\
OFI_h=\sum_{n=N\left(t_{h-1}\right)+1}^{N\left(t_h\right)}e_n
\end{gather*}
To provide a concrete example, in the most granular scenario, we calculate the indicator \( OFI_1 = e_n \) for each individual tick. This indicator serves as the independent variable, while the subsequent average mid-price change after one tick (the minimum time interval) becomes the dependent variable in our regression.

After completing the regression training, we evaluate the results by performing regression calculations for each individual data point. In equity index futures markets, both long and short positions can be implemented. We employ a single model for regression predictions across both long and short positions. The result computation process separates into positive and negative groups:
\begin{enumerate}
    \item When a data point's regression output is positive , we interpret this as the metric predicting positive price movement and include the corresponding actual price change in the positive group.
    \item Conversely, for negative regression outputs, we aggregate the actual price changes in the negative group.
\end{enumerate}
Intuitively, the positive/negative group's aggregate represents the hypothetical total profit and loss from taking long/short positions based on the metric. We later visualize the regression results through scatter plots where:
\begin{itemize}
    \item The x-axis shows the predicted mid-price change for each tick's metric value.
    \item The y-axis displays the actual mid-price change.
\end{itemize}
Furthermore, we can generate alternative indicators by aggregating historical $OFI$ values over varying time windows and perform regression predictions for different forecast horizons. This creates a two-dimensional parameter space that we systematically explore. Below, we present these exhaustive search results and compare them with our theoretical model predictions.

Next, we select the regression algorithm. After comparing various methods, we employ the least absolute shrinkage and selection operator (LASSO) algorithm for regression, based on the following considerations:
\begin{enumerate}
    \item We prioritize the model's predictive capability, particularly its out-of-sample generalization performance. This enables more robust evaluation of both the indicator's effectiveness and our price drift model. The LASSO algorithm provides regularization that proves crucial for enhancing the model's out-of-sample generalization. As demonstrated in subsequent results, the combined use of our indicators and LASSO yields satisfactory generalization across all significant time scales.
    \item We require analytical capability for metric combinations. LASSO performs variable selection that effectively handles potential multicollinearity among metrics. A comprehensive analysis of LASSO for various metrics can be found in \cite{sahalia1}.
\end{enumerate}

Regarding the algorithmic implementation, we conduct backtesting using LASSO regression on one year of tick data. To optimize hyperparameter selection, we partition each year's data into training and test sets. For the training set, we employ 5-fold cross-validation to determine the regularization hyperparameter. This represents a robust yet straightforward approach that enables comprehensive evaluation of the indicator's predictive power across varying forecast windows.

\subsection{Data analysis results}
\begin{figure}[H]
  \centering
  % 上图
  \begin{minipage}[t]{\textwidth}
    \centering
    \includegraphics[width=0.8\linewidth]{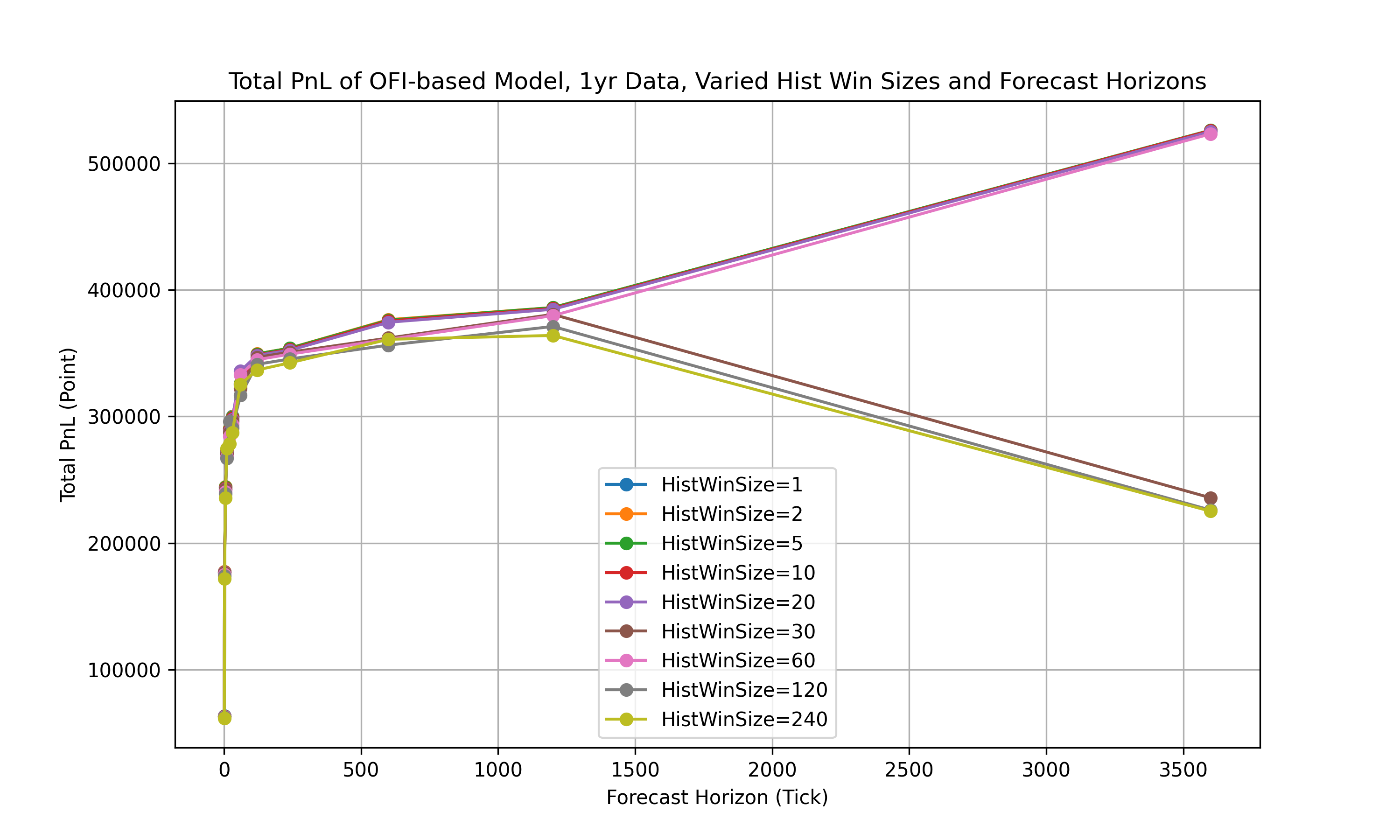}
    \caption*{(a) Empirical regression prediction total profit}  % 使用caption*取消编号，手动添加(a)标签
  \end{minipage}
  \vspace{0.5cm} % 调整图片间距
  % 下图
  \begin{minipage}[t]{\textwidth}
    \centering
    \includegraphics[width=0.8\linewidth]{figures/expectedlogreturn_mu0_10.PNG}
    \caption*{(b) Model prediction total profit}  % 使用caption*取消编号，手动添加(b)标签
  \end{minipage}
  \caption{The annual total profit and loss of the $OFI$-based regression model as a function of forecast horizon, a comparison between (a) Empirical regression prediction  (b) Model prediction}
  \label{fig:totalprofit}
\end{figure}

Figure~\ref{fig:totalprofit} shows the overall performance plot, which includes the total points of predicted returns from $OFI$ regression calculations across different historical window lengths as they vary with prediction time, with detailed data available in table of Appendix~\ref{app:regression_results}.  The x-axis in the plot represents different forecast horizon measured in tick counts, where a value of 500 corresponds to 250 seconds given a fixed rate of 2 ticks per second. We tested prediction horizons ranging from 1 tick (0.5 seconds later) to 3600 ticks (30 minutes later), with each distinct curve in the plot representing the use of different historical time window sizes to aggregate the \( e_n \) values into $OFI$ indicators, perform regressions using these $OFI$ indicators, and calculate the overall profitability based on the regression results. The overall profitability is measured in points, where one point in CSI 300 stock index futures represents 300 Chinese yuan (CNY 300).

The figure above shows that the regression predictions from the data exhibit similar patterns to the model-predicted evolution of return means over time, providing empirical validation for our theoretical model. Figure \ref{fig:totalprofit} demonstrates that prediction results show negligible differences across various historical windows when the prediction horizon ranges from 0.5 seconds to 10 minutes, while significant variations emerge when the prediction window exceeds 1500 ticks (12 minutes). Given that our maximum historical window is 240 ticks (2 minutes), the metric maintains stable and gradually improving performance as the prediction window extends up to 10 minutes, confirming the robustness of our approach within this specific temporal configuration where the historical window spans 2 minutes and prediction horizons remain below 10 minutes.

\begin{figure}[H]
  \centering
  \includegraphics[width=0.8\textwidth]{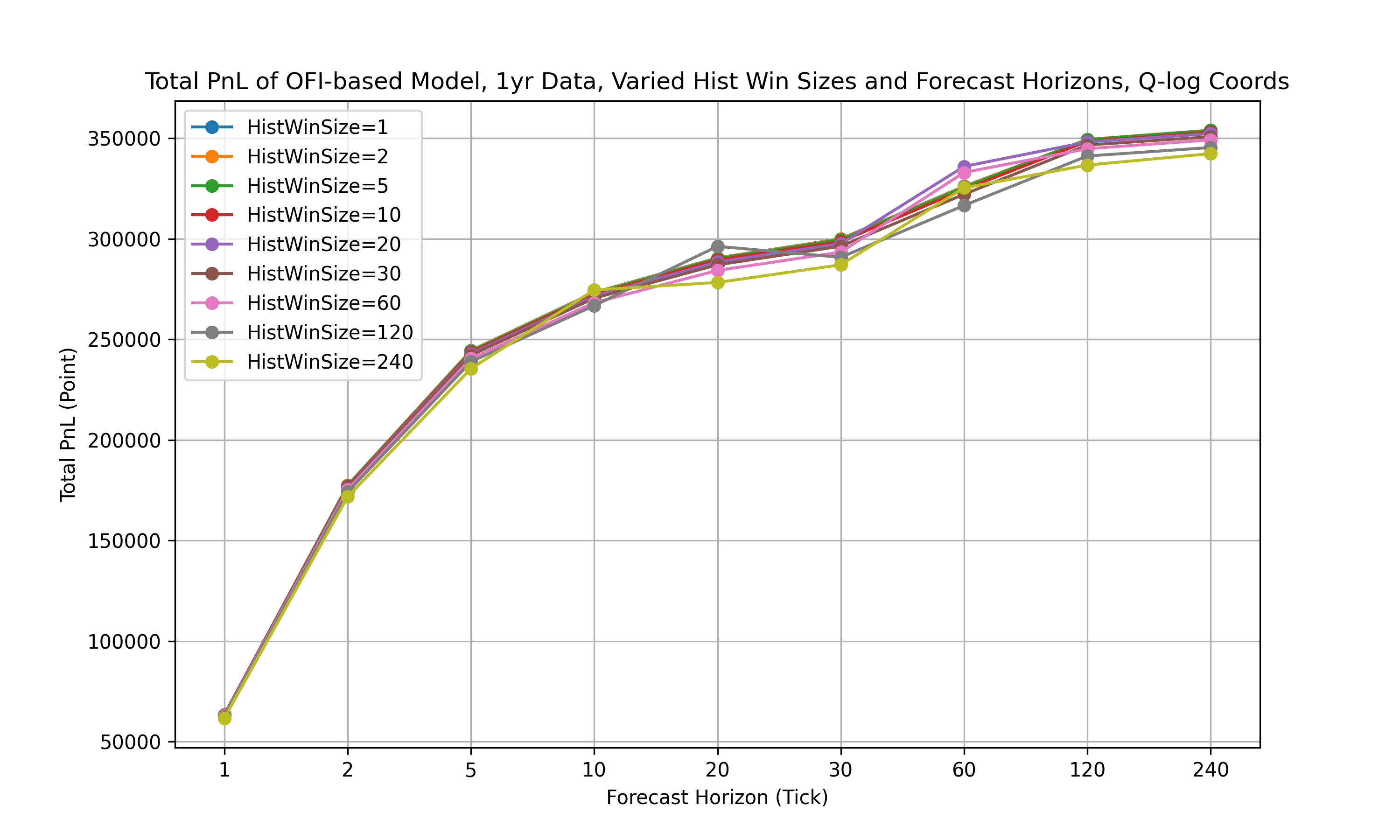} % 调整width控制大小
  \caption{The annual total PnL of the $OFI$-based regression model as a function of historical window sizes and forecast horizon under local quasi-logarithmic coordinates} % 自动编号
  \label{fig:totalprofit_log} % 标签需放在caption之后
\end{figure}

Figure \ref{fig:totalprofit_log} presents a magnified view of the short-term forecast horizon from the previous plot, with the x-axis transformed to a quasi-logarithmic scale that accommodates conventional time intervals such as half-minute and one-minute increments. The results demonstrate that for short-term predictions within 240 ticks (2 minutes), the predictive returns show significant growth with increasing forecast horizon under this modified coordinate system. The observed pattern exhibits characteristics approximating a saturated logarithmic cumulative relationship, which further corroborates the temporal dependence of aggregate returns predicted in the preceding section.

We now present graphical representations of the regression data points and fitted lines, beginning with a set of plots showing 1-tick-ahead predictions using different historical windows of 1-tick, 2-tick, 10-tick and 30-tick intervals, where the x-axis displays the predicted mid-price change for the next 1-tick period calculated from the regression formula and the y-axis shows the actual mid-price change.
\begin{figure}[H]
  \centering
  % 第一行
  \begin{minipage}[b]{0.48\textwidth}
    \centering
    \includegraphics[width=\linewidth]{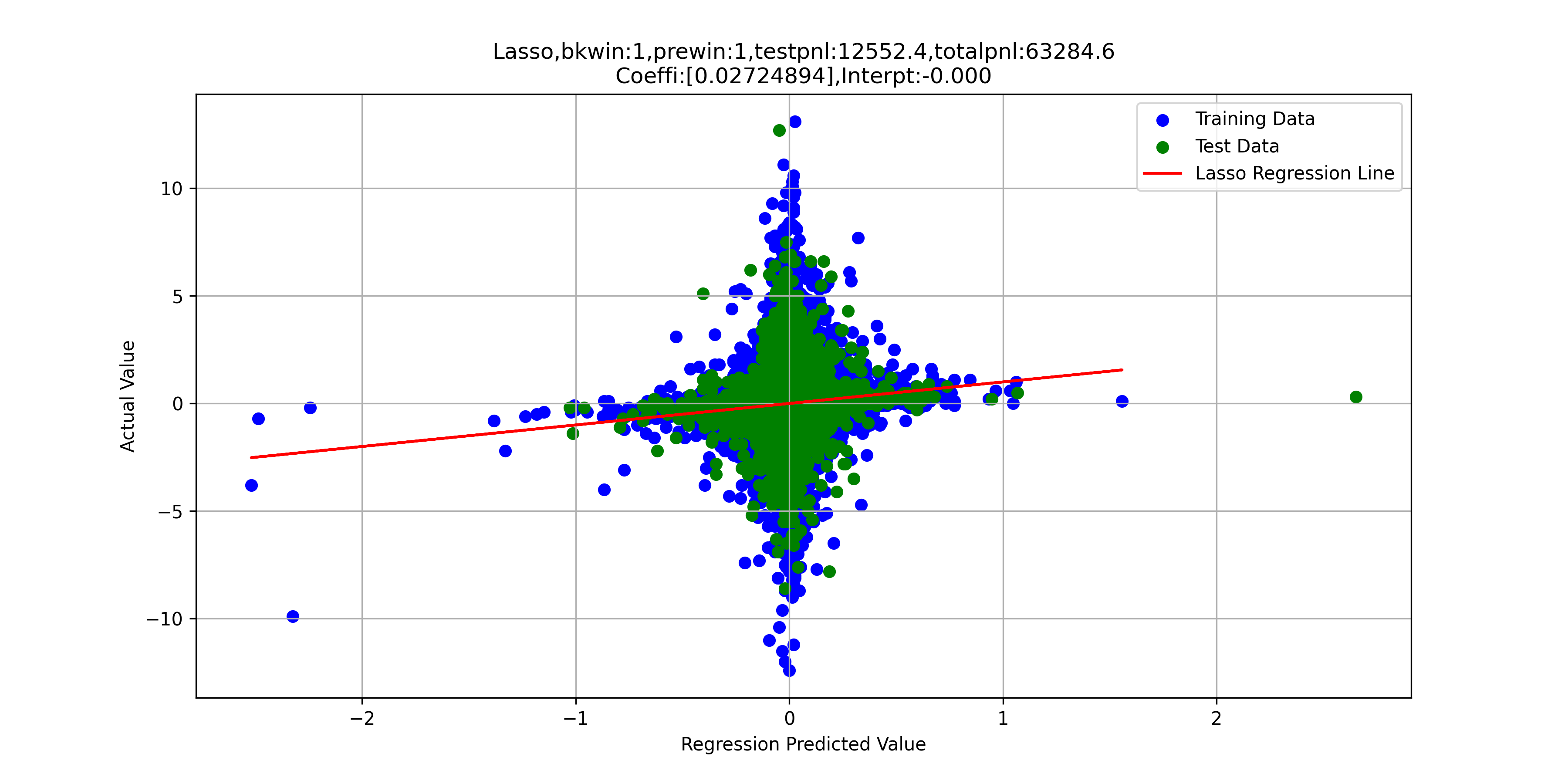}
    \caption*{(a) Hist win = 1 tick}
  \end{minipage}
  \hfill
  \begin{minipage}[b]{0.48\textwidth}
    \centering
    \includegraphics[width=\linewidth]{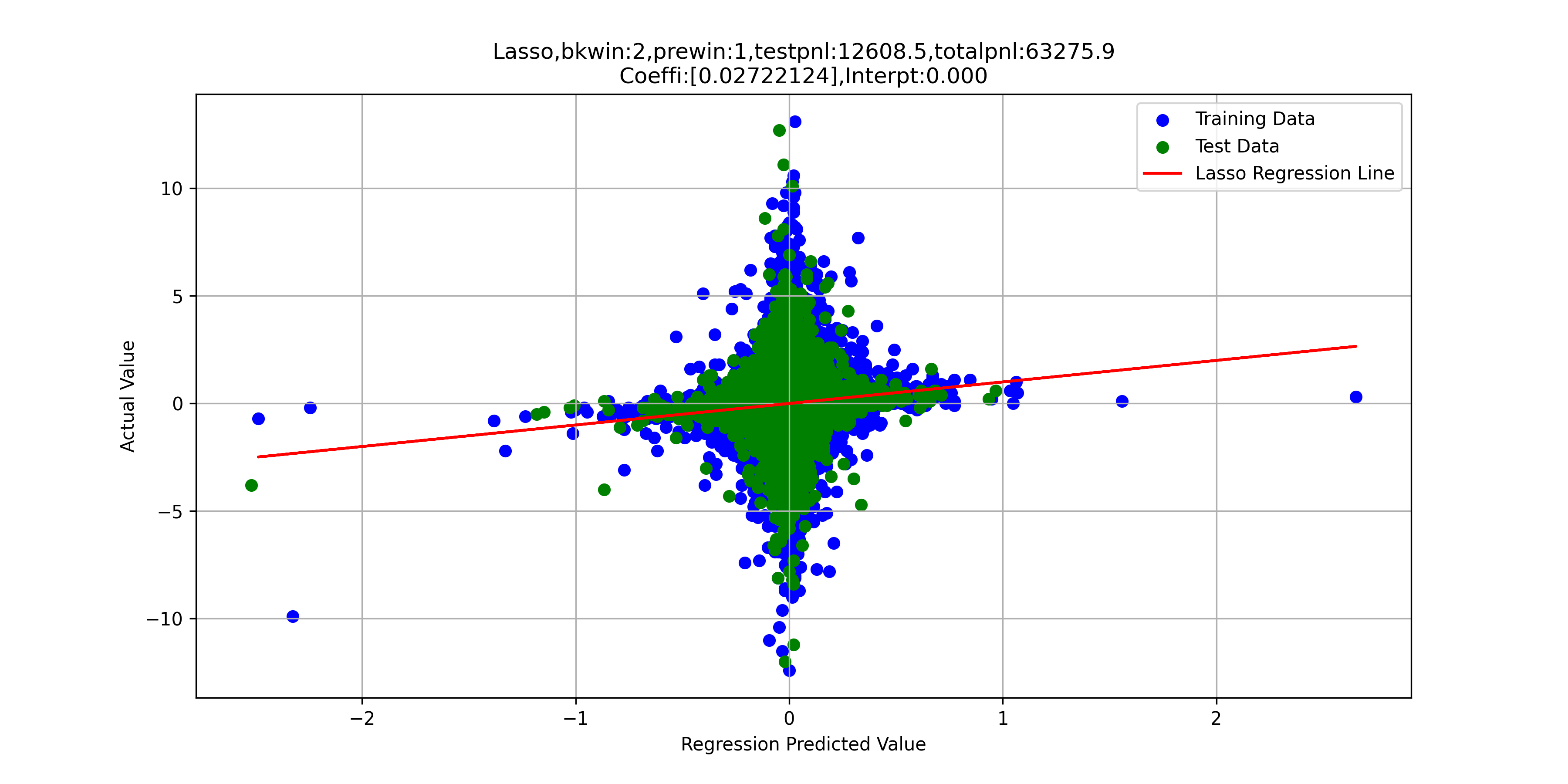}
    \caption*{(b) Hist win = 2 tick}
  \end{minipage}
  
  % 第二行
  \begin{minipage}[b]{0.48\textwidth}
    \centering
    \includegraphics[width=\linewidth]{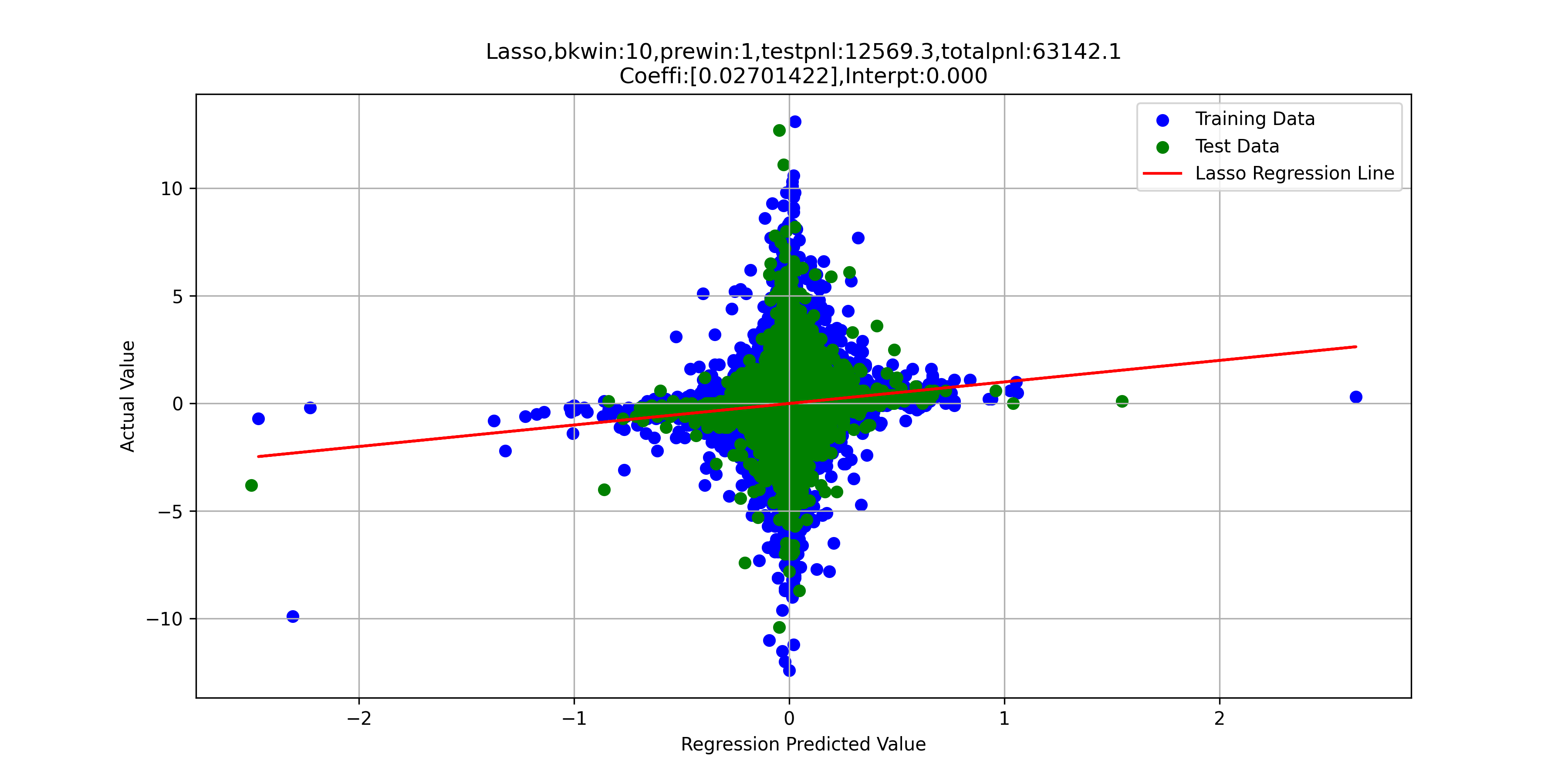}
    \caption*{(c) Hist win = 10 tick}
  \end{minipage}
  \hfill
  \begin{minipage}[b]{0.48\textwidth}
    \centering
    \includegraphics[width=\linewidth]{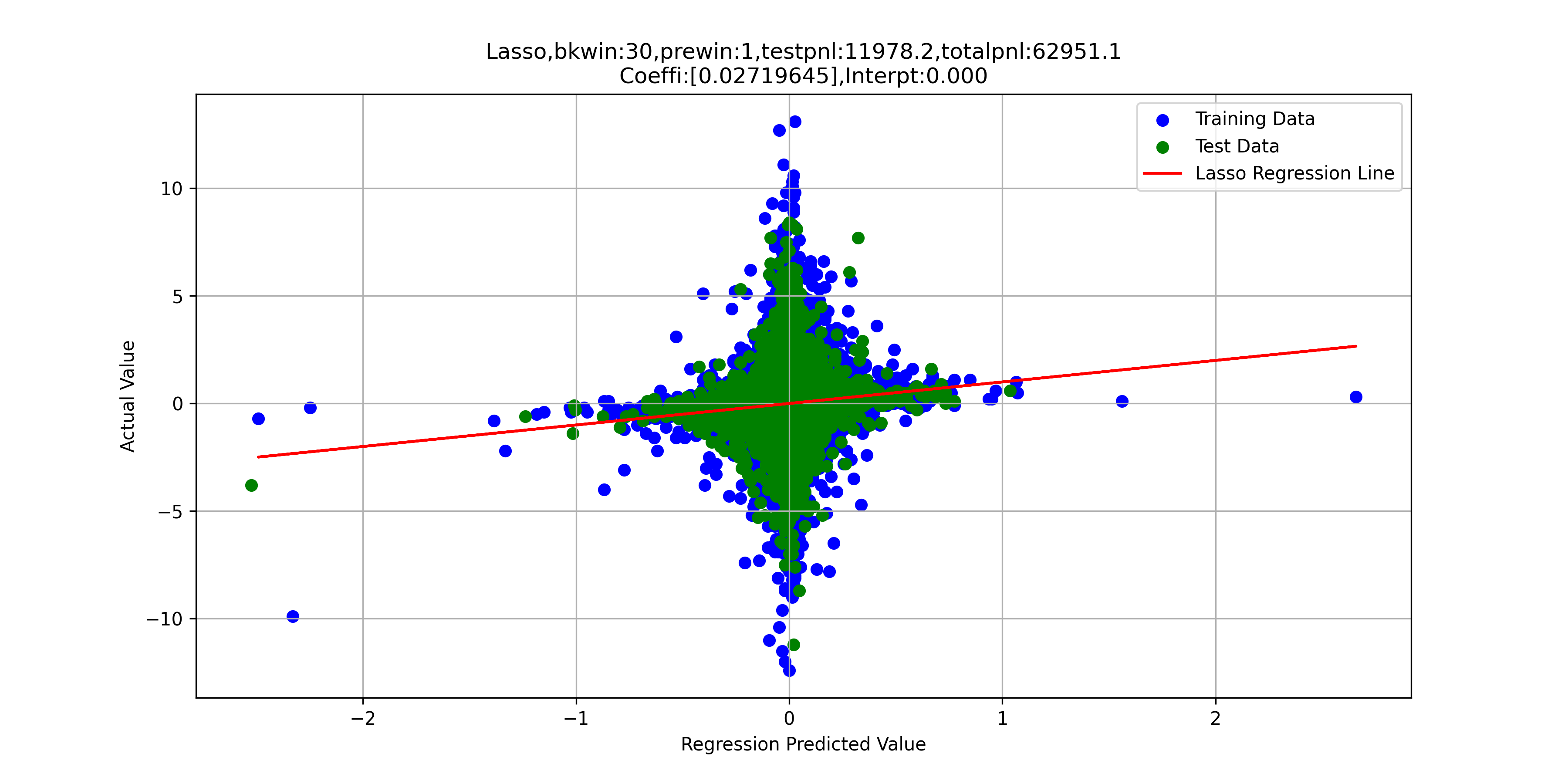}
    \caption*{(d) Hist win = 30 tick}
  \end{minipage}
  
  \caption{Regression results of different historic window sizes for 1 tick forecast horizon (a) hist window size 1 tick; (b) hist window size 2 tick; (c) hist window size 10 tick; (d) hist window size 30 tick}
  \label{fig:regression_pre1tick}
\end{figure}
Next, we present regression analysis using a historical window of 10 ticks (5 seconds) to predict price movements at two distinct horizons: 60 ticks (30 seconds) and 240 ticks (2 minutes).
\begin{figure}[H] 
  \centering
  \begin{minipage}[b]{0.48\textwidth}
    \centering
    \includegraphics[width=\linewidth]{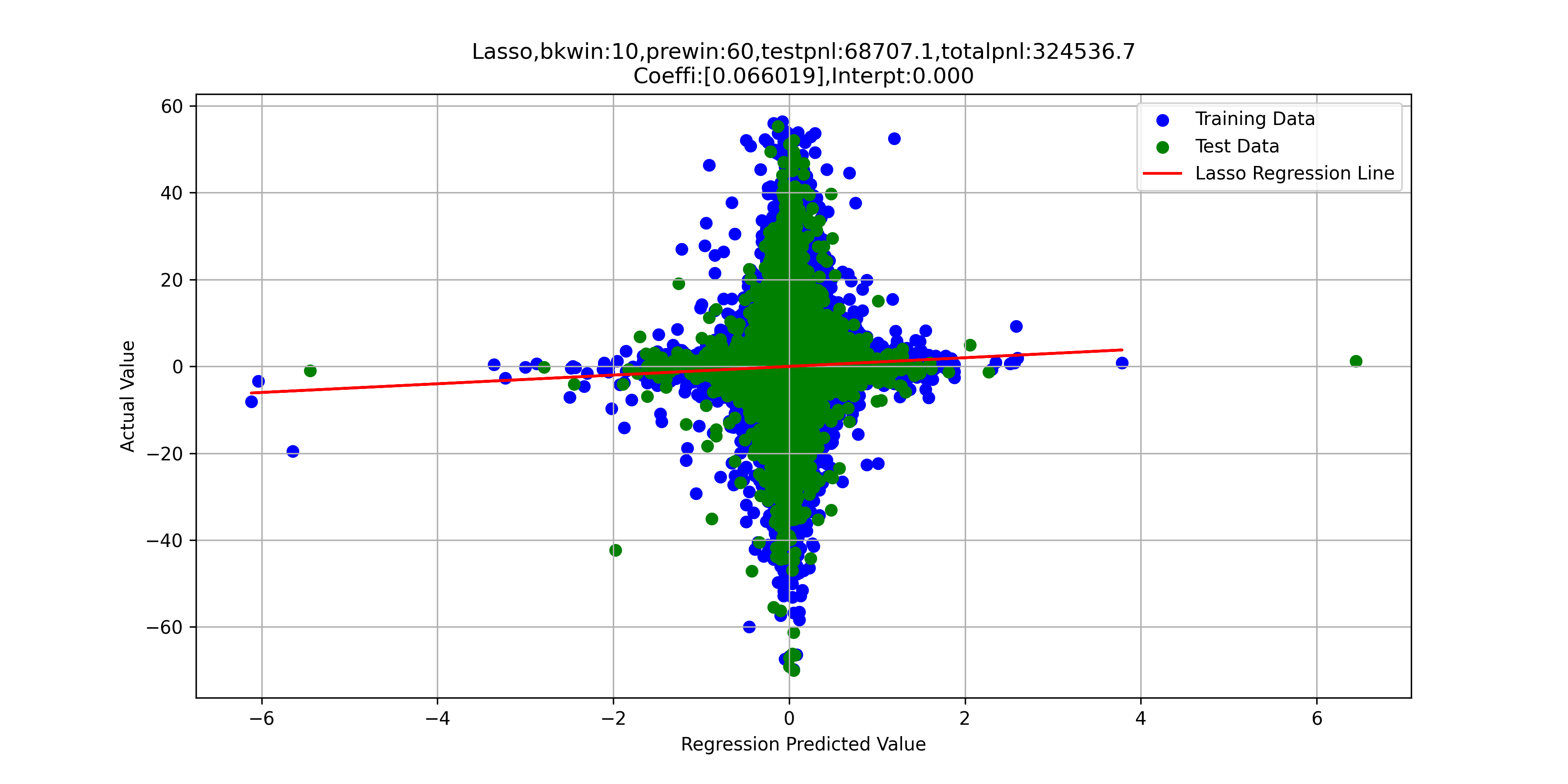}
    \caption*{(a) FCH = 60 ticks}  % 使用caption*取消编号，手动添加(a)标签
  \end{minipage}
  \hfill
  \begin{minipage}[b]{0.48\textwidth}
    \centering
    \includegraphics[width=\linewidth]{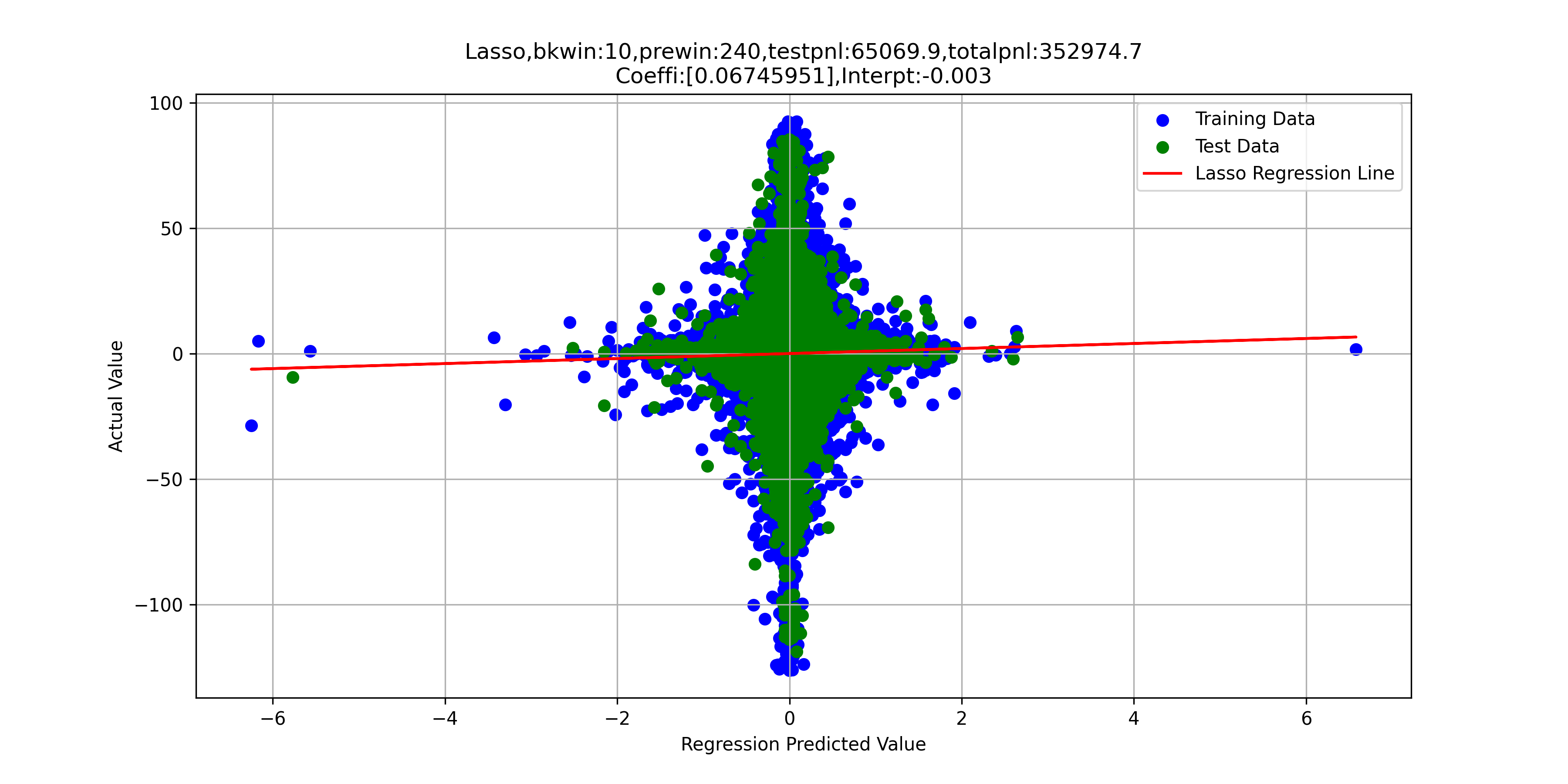}
    \caption*{(b) FCH = 240 ticks}  % 使用caption*取消编号，手动添加(b)标签
  \end{minipage}
  \caption{Regression results of 10 tick historic window sizes for different forecast horizons (a) forecast horizon = 60 ticks; (b) forecast horizon = 240 ticks;}
  \label{fig:regression_back10tick}
\end{figure}
The data in the figure demonstrate that while the overall distributions of data points show minimal morphological differences, the aggregate profits increase significantly from 64,000 points at the 0.5-second horizon to approximately 324,537 points at 30 seconds and 352,975 points at the 2-minute horizon. The regression coefficients exhibit a parallel enhancement, rising from \( 0.027 \) for the 0.5-second prediction to \( 0.066 \) at 30 seconds and stabilizing at \( 0.067 \) for the 2-minute forecast, indicating statistically significant improvements in predictive efficacy across extended time horizons.

In the six regression plots above, the blue data points represent in-sample training data while the green points denote out-of-sample test data, with both distributions exhibiting fundamental consistency that validates the predictive capability of the regression model; detailed quantitative results will be presented in the following section.

\begin{figure}[H] 
  \centering
  \begin{minipage}[b]{0.48\textwidth}
    \centering
    \includegraphics[width=\linewidth]{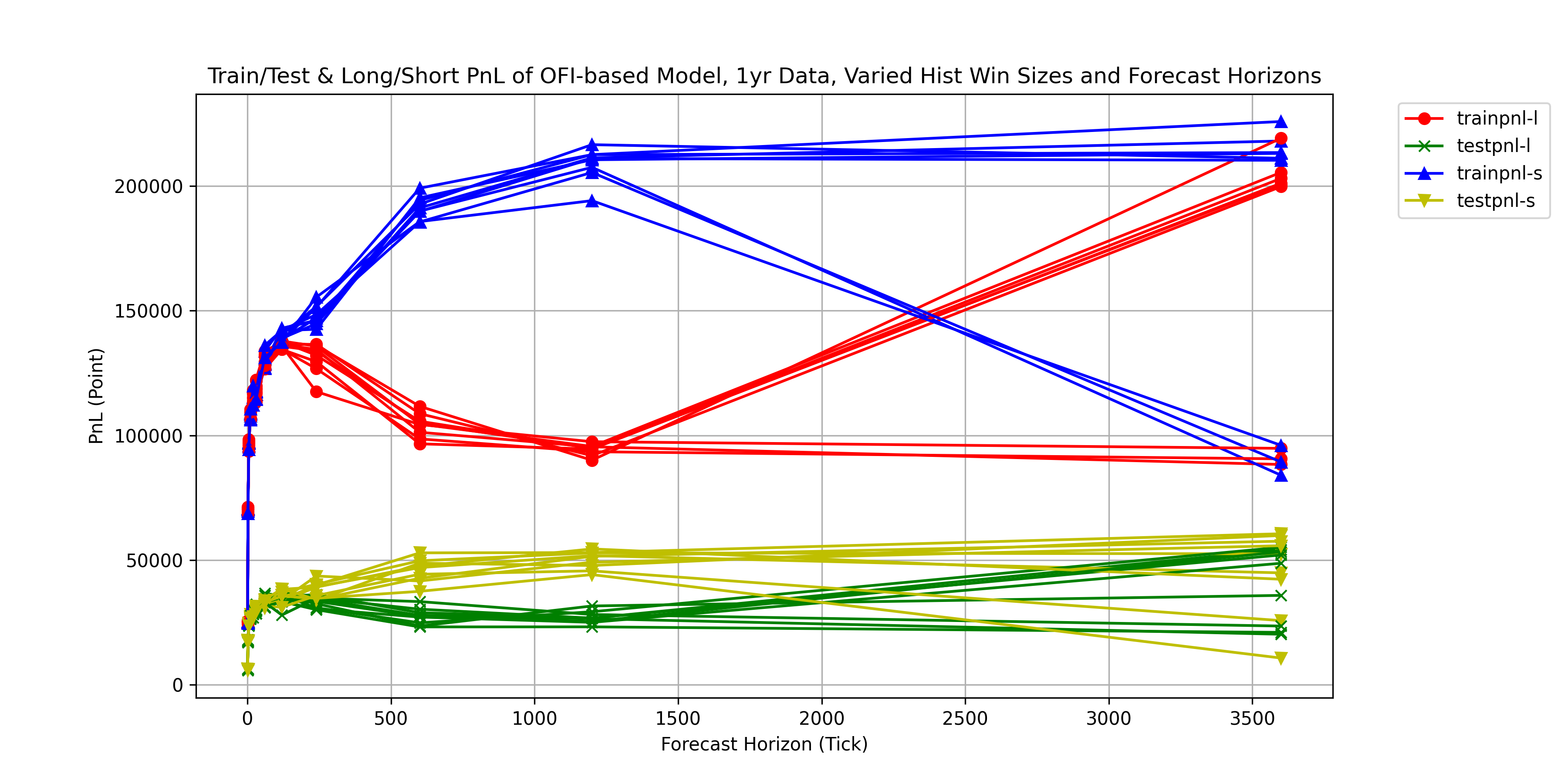}
    \caption*{(a) FCH = [0,3600]}  % 使用caption*取消编号，手动添加(a)标签
  \end{minipage}
  \hfill
  \begin{minipage}[b]{0.48\textwidth}
    \centering
    \includegraphics[width=\linewidth]{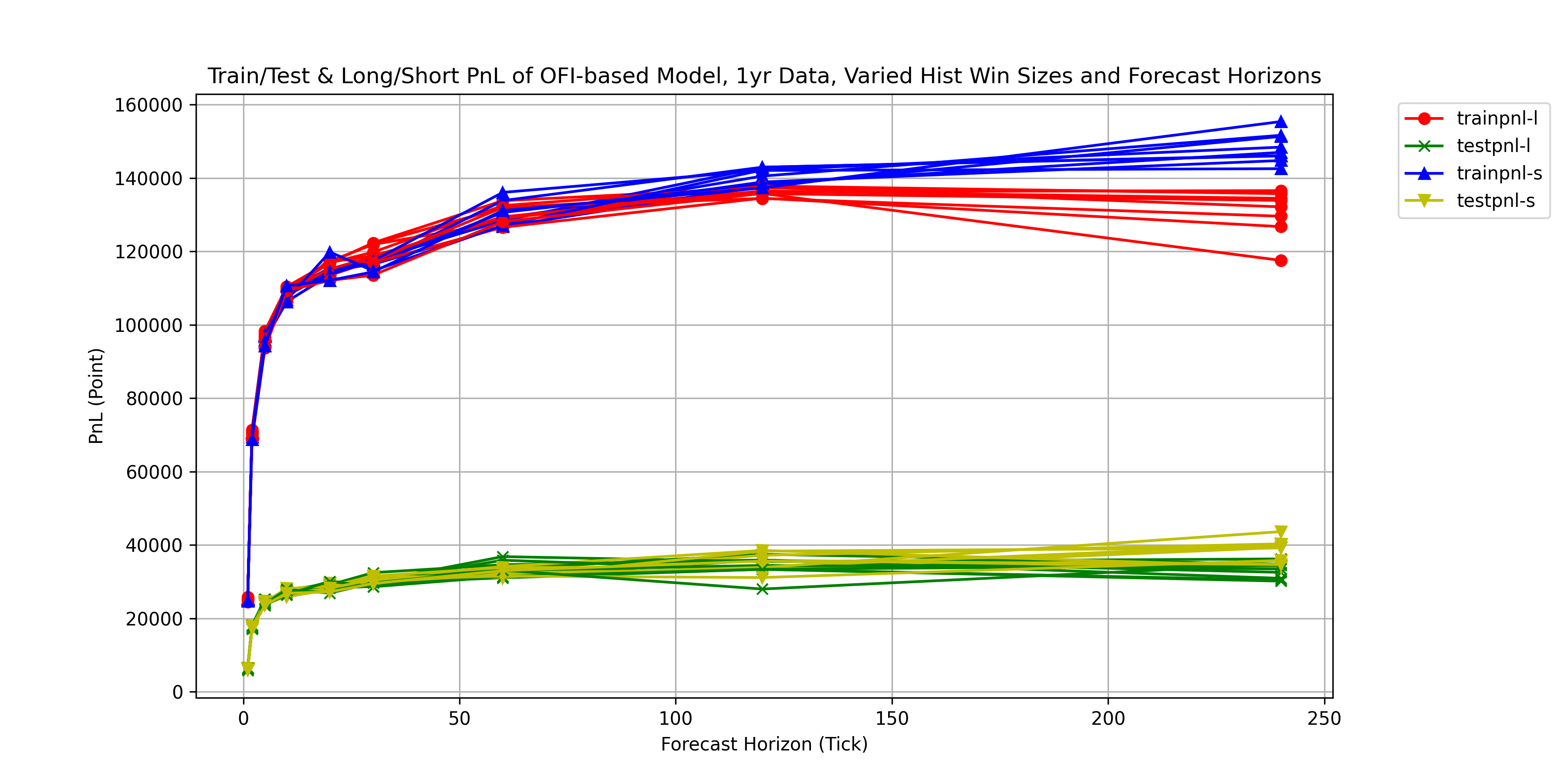}
    \caption*{(b) FCH = [0,240]}  % 使用caption*取消编号，手动添加(b)标签
  \end{minipage}
  \caption{Prediction performance of the $OFI$-based regression model: In-sample vs. out-of-sample, long vs. short (a) Forecast horizon = [0,3600] (b) Forecast horizon = [0,240]}
  \label{fig:regression_in_out}
\end{figure}

We now present the training and test samples separately, with long and short positions also displayed independently in Figure\ref{fig:regression_in_out} to examine the out-of-sample predictive performance of the metrics and regression, as well as the symmetry between long and short positions. The two subplots feature different x-axis ranges: the left panel shows the complete view for 0 to 3600 ticks (30 minutes), while the right panel provides a magnified view of 0 to 240 ticks (2 minutes). The curves are defined as: trainpnl-l (training set long positions), testpnl-l (test set long positions), trainpnl-s (training set short positions), and testpnl-s (test set short positions), with each curve corresponding to different historical windows while maintaining the forecast horizon on the x-axis. Note that the in-sample data volume is four times that of the out-of-sample period.

\begin{figure}[H]
  \centering
  \includegraphics[width=0.8\textwidth]{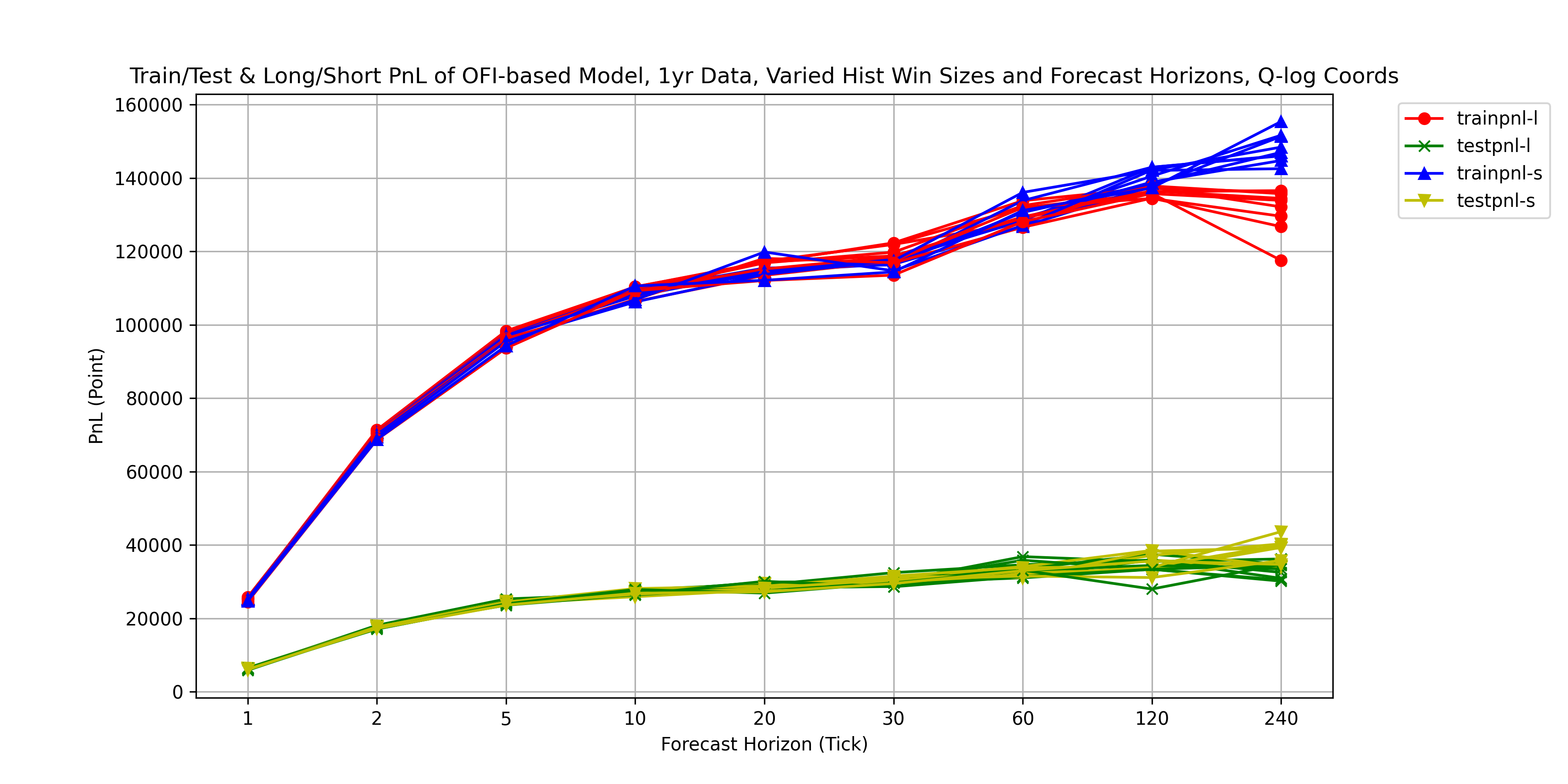} % 调整width控制大小 
  \caption{Regression profit and loss performance within 2-minute horizon under quasi-logarithmic coordinates: In-sample vs. out-of-sample and long-short comparisons} % 自动编号
  \label{fig:regression_in_out_2m} % 标签需放在caption之后
\end{figure}

Figure \ref{fig:regression_in_out} shows that out-of-sample predictions largely maintain the profit trends observed in-sample. The average out-of-sample returns are approximately one-quarter or slightly less of the in-sample profits due to the 4:1 sample size ratio between them. Both long and short positions exhibit stable profitability in out-of-sample tests, following saturated exponential patterns consistent with the model's theoretical predictions. Within the 240-tick (2-minute) horizon shown in the right panel, the forecasting performance remains robust, while beyond this threshold significant divergences emerge between long and short positions, confirming the indicator's short-term predictive nature concentrated below 2 minutes. The similar fitting and forecasting accuracy between long and short positions across both sample types aligns with expectations for short-horizon predictive models. Figure \ref{fig:regression_in_out_2m}'s quasi-logarithmic coordinate transformation of the sub-240-tick range further verifies the stability of out-of-sample performance, with consistent profit ratios across prediction horizons matching the model's derivation in the preceding section. Minimal variations across different historical window lengths preclude further segmentation, as evidenced by the marginal differences between their corresponding curves.

\section{The combinatorial performance of order flow imbalance with other metrics}
\begin{figure}[H]
  \centering
  \includegraphics[width=0.8\textwidth]{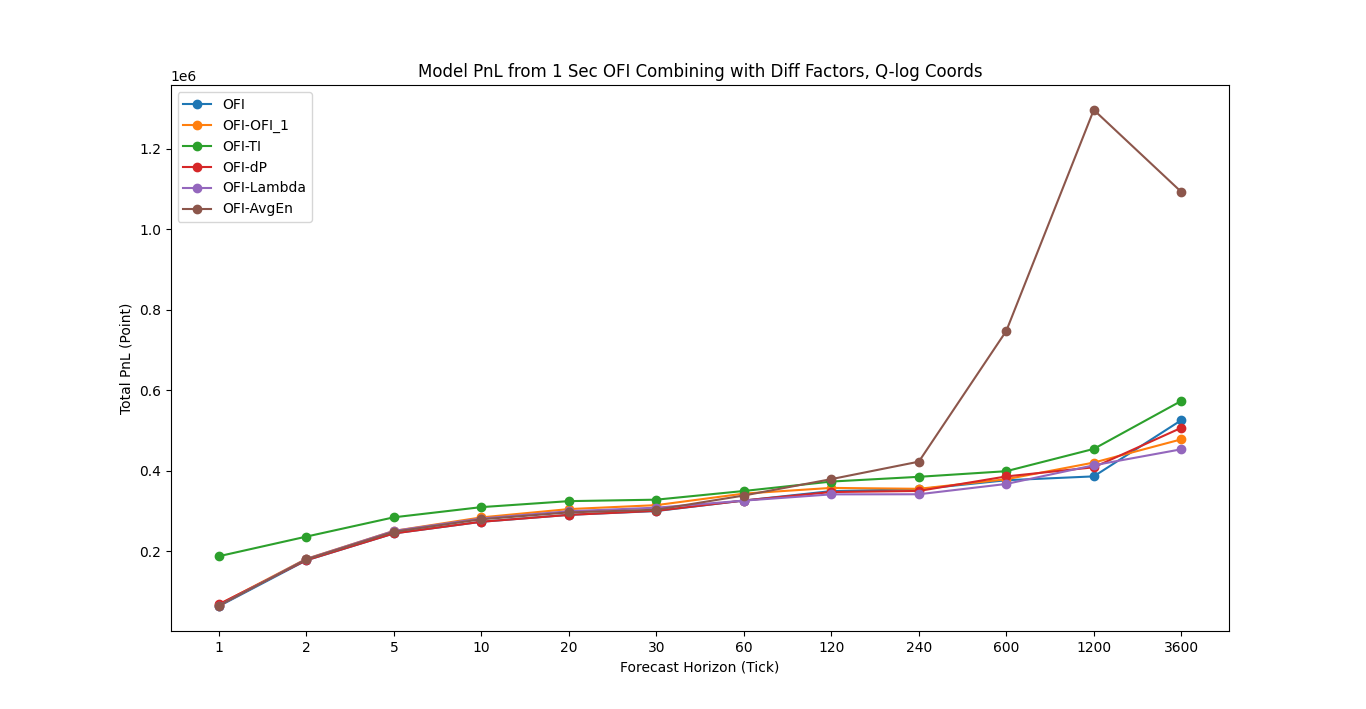} % 调整width控制大小 
  \caption{Prediction performance of combined metric with $OFI$} % 自动编号
  \label{fig:combined_metric} % 标签需放在caption之后
\end{figure}

We proceed to extend our analysis to identify additional metrics that can enhance predictive performance when combined with the core $OFI$ metric while minimizing factor interference. To preserve the dominance of the $OFI$ metric, we perform two-factor LASSO regressions pairing $OFI$ with five distinct co-variates: (1) a lagged $OFI$ value \( OFI_{t-1} \) from the preceding time window, (2) the $TI$ (trade imbalance) metric introduced in Section 2, (3) concurrent mid-price changes \( \Delta P_t \), (4) the Lambda metric measuring price impact per unit trade volume as previously defined, and (5) \( AvgEn \), the cumulative average of all \( e_n \) values from the beginning of each continuous trading interval. Historical analysis reveals that the predictive power of $OFI$ is insensitive to extended historical window sizes but exhibits strong short-term efficacy, thus we standardize all calculations using a 2-tick (1-second) historical window for consistent comparison. Figure \ref{fig:combined_metric} presents the regression outcomes across varying forecast horizons, with the x-axis following a quasi-logarithmic scale where 10 ticks correspond to 5-second forecast and 3600 ticks represent 30-minute forecasts under the long-position model. This transformed coordinate system particularly accentuates short-term performance within critical sub-240-tick (2-minute) windows while maintaining comparability across time horizons.
Figure \ref{fig:combined_metric} shows that different metric exhibits varying predictive performance in different forecast horizons, highlighting the importance of selecting window-specific factors. For ultra-short-term predictions below 2 ticks (1 second), the $TI$ metric provides significant enhancement with average \( \Delta PnL \approx 0.10 \times 10^{6} \) points while maintaining measurable contributions with average \( \Delta PnL \approx 0.05 \times 10^{6} \) points at extended horizons. Conversely, the $AvgEn$ metric becomes increasingly dominant beyond 240 ticks (2 minutes), with its explanatory power growing monotonically as \( AvgEn = \frac{1}{n}\sum_{k=1}^n e_k \) effectively captures the cumulative $OFI$ effect over extended periods. This aligns with the theoretical expectation that long-window $OFI$ aggregates (represented by $AvgEn$) substantially improve predictions in the 2-10 minute range \( t \in [240, 3600] \) ticks, whereas high-frequency signals like $TI$ excel at sub-5-tick horizons.

\section{Time-varying efficacy of metrics with regime-switching characteristics}
Our prior analysis examined factor efficacy evaluation, yet empirical trading reveals substantial temporal variability in factor effectiveness - to the extent that certain markets become intermittently tradable. Table \ref{tab:correlation_mon} presents monthly autocorrelation coefficients for lagged \( e_n \) values, where each row corresponds to a specific month's delayed autocorrelation. The data demonstrate significant cross-month divergence in same-lag autocorrelation, particularly beyond the 3rd lag: October 2024 (row 2410) and February 2024 exhibit strong 3rd-order autocorrelation of 0.074 and 0.08 respectively, whereas August 2024 shows merely 0.019 - a multi-fold difference that underscores the temporal non-stationarity of \( e_n \)'s predictive structure.

\begin{figure}[H] 
  \centering
  \begin{minipage}[b]{0.48\textwidth}
    \centering
    %\includegraphics[width=\linewidth,adjust=height*1.5]{figures/combined1yr_en_mt_lag10_autocorr_12months1.PNG}
    %\includegraphics[width=\linewidth, height=1.5\totalheight]{figures/combined1yr_en_mt_lag10_autocorr_12months1.PNG}
    %\adjustbox{width=\linewidth, height=0.33\totalheight} {\includegraphics{figures/combined1yr_en_mt_lag10_autocorr_12months1.PNG}}
    %\adjustbox{width=\linewidth}{\includegraphics{figures/combined1yr_en_mt_lag10_autocorr_12months1.PNG}}
    \includegraphics[width=\linewidth, height=0.21\textheight]{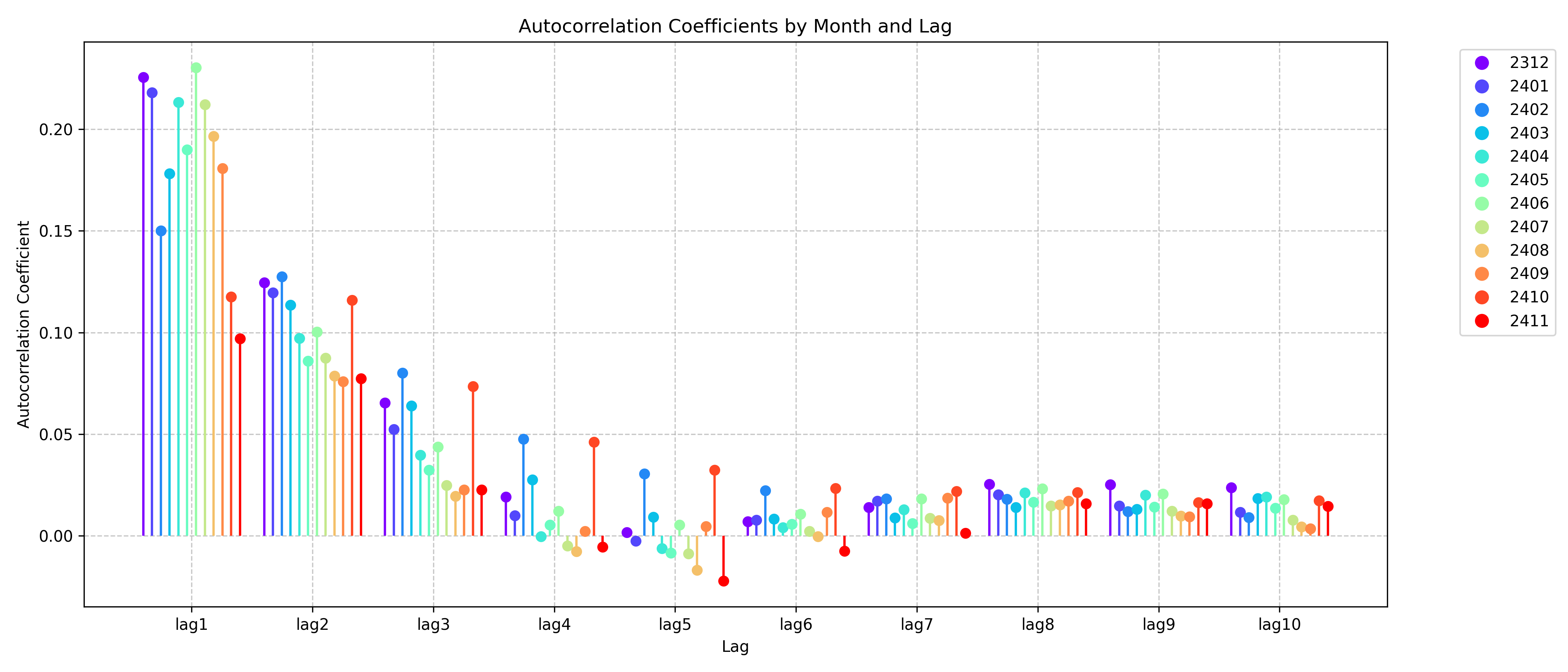} 
    \caption*{(a) Autocorrelation}  % 使用caption*取消编号，手动添加(a)标签
  \end{minipage}
  \hfill 
  \begin{minipage}[b]{0.48\textwidth}
    \centering
    \includegraphics[width=\linewidth]{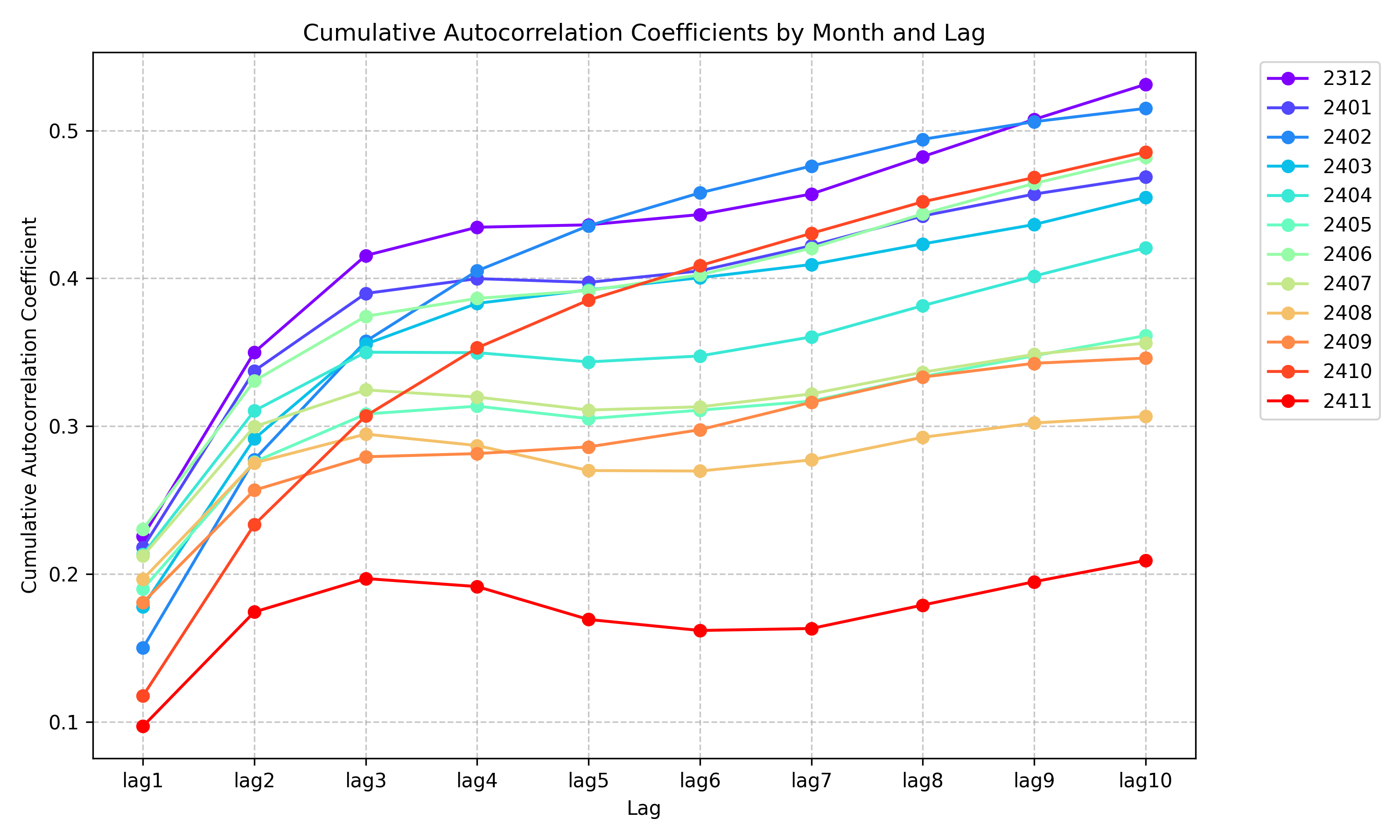}
    \caption*{(b) Cumulative autocorrelation}  % 使用caption*取消编号，手动添加(b)标签
  \end{minipage}
  \caption{Autocorrelation and cumulative autocorrelation of $e_n$ from lag 1 to lag 10 in different month(a) autocorrelation  (b) cumulative autocorrelation}
  \label{fig:en_corr_month}
\end{figure}

Figure \ref{fig:en_corr_month} visualizes the tabular data with each month's time series represented by distinct rainbow-color-coded curves, demonstrating the chromatic segregation of monthly autocorrelation patterns for immediate visual comparison.
%\begin{table}[htbp]
\begin{table}[H]
  \centering
{\footnotesize
  \begin{tabularx}{\textwidth}{|*{11}{X|}}  % 15列自动调整宽度
    \hline
    Mon &	lag1 & lag2 & lag3 & lag4 & lag5 &	lag6 &	lag7 & lag8 & lag9 &	lag10 \\
    \hline
2312 & 0.226& 0.124& \textbf{0.065}& 	0.019& 	0.002& 	0.007& 	0.014& 	0.025& 	0.025& 	0.024 \\
2401& 0.218 &	0.120& 	0.052& 	0.010& 	-0.003& 0.008& 	0.017 &	0.020 &	0.015& 	0.012 \\
2402 & 0.150 & 0.127 & \textbf{0.080} & \textbf{0.048} & \textbf{0.030} & 0.022 & 0.018 & 0.018 & 0.012 & 0.009 \\
2403 & 0.178 & 0.113 & 0.064 & 0.028 & 0.009 & 0.008 & 0.009 & 0.014 & 0.013 & 0.018 \\
2404 & 0.213 & 0.097 & 0.040 & 0.000 & -0.006 & 0.004 & 0.013 & 0.021 & 0.020 & 0.019 \\
2405 & 0.190 & 0.086 & 0.032 & 0.005 & -0.008 & 0.006 & 0.006 & 0.017 & 0.014 & 0.014 \\
2406 & 0.230 & 0.100 & 0.044 & 0.012 & 0.005 & 0.011 & 0.018 & 0.023 & 0.021 & 0.018 \\
2407 & 0.212 & 0.088 & 0.025 & -0.005 & -0.009 & 0.002 & 0.009 & 0.015 & 0.012 & 0.008 \\
2408 & 0.196 & 0.079 & 0.019 & -0.008 & -0.017 & 0.000 & 0.008 & 0.015 & 0.010 & 0.004 \\
2409 & 0.181 & 0.076 & 0.023 & 0.002 & 0.005 & 0.012 & 0.019 & 0.017 & 0.009 & 0.004 \\
2410 & 0.118 & 0.116 & \textbf{0.074} & \textbf{0.046} & \textbf{0.032} & 0.023 & 0.022 & 0.021 & 0.016 & 0.017 \\
2411 & 0.097 & 0.077 & 0.023 & -0.005 & -0.022 & -0.007 & 0.001 & 0.016 & 0.016 & 0.014 \\
    \hline
    % 更多行...
  \end{tabularx}
}
\caption{Autocorrelation coefficients of the \( e_n \) across different months from lag 1 to lag 10}  % 标题在表格后
  \label{tab:correlation_mon}  % 标签用于引用
\end{table}

Figure \ref{fig:en_corr_month} reveals significant variations in the memory effects of the accumulated \( e_n \) metric across different months. The metric demonstrates superior persistence during December 2023, January 2024, and October 2024 with the cumulative autocorrelation around 0.5, while exhibiting weaker performance in August and November 2024 with the cumulative autocorrelation approximately 0.2. This is a more than twofold difference. This substantial temporal variability in the metric's efficacy directly impacts regression performance, as evidenced by more empirical tests with improved model accuracy during periods of higher metric effectiveness. Consequently, practical applications demonstrate that longer back testing windows do not necessarily enhance model quality; rather, optimal window selection should align with each asset's unique characteristic timescales for \( e_n \)'s memory effects.

\begin{figure}[H] 
  \centering
  \begin{minipage}[b]{0.48\textwidth}
    \centering
    \includegraphics[width=\linewidth, height=0.18\textheight]{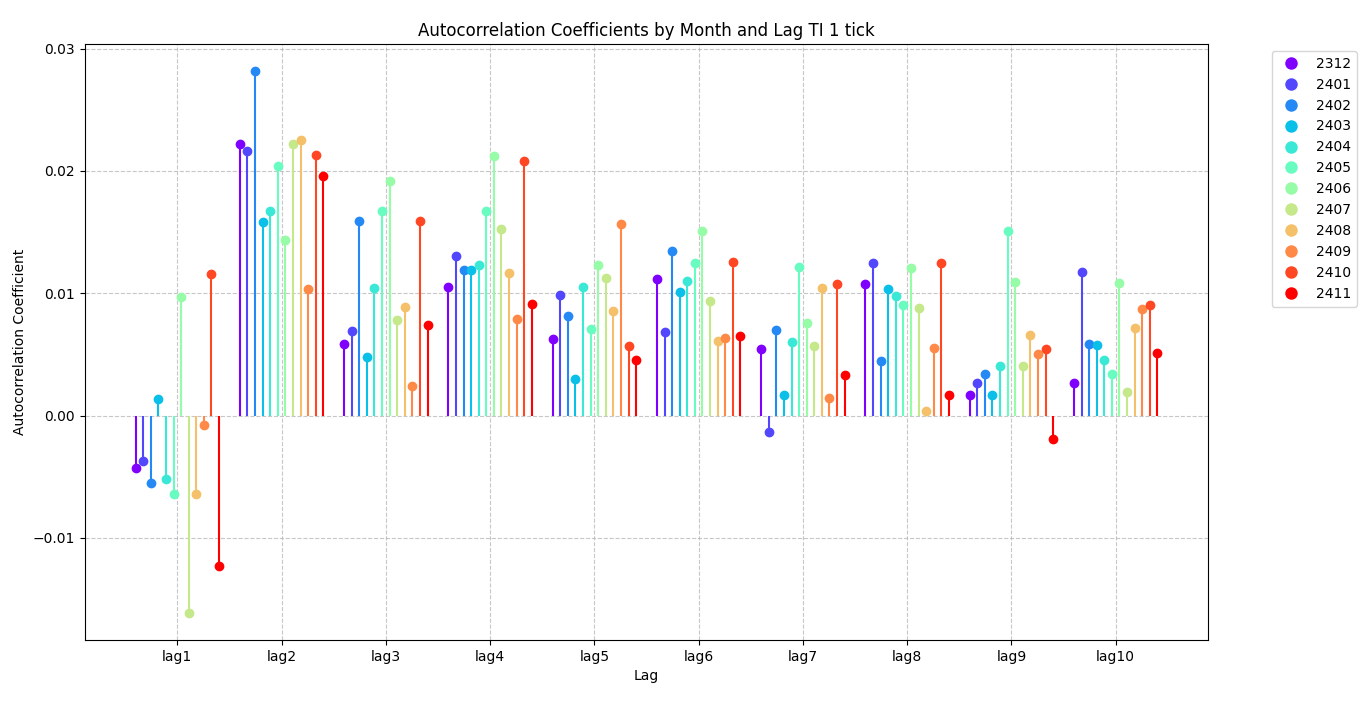} 
    \caption*{(a) Autocorrelation}  % 使用caption*取消编号，手动添加(a)标签
  \end{minipage}
  \hfill 
  \begin{minipage}[b]{0.48\textwidth}
    \centering
    \includegraphics[width=\linewidth]{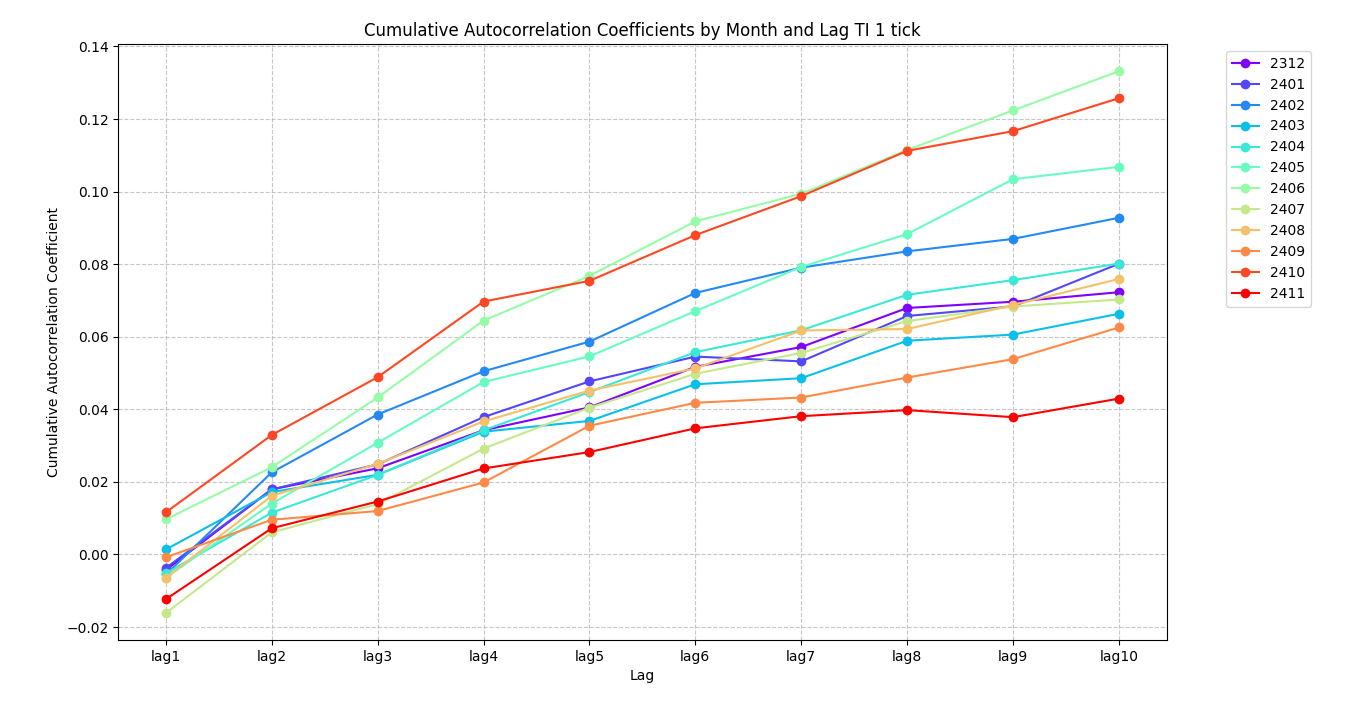}
    \caption*{(b) Cumulative autocorrelation}  % 使用caption*取消编号，手动添加(b)标签
  \end{minipage}
  \caption{Autocorrelation and cumulative autocorrelation of $\omega_n$ from lag 1 to lag 10 in different month(a) autocorrelation  (b) cumulative autocorrelation}
  \label{fig:omegan_corr_month}
\end{figure}

Similarly, we calculate the autocorrelation and cumulative autocorrelation of 1 tick \( TI \), that is, $\omega_n$ in 12 consecutive months and compare the results of different month in Figure \ref{fig:omegan_corr_month}.
The figure shows that the autocorrelation of the \( \omega_n \) at all lags decrease significantly more than those of the \( e_n \), with the first lag even exhibiting strong negative correlation. The calculations in Section 2 also show that the \( e_n \) metric has much higher contemporaneous correlation with mid-price changes than the \( \omega_n \) metric. Therefore, overall, this again verifies that the \( e_n \) is more effective than the \( \omega_n \) in microstructure price move prediction. As a result, the $OFI$ metric is more effective than $TI$ in price predictions. This also confirms our consistent method for testing various microstructure metrics, namely using both autocorrelation and the sum of autocorrelation to calculate the memory effect of metrics.

\begin{figure}[H] 
  \centering
  \begin{minipage}[b]{0.48\textwidth}
    \centering
    \includegraphics[width=\linewidth, height=0.20\textheight]{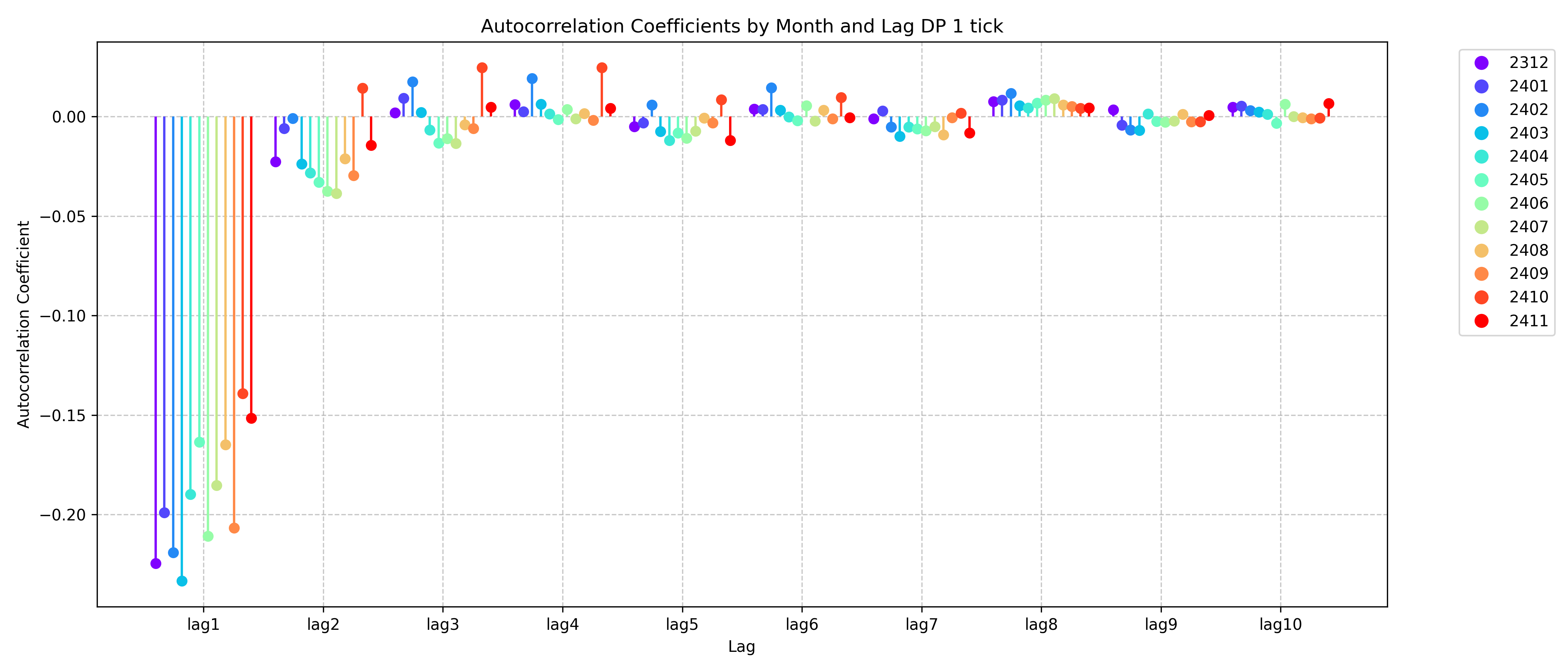} 
    \caption*{(a) Autocorrelation}  % 使用caption*取消编号，手动添加(a)标签
  \end{minipage}
  \hfill 
  \begin{minipage}[b]{0.48\textwidth}
    \centering
    \includegraphics[width=\linewidth]{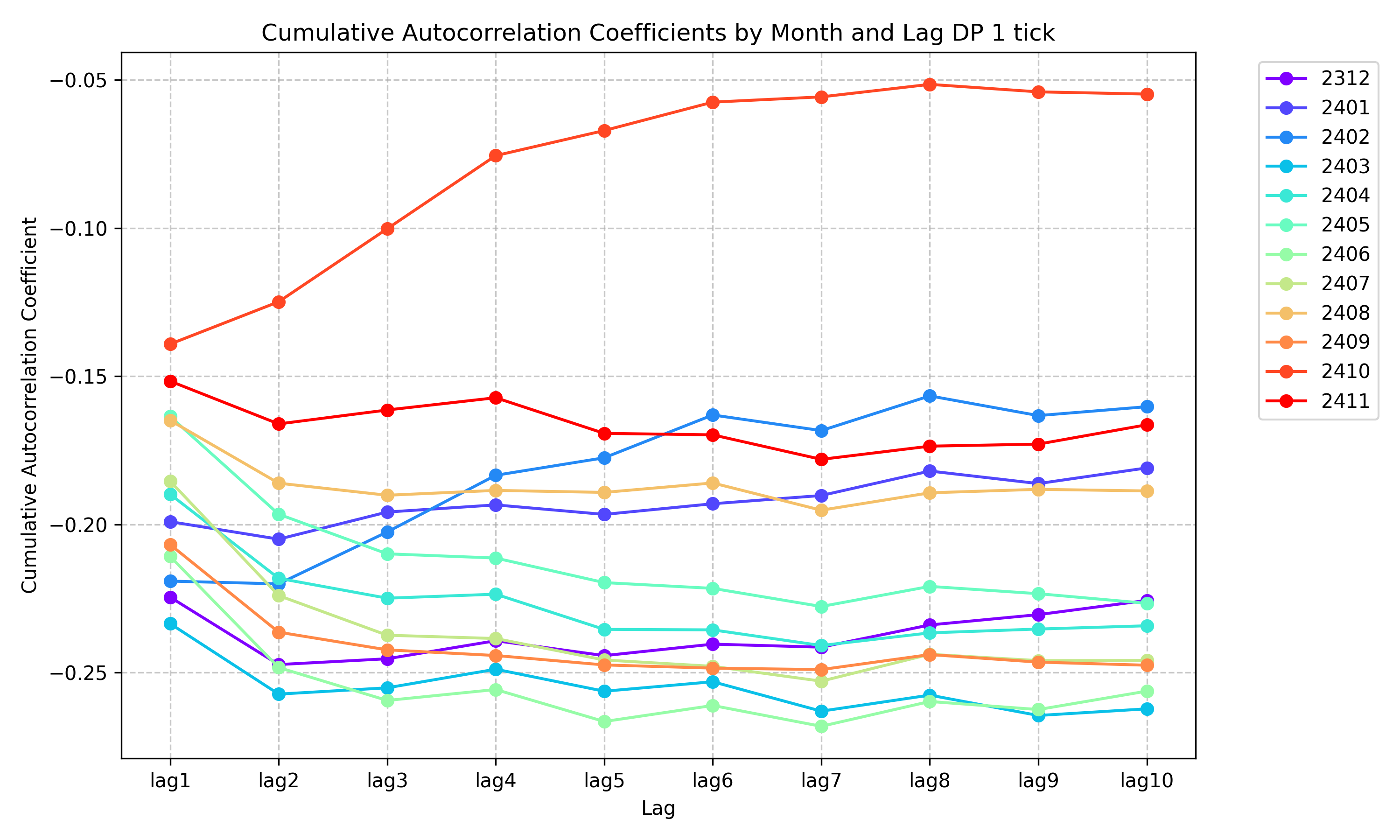}
    \caption*{(b) Cumulative autocorrelation}  % 使用caption*取消编号，手动添加(b)标签
  \end{minipage}
  \caption{Autocorrelation and cumulative autocorrelation of mid-prices changes from lag 1 to lag 10 in different month(a) autocorrelation  (b) cumulative autocorrelation}
  \label{fig:dp_corr_month}
\end{figure}

We now further examine whether the mid-price changes itself exhibit memory effects. Figure \ref{fig:dp_corr_month} displays the autocorrelation and cumulative autocorrelation of mid-price changes for each month.
The figure shows that tick-level mid-price changes exhibit weak memory effects. In the short term, their first-order autocorrelation is negative, and long-term cumulative autocorrelation also tends to be negative, indicating the mid-price changes factor has a negative directional effect on overall price prediction. Notably, during October 2024 - a high-volatility month - the higher-order autocorrelation of this indicator show predominantly positive values, demonstrating clear momentum characteristics. This coincides with significant fluctuations in the CSI 300 index, suggesting potential market efficiency degradation where price momentum factors may undergo qualitative changes in autocorrelation patterns. This further confirms the instability of this factor. In contrast, the $OFI$ metric maintains almost all positive autocorrelation across all months and all lags, demonstrating robust correlation stability. We have following conclusions for this section.
\begin{enumerate}
    \item Different metric exhibits significant variations across different months, which we refer to as regime changes, potentially due to differences in market efficiency. Therefore, rolling calculations are necessary during backtesting and model application, as using all historical data for backtesting and prediction may yield suboptimal results.
    \item A robust metric maintains memory effects in its autocorrelation in every month, though the strength of these memory effects varies significantly between periods. Taking the $OFI$ metric as an example, these variations represent quantitative rather than qualitative changes - the metric retains consistent directional effects (remaining positive across all time periods) while exhibiting different degrees of autocorrelation and memory persistence.
    \item Unstable metrics may exhibit sign variations in their autocorrelation across different months, coupled with poor memory persistence - as exemplified by mid-price changes metric. While such metrics can potentially be used in combination with other metrics, their standalone predictive performance remains suboptimal.
\end{enumerate}

\section{Future research directions} 
First, this study does not incorporate the influence of market depth. Research in \cite{cont1} shows that the depth of the market exhibits significant characteristics that vary over time and jointly affects the price dynamics with the $OFI$. Since our modeling framework primarily focuses on the stochastic process properties of the $OFI$ constructed from the \( e_n \) metric, we exclusively examine the average characteristics of this metric. Future research should integrate market depth into the analytical scope.

Second, this study does not conduct an in-depth investigation of market volatility factors. As evident from the O-U process modeling in Section 2, volatility significantly impacts both the market's average return rate and the expected Sharpe ratio. It can be anticipated that to optimize the trading Sharpe ratio, position-holding periods should be extended in low-volatility environments and shortened during high-volatility regimes. This process also requires quantitative modeling for precise implementation.

Third, more empirical analysis is needed for the logarithmic return variance. It is necessary to optimize, model, and conduct empirical research on the Quasi-Sharpe ratio.

\section{Conclusion}
In this study, we investigate and model the order flow imbalance $OFI$ metric in CSI 300 index futures. We find that order flow imbalance can be modeled as a market shock characterized by rapid initiation, prolonged response duration, memory effects, and temporal asymmetry. The shock response can be represented by a mean-reverting Ornstein-Uhlenbeck process. By coupling this response with geometric Brownian motion, we derive the temporal evolution of post-shock market return, mean, variance, and Sharpe ratios. Empirical data validates the derived return mean dynamics. Furthermore, we test the combined effectiveness of $OFI$ with other common metrics, revealing significantly varying predictive performance across different time horizons. Our analysis of $OFI$'s time-varying efficacy identifies distinct market regimes of high and low efficiency, with the latter presenting trading opportunities. Notably, robust metrics like $OFI$ exhibit only quantitative variations (no sign reversals) across regimes. This memory persistence and sign invariance can serve as selection criteria for predictive metrics. An effective metric demonstrates the following characteristics:
\begin{enumerate}
    \item The metric exhibits significant contemporaneous correlation with price movements, with stable correlation coefficients across different time windows.
    \item The metric demonstrates memory effects, characterized by high autocorrelation with slow coefficient decay.
    \item Across varying market regimes, the metric maintains stability in both aforementioned properties, exhibiting only quantitative variations while preserving both the sign consistency and fundamental stability of its numerical values.
\end{enumerate}

% References

\section*{Appendix}  % 不编号的总标题
\addcontentsline{toc}{section}{Appendix}  % 添加到目录中

\appendix  % 开始附录模式，后面的 section 自动变成 A.1, A.2...

\renewcommand{\thesubsection}{\Alph{section}.\arabic{subsection}} % 添加这行
\setcounter{section}{0} % 重置section计数器
\setcounter{subsection}{0} % 重置subsection计数器

\section{Logarithmic return process derivations}
\subsection{The variance of a certain O-U process} \label{app:mut_var}
A zero-mean-reverting Ornstein-Uhlenbeck (OU) process driven by a Lévy process is described by the following stochastic differential equation:
\begin{equation*}
    d\mu_t=-\theta\mu_t dt+dL_t
\end{equation*}
The term \( \sigma_L^2 = E\left[L_t^2\right]/t \) represents the per-unit-time variance of the Lévy process. The initial value of \( \mu_t \) is \( \mu_0 \), that is, \( \mu_t = \mu_0 \) when \( t = 0 \). Derive the time-dependent variance formula for this process \( \mu_t \).
\begin{equation*}
    \mathrm{Var}\left(\mu_t\right)=\frac{\sigma_L^2}{2\theta}\left(1-e^{-2\theta t}\right)
\end{equation*}

Proof: Step 1, Solve the stochastic differential equation for \( \mu_t \) using the integrating factor method. The solution can be expressed as:
\begin{equation*}
    \mu_t=\mu_0e^{-\theta t}+\int_{0}^{t}e^{-\theta\left(t-s\right)} dL_s
\end{equation*}
It's well known that:
\begin{equation*}
    \mathrm{Var}\left(\mu_t\right)=E\left[\mu_t^2\right]-\left(E\left[\mu_t\right]\right)^2
\end{equation*}
Step 2, Compute \( \mathrm{E}\left[\mu_t^2\right] \) and \( \mathrm{E}\left[\mu_t\right] \) separately:
\begin{equation*}
    E\left[\mu_t\right]=E\left[\mu_0e^{-\theta t}+\int_{0}^{t}e^{-\theta\left(t-s\right)} d L_s\right]=\mu_0e^{-\theta t}+\int_{0}^{t}e^{-\theta\left(t-s\right)} E\left[dL_s\right]=\mu_0e^{-\theta t}
\end{equation*}
Continue to calculate $\mathrm{E}\left[\mu_t^2\right]$, substitute the expression of $\mu_t$, we have:
\begin{equation*}
    E\left[\mu_t^2\right]=E\left[\left(\mu_0e^{-\theta t}+\int_{0}^{t}e^{-\theta\left(t-s\right)} d L_s\right)^2\right]
\end{equation*}
Expanding the expression, we have:
\begin{equation*}
    E\left[\mu_t^2\right]=E\left[\mu_0^2e^{-2\theta t}+2\mu_0e^{-\theta t}\int_{0}^{t}e^{-\theta\left(t-s\right)} d L_s+\left(\int_{0}^{t}e^{-\theta\left(t-s\right)} d L_s\right)^2\right]
\end{equation*}
Since the increments of the Lévy process are independent with zero mean (\( \mathrm{E}\left[dL_s\right] = 0 \)), the cross-term in the above expression vanishes, yielding:
\begin{equation}
    E\left[\mu_t^2\right]=\mu_0^2e^{-2\theta t}+E\left[\left(\int_{0}^{t}e^{-\theta\left(t-s\right)} d L_s\right)^2\right] \label{eq:emu2}
\end{equation}
According to the properties of stochastic integrals, we have:
\begin{equation*}
    E\left[\left(\int_{0}^{t}e^{-\theta\left(t-s\right)} d\ L_s\right)^2\right]=\int_{0}^{t}\int_{0}^{t}{e^{-\theta\left(t-s\right)}e^{-\theta\left(t-u\right)}E\left[dL_sdL_u\right]}
\end{equation*}
As $L_t$ carries the properties of Lévy process:
\begin{equation*}
    E\left[dL_sdL_u\right]=\sigma_L^2\delta\left(s-u\right) ds du
\end{equation*}
where \( \delta \) denotes the Dirac delta function, the integral simplifies to:
\begin{equation*}
    E\left[\left(\int_{0}^{t}e^{-\theta\left(t-s\right)} d\ L_s\right)^2\right]=\sigma_L^2\int_{0}^{t}e^{-2\theta\left(t-s\right)} ds
\end{equation*}
Calculate the integrals, we have
\begin{equation*}
    \int_{0}^{t}e^{-2\theta\left(t-s\right)} ds=e^{-2\theta t}\int_{0}^{t}e^{2\theta s} ds=e^{-2\theta t}\left[\frac{e^{2\theta s}}{2\theta}\right]_0^t=e^{-2\theta t}\left(\frac{e^{2\theta t}-1}{2\theta}\right)=\frac{1-e^{-2\theta t}}{2\theta}
\end{equation*}
Therefore:
\begin{equation*}
    E\left[\left(\int_{0}^{t}e^{-\theta\left(t-s\right)} d\ L_s\right)^2\right]=\sigma_L^2\frac{1-e^{-2\theta t}}{2\theta}
\end{equation*}
Substitute into \eqref{eq:emu2}, we have:
\begin{equation*}
    E\left[\mu_t^2\right]=\mu_0^2e^{-2\theta t}+\sigma_L^2\frac{1-e^{-2\theta t}}{2\theta}
\end{equation*}
Step 3, combining the above two items, we have:
\begin{equation*}
    \mathrm{Var}\left(\mu_t\right)=E\left[\mu_t^2\right]-\left(E\left[\mu_t\right]\right)^2=\mu_0^2e^{-2\theta t}+\sigma_L^2\frac{1-e^{-2\theta t}}{2\theta}-\mu_0^2e^{-2\theta t}=\sigma_L^2\frac{1-e^{-2\theta t}}{2\theta}
\end{equation*}
Rearrange, we obtain:
\begin{equation*}
    \mathrm{Var}\left(\mu_t\right)=\frac{\sigma_L^2}{2\theta}\left(1-e^{-2\theta t}\right)
\end{equation*}
\begin{flushright}
Q.E.D.
\end{flushright}
\subsection{The variance of logarithmic return}  \label{app:logarithmic_ret_var}
The temporal evolution of log-returns follows the process:
\begin{equation}
    \ln{\left(\frac{S_t}{S_0}\right)}=\frac{\mu_0}{\theta}\left(1-e^{-\theta t}\right)+\frac{1}{\theta}\int_{0}^{t}{\left(1-e^{-\theta\left(t-u\right)}\right)dL_u}-\frac{1}{2}\sigma^2t+\sigma W_t \label{eq:logret_expression}
\end{equation}
Prove that the variance of logarithmic return is the following:
\begin{equation*}
    \mathrm{Var}\left[\ln{\left(\frac{S_t}{S_0}\right)}\right]=\sigma^2t+\frac{\sigma_L^2}{\theta^2}\left[t-\frac{2}{\theta}\left(1-e^{-\theta t}\right)+\frac{1}{2\theta}\left(1-e^{-2\theta t}\right)\right]
\end{equation*}
Where:
\begin{itemize}
    \item $W_t$: The standard Brownian motion.
    \item $\sigma_L^2$: \( \sigma_L^2 = \text{E}\left[L_t^2\right]/t \) represents the per-unit-time variance of the Lévy process. and $\sigma_L^2 := \lim_{t\rightarrow0} \frac{\text{E}[L_t^2]}{t}$
    \item $\mu_t$: A zero-mean O-U process driven by a Lévy process is described by the stochastic differential equation: $d\mu_t = -\theta \mu_t dt + dL_t$
    \item \( L_t \): A zero-mean, finite second-moment Lévy process with symmetric distribution.
\end{itemize}
In addition, it's under the assumption that the Lévy process driving the drift term is independent of the Wiener process in the geometric Brownian motion component.

Proof: Step 1, decompose the log-return process in \eqref{eq:logret_expression} into four components, where the variance is exclusively determined by the stochastic terms (second and fourth terms). Define:
\begin{equation*}
    \mathrm{A}_\mathrm{t}=\frac{1}{\theta}\int_{0}^{t}{\left(1-e^{-\theta\left(t-u\right)}\right)dL_u}
\end{equation*}
Given the independence between the Lévy process driving the drift term and the Wiener process in the geometric Brownian motion component, we derive:
\begin{equation}
    \mathrm{Var}\left[\ln{\left(\frac{S_t}{S_0}\right)}\right]=\mathrm{Var}\left(\mathrm{A}_\mathrm{t}\right)+\sigma^2t
\end{equation}
Step 2, We compute \( \mathrm{Var}\left(\mathrm{A}_\mathrm{t}\right) \) by applying the isometry property of Lévy integrals, define:
\begin{equation*}
    B\left(u\right)=\frac{1-e^{-\theta\left(t-u\right)}}{\theta}
\end{equation*}
We have:
\begin{equation*}
    \mathrm{Var}\left[A_t\right]=E\left[\left(\int_{0}^{t}{B\left(u\right)dL_u}\right)^2\right]=\int_{0}^{t}{B\left(u\right)^2\sigma_L^2du}
\end{equation*}
where \( \sigma_L^2 = \text{E}\left[L_t^2\right]/t \) denotes the per-unit-time variance of the Lévy process. Substituting and expanding yields:
\begin{equation*}
    \mathrm{Var}\left[A_t\right]=\sigma_L^2\int_{0}^{t}{\left(\frac{1-e^{-\theta\left(t-u\right)}}{\theta}\right)^2du}
\end{equation*}
By substituting \( s = t - u \) and changing the integration variable (where \( u \) ranges from 0 to \( t \), corresponding to \( s \) from \( t \) to 0), we obtain:
\begin{equation*}
    \mathrm{Var}\left[A_t\right]=\sigma_L^2\int_{0}^{t}{\left(\frac{1-e^{-\theta s}}{\theta}\right)^2ds}=\frac{\sigma_L^2}{\theta^2}\int_{0}^{t}{\left(1-e^{-\theta s}\right)^2ds}
\end{equation*}
Expand the integrand:
\begin{equation*}
    \left(1-e^{-\theta s}\right)^2=1-2e^{-\theta s}+e^{-2\theta s}
\end{equation*}
After integrating term by term, we obtain:
\begin{equation*}
    \int_{0}^{t}{\left(1-e^{-\theta s}\right)^2ds}=t-\frac{2}{\theta}\left(1-e^{-\theta t}\right)+\frac{1}{2\theta}\left(1-e^{-2\theta t}\right)
\end{equation*}
Therefore:
\begin{equation*}
    \mathrm{Var}\left[A_t\right]=\frac{\sigma_L^2}{\theta^2}\left[t-\frac{2}{\theta}\left(1-e^{-\theta t}\right)+\frac{1}{2\theta}\left(1-e^{-2\theta t}\right)\right]
\end{equation*}
Combining with the variance from the Brownian motion term, we obtain:
\begin{equation*}
    \mathrm{Var}\left[\ln{\left(\frac{S_t}{S_0}\right)}\right]=\sigma^2t+\frac{\sigma_L^2}{\theta^2}\left[t-\frac{2}{\theta}\left(1-e^{-\theta t}\right)+\frac{1}{2\theta}\left(1-e^{-2\theta t}\right)\right]
\end{equation*}
\begin{flushright}
Q.E.D.
\end{flushright}

\subsection{Quasi-Sharpe ratio asymptotic value} \label{app:quasi_sharpe}
The quasi-Sharpe ratio formula based on the expected log-return and standard deviation is given by:
\begin{equation*}
\mathrm{QuasiSharpe}\left(t\right)=\mathrm{\mathrm{QS}}\left(t\right)=\frac{\mu_0\cdot\frac{1-e^{-\theta t}}{\theta}-\frac{1}{2}\sigma^2t}{\sqrt{\sigma^2t+\frac{\sigma_L^2}{\theta^2}\left(t-\frac{2}{\theta}\left(1-e^{-\theta t}\right)+\frac{1}{2\theta}\left(1-e^{-2\theta t}\right)\right)}}
\end{equation*}
As \( t \rightarrow 0^+ \), prove the following equality:
\begin{gather*}
    \lim_{t\rightarrow0^+}{\mathrm{QS}}\left(t\right)=0 \\
    \mathrm{QS}\left(t\right)\sim\left(\frac{\mu_0-\frac{1}{2}\sigma^2}{\sigma}\right)\cdot\sqrt t
\end{gather*}
Proof: Observe that as \( t \rightarrow 0^+ \), both the numerator and denominator approach 0, resulting in an indeterminate form. We investigate its asymptotic behavior near \( t \rightarrow 0^+ \) through Taylor expansion. First, expand the fractional component of the numerator:
\begin{equation*}
    \frac{1-e^{-\theta t}}{\theta}=t-\frac{1}{2}\theta t^2+\mathcal{O}\left(t^3\right)
\end{equation*}
Therefore, the expected log-return represented by the numerator is:
\begin{equation*}
    E\left[\ln{\left(S_t/S_0\right)}\right]=\left(\mu_0-\frac{1}{2}\sigma^2\right)t-\frac{1}{2}\mu_0\theta t^2+\mathcal{O}\left(t^3\right)
\end{equation*}
Expand the variance term in the denominator by decomposing it as follows:
\begin{gather*}
    1-e^{-\theta t}=\theta\ t-\frac{1}{2}\theta^2t^2+\mathcal{O}\left(t^3\right) \\
    1-e^{-2\theta t}=2\theta t-2\theta^2t^2+\mathcal{O}\left(t^3\right)
\end{gather*}
Substitute into the variance equation:
\begin{gather*}
    \mathrm{Var}\left[\ln{\left(S_t/S_0\right)}\right]=\sigma^2t+\frac{\sigma_L^2}{\theta^2}\left[t-\frac{2}{\theta}\left(\theta t-\frac{1}{2}\theta^2t^2\right)+\frac{1}{2\theta}\left(2\theta t-2\theta^2t^2\right)+\mathcal{O}\left(t^3\right)\right]\\
    \mathrm{Var}=\sigma^2t+\frac{\sigma_L^2}{\theta^2}\left(t-\left(2t-\theta t^2\right)+\left(t-\theta t^2\right)+\mathcal{O}\left(t^3\right)\right)=\sigma^2t+\frac{\sigma_L^2}{\theta^2}\left(0+0+\mathcal{O}\left(t^3\right)\right)=\sigma^2t+\mathcal{O}\left(t^3\right)\\
    \mathrm{Var}\left[\ln{\left(S_t/S_0\right)}\right]=\sigma^2t+\mathcal{O}\left(t^3\right)
\end{gather*}
The square root of above is:
\begin{equation*}
\sqrt{\mathrm{Var}\left[\ln{\left(S_t/S_0\right)}\right]}=\sigma\sqrt t+\mathcal{O}\left(t^{5/2}\right)
\end{equation*}
Substitute into the quasi-Sharpe ratio equation:
\begin{equation*}
    \mathrm{QS}\left(t\right)=\frac{\left(\mu_0-\frac{1}{2}\sigma^2\right)t-\frac{1}{2}\mu_0\theta t^2+\mathcal{O}\left(t^3\right)}{\sigma\sqrt t+\mathcal{O}\left(t^{5/2}\right)}=\frac{\mu_0 - \frac{1}{2} \sigma^2}{\sigma} \sqrt{t} - \frac{1}{2} \frac{\mu_0 \theta}{\sigma} t^{3/2} + \mathcal{O}(t^{5/2})
\end{equation*}
\begin{flushright}
Q.E.D.
\end{flushright}

\subsection{The maximum of the logarithmic-return mean process}  \label{app:max_main}
The log-return process is described by following:
\begin{equation*}
    E\left[R_t\right]=\mu_0\frac{1-e^{-\theta t}}{\theta}-\frac{\sigma^2}{2}t
\end{equation*}
Prove that, the maximum is obtained when t is:
\begin{equation*}
    t=-\frac{1}{\theta}\ln{\left(\frac{\sigma^2}{2\mu_0}\right)}
\end{equation*}
With the maximum being:
\begin{equation*}
    E\left[R_t\right]_{max}=\frac{\mu_0}{\theta}+\frac{\sigma^2}{2\theta}\ln{\left(\frac{\sigma^2}{2e\mu_0}\right)}
\end{equation*}
Proof:The results demonstrate that when \( t = 0 \), the expected return is zero. The expected return initially increases with \( t \), reaches a maximum, and subsequently decreases. The first term of the expression is a concave function of \( t \), while the second term is linear, resulting in an overall concave function with global maximum. We now derive this extrema by differentiating \( \text{E}\left[R_t\right] \) with respect to \( t \):
\begin{equation*}
    \frac{d}{dt}E\left[R_t\right]=\frac{\mu_0}{\theta}\cdot\frac{d}{dt}\left(1-e^{-\theta t}\right)-\frac{\sigma^2}{2}
\end{equation*}
and:
\begin{equation*}
    \frac{d}{dt}\left(1-e^{-\theta t}\right)=\theta\ e^{-\theta t}
\end{equation*}
Therefore:
\begin{equation*}
    \frac{d}{dt}E\left[R_t\right]=\mu_0e^{-\theta t}-\frac{\sigma^2}{2}
\end{equation*}
Setting the above expression equal to zero yields:
\begin{equation*}
    \mu_0e^{-\theta t}=\frac{\sigma^2}{2}
\end{equation*}
Solve it, we have:
\begin{equation*}
    t=-\frac{1}{\theta}\ln{\left(\frac{\sigma^2}{2\mu_0}\right)}
\end{equation*}
The equation further reveals that to ensure \( t \geq 0 \) (and thus maintain mathematical meaningfulness), the condition \( \frac{\sigma^2}{2\mu_0} < 1 \) must hold, which equivalently implies:
\begin{equation*}
    \mu_0>\frac{\sigma^2}{2}
\end{equation*}
This implies that the initial drift acting as a shock must be sufficiently large—specifically, large relative to the volatility—to be economically meaningful. We now substitute the derived time value \( t^* \) that maximizes the expected return into the mean return formula. Substituting \( t = -\frac{1}{\theta}\ln{\left(\frac{\sigma^2}{2\mu_0}\right)} \), the left-hand term becomes:
\begin{equation*}
    \mu_0\frac{1-e^{-\theta t}}{\theta}=\mu_0\frac{1-\frac{\sigma^2}{2\mu_0}}{\theta}=\mu_0\frac{\frac{2\mu_0-\sigma^2}{2\mu_0}}{\theta}=\frac{2\mu_0-\sigma^2}{2\theta}
\end{equation*}
The right-hand term:
\begin{equation*}
    -\frac{\sigma^2}{2}t=-\frac{\sigma^2}{2}\cdot\left(-\frac{1}{\theta}\ln{\left(\frac{\sigma^2}{2\mu_0}\right)}\right)=\frac{\sigma^2}{2\theta}\ln{\left(\frac{\sigma^2}{2\mu_0}\right)}
\end{equation*}
Combining these terms yields the maximum expected return:
\begin{equation*}
    E\left[R_t\right]_{max}=\frac{\mu_0}{\theta}+\frac{\sigma^2}{2\theta}\ln{\left(\frac{\sigma^2}{2e\mu_0}\right)}
\end{equation*}
\begin{flushright}
Q.E.D.
\end{flushright}

\clearpage  % 结束当前页并刷新所有浮动体
\section{Regression results} \label{app:regression_results}
\begin{table}[H]
  \centering
  \caption{LASSO regression results for different historical windows sizes and forecast horizons in ticks}  % 标题在表格后
  \label{tab:regression_result}  % 标签用于引用
{\footnotesize
  %\begin{tabularx}{\textwidth}{|*{11}{X|}}  % 15列自动调整宽度
  \begin{tabular}{|c|c|c c c c|c c c c|c|}
    \hline
    HistWinSize &	FcastHorzn & MSE&	$R^2$ &	Coef0&Intcpt&Trainpnl-l&Testpnl-l&Trainpnl-s&Testpnl-s&	Totalpnl \\
    \hline
1 & 1 & 0.064 & 1.123\% & 0.027 & 0.000 & \num{25246.3} & \num{6485.7} & \num{25485.9} & \num{6066.7} & \num{63284.6} \\
1 & 2 & 0.104 & 1.997\% & 0.046 & 0.000 & \num{71084.8} & \num{17779.2} & \num{70625.6} & \num{17924.4} & \num{177414.0} \\
1 & 5 & 0.275 & 1.094\% & 0.054 & 0.000 & \num{97663.2} & \num{24955.2} & \num{97110.1} & \num{24707.2} & \num{244435.7} \\
1 & 10 & 0.574 & 0.613\% & 0.060 & 0.000 & \num{110191.7} & \num{27265.3} & \num{108714.1} & \num{27145.7} & \num{273316.8} \\
1 & 20 & 1.189 & 0.309\% & 0.062 & 0.000 & \num{117124.4} & \num{29776.2} & \num{115132.6} & \num{28461.6} & \num{290494.8} \\
1 & 30 & 1.807 & 0.224\% & 0.062 & 0.001 & \num{122038.3} & \num{30288.3} & \num{117840.3} & \num{29812.8} & \num{299979.7} \\
1 & 60 & 3.737 & 0.127\% & 0.067 & 0.000 & \num{131908.9} & \num{33446.3} & \num{129166.0} & \num{31551.7} & \num{326072.9} \\
1 & 120 & 7.564 & 0.067\% & 0.066 & -0.001 & \num{135736.8} & \num{37459.2} & \num{138905.2} & \num{37165.2} & \num{349266.4} \\
1 & 240 & 14.760 & 0.030\% & 0.067 & -0.002 & \num{133894.5} & \num{34757.9} & \num{144774.1} & \num{40293.4} & \num{353719.9} \\
1 & 600 & 36.334 & 0.014\% & 0.066 & -0.016 & \num{104918.0} & \num{30332.1} & \num{191184.4} & \num{49743.7} & \num{376178.2} \\
1 & 1200 & 69.466 & 0.008\% & 0.074 & -0.023 & \num{94829.1} & \num{26753.8} & \num{211261.8} & \num{52895.2} & \num{385739.9} \\
1 & 3600 & 208.275 & 0.004\% & 0.106 & -0.004 & \num{200756.0} & \num{54234.0} & \num{218014.4} & \num{52529.4} & \num{525533.8} \\
    \hline
2 & 1 & 0.064 & 1.135\% & 0.027 & 0.000 & \num{25435.6} & \num{6269.4} & \num{25231.8} & \num{6339.1} & \num{63275.9} \\
2 & 2 & 0.105 & 1.937\% & 0.046 & 0.000 & \num{71353.2} & \num{17457.5} & \num{70840.5} & \num{17697.4} & \num{177348.6} \\
2 & 5 & 0.275 & 1.063\% & 0.054 & 0.000 & \num{98327.4} & \num{24214.9} & \num{97179.3} & \num{24604.4} & \num{244326.0} \\
2 & 10 & 0.574 & 0.663\% & 0.060 & 0.000 & \num{110370.0} & \num{27016.7} & \num{108601.1} & \num{27220.7} & \num{273208.5} \\
2 & 20 & 1.188 & 0.329\% & 0.061 & 0.000 & \num{116909.2} & \num{29911.1} & \num{114285.5} & \num{29312.5} & \num{290418.3} \\
2 & 30 & 1.819 & 0.218\% & 0.062 & 0.001 & \num{122331.8} & \num{29883.9} & \num{117752.4} & \num{29893.5} & \num{299861.6} \\
2 & 60 & 3.697 & 0.129\% & 0.067 & 0.001 & \num{133819.5} & \num{31334.6} & \num{128362.9} & \num{32413.1} & \num{325930.1} \\
2 & 120 & 7.560 & 0.062\% & 0.067 & -0.001 & \num{137301.8} & \num{35879.5} & \num{142937.1} & \num{33309.6} & \num{349428.0} \\
2 & 240 & 14.774 & 0.032\% & 0.067 & -0.002 & \num{136135.7} & \num{32521.0} & \num{146027.9} & \num{39315.9} & \num{354000.5} \\
2 & 600 & 36.231 & 0.014\% & 0.066 & -0.016 & \num{111659.6} & \num{23556.7} & \num{194227.8} & \num{47011.3} & \num{376455.4} \\
2 & 1200 & 69.241 & 0.009\% & 0.073 & -0.024 & \num{90042.5} & \num{31627.6} & \num{212555.0} & \num{51804.8} & \num{386029.9} \\
2 & 3600 & 207.281 & 0.001\% & 0.112 & -0.002 & \num{219157.2} & \num{35865.1} & \num{225822.0} & \num{44893.4} & \num{525737.7} \\
    \hline
5 & 1 & 0.065 & 1.119\% & 0.027 & 0.000 & \num{25267.4} & \num{6416.6} & \num{25307.6} & \num{6239.5} & \num{63231.1} \\
5 & 2 & 0.104 & 1.991\% & 0.046 & 0.000 & \num{71097.3} & \num{17628.2} & \num{70738.3} & \num{17712.9} & \num{177176.7} \\
5 & 5 & 0.274 & 1.075\% & 0.054 & 0.000 & \num{97928.0} & \num{24484.3} & \num{97447.4} & \num{24186.8} & \num{244046.5} \\
5 & 10 & 0.578 & 0.605\% & 0.060 & 0.000 & \num{109465.4} & \num{27821.9} & \num{108576.7} & \num{27069.5} & \num{272933.5} \\
5 & 20 & 1.190 & 0.308\% & 0.062 & 0.001 & \num{117330.6} & \num{29333.7} & \num{113958.0} & \num{29568.6} & \num{290190.9} \\
5 & 30 & 1.807 & 0.229\% & 0.062 & 0.001 & \num{121899.5} & \num{30069.5} & \num{117415.7} & \num{30182.4} & \num{299567.1} \\
5 & 60 & 3.699 & 0.136\% & 0.067 & 0.000 & \num{127878.9} & \num{36816.6} & \num{128678.3} & \num{32238.4} & \num{325612.2} \\
5 & 120 & 7.542 & 0.063\% & 0.067 & -0.001 & \num{137486.9} & \num{35453.2} & \num{140523.8} & \num{35872.1} & \num{349336.0} \\
5 & 240 & 14.815 & 0.030\% & 0.067 & -0.004 & \num{132138.1} & \num{36187.6} & \num{151653.8} & \num{33933.8} & \num{353913.3} \\
5 & 600 & 36.635 & 0.010\% & 0.069 & -0.016 & \num{105665.9} & \num{28929.6} & \num{192684.2} & \num{48780.5} & \num{376060.2} \\
5 & 1200 & 69.408 & 0.007\% & 0.075 & -0.024 & \num{94846.1} & \num{26763.0} & \num{216544.9} & \num{47842.3} & \num{385996.3} \\
5 & 3600 & 207.147 & 0.007\% & 0.097 & -0.002 & \num{203049.9} & \num{52077.2} & \num{211114.1} & \num{59881.9} & \num{526123.1} \\
  \hline
10 & 1 & 0.064 & 1.175\% & 0.027 & 0.000 & \num{25391.0} & \num{6255.9} & \num{25181.8} & \num{6313.4} & \num{63142.1} \\
10 & 2 & 0.105 & 2.009\% & 0.046 & 0.000 & \num{70491.9} & \num{18092.8} & \num{70487.9} & \num{17797.5} & \num{176870.1} \\
10 & 5 & 0.275 & 1.111\% & 0.054 & 0.000 & \num{96903.8} & \num{25280.2} & \num{96760.1} & \num{24564.1} & \num{243508.2} \\
10 & 10 & 0.575 & 0.593\% & 0.060 & 0.000 & \num{110327.9} & \num{26638.3} & \num{108884.1} & \num{26453.5} & \num{272303.8} \\
10 & 20 & 1.188 & 0.324\% & 0.061 & 0.000 & \num{117132.6} & \num{29123.6} & \num{115117.4} & \num{28172.0} & \num{289545.6} \\
10 & 30 & 1.798 & 0.229\% & 0.062 & 0.000 & \num{118770.9} & \num{32443.7} & \num{116479.8} & \num{30929.6} & \num{298624.0} \\
10 & 60 & 3.695 & 0.136\% & 0.066 & 0.000 & \num{129056.0} & \num{34604.7} & \num{126773.6} & \num{34102.4} & \num{324536.7} \\
10 & 120 & 7.536 & 0.066\% & 0.066 & 0.000 & \num{136286.9} & \num{35888.9} & \num{137816.3} & \num{38508.3} & \num{348500.4} \\
10 & 240 & 14.833 & 0.027\% & 0.067 & -0.003 & \num{136561.1} & \num{30906.2} & \num{151343.7} & \num{34163.7} & \num{352974.7} \\
10 & 600 & 36.346 & 0.010\% & 0.068 & -0.017 & \num{108717.2} & \num{24872.7} & \num{199096.6} & \num{42722.2} & \num{375408.7} \\
10 & 1200 & 69.384 & 0.010\% & 0.071 & -0.024 & \num{91870.2} & \num{29312.9} & \num{212556.5} & \num{51546.9} & \num{385286.5} \\
10 & 3600 & 205.925 & 0.005\% & 0.104 & -0.003 & \num{199741.9} & \num{55168.2} & \num{213376.1} & \num{57590.2} & \num{525876.4} \\
    \hline
20 & 1 & 0.065 & 1.163\% & 0.027 & 0.000 & \num{25754.2} & \num{5826.8} & \num{25199.6} & \num{6271.9} & \num{63052.5} \\
20 & 2 & 0.104 & 1.963\% & 0.046 & 0.000 & \num{70360.1} & \num{18001.3} & \num{70474.8} & \num{17681.3} & \num{176517.5} \\
20 & 5 & 0.272 & 1.086\% & 0.054 & 0.000 & \num{97351.6} & \num{24351.9} & \num{97083.6} & \num{24014.9} & \num{242802.0} \\
20 & 10 & 0.576 & 0.602\% & 0.060 & 0.000 & \num{109271.8} & \num{27081.7} & \num{109247.8} & \num{25890.3} & \num{271491.6} \\
20 & 20 & 1.172 & 0.314\% & 0.061 & 0.000 & \num{116864.7} & \num{28385.7} & \num{115453.0} & \num{27783.5} & \num{288486.9} \\
20 & 30 & 1.779 & 0.218\% & 0.062 & 0.001 & \num{119787.0} & \num{29930.2} & \num{116331.8} & \num{31526.7} & \num{297575.7} \\
20 & 60 & 3.673 & 0.137\% & 0.066 & 0.000 & \num{132471.5} & \num{35836.2} & \num{133840.6} & \num{33836.2} & \num{335984.5} \\
20 & 120 & 7.487 & 0.062\% & 0.066 & -0.001 & \num{137824.8} & \num{33354.6} & \num{142951.0} & \num{34041.6} & \num{348172.0} \\
20 & 240 & 14.791 & 0.029\% & 0.067 & -0.002 & \num{135729.2} & \num{30458.9} & \num{146113.1} & \num{39987.9} & \num{352289.1} \\
20 & 600 & 36.659 & 0.013\% & 0.066 & -0.016 & \num{104535.2} & \num{27168.9} & \num{189787.5} & \num{52917.9} & \num{374409.5} \\
20 & 1200 & 69.308 & 0.007\% & 0.074 & -0.023 & \num{95570.6} & \num{25064.6} & \num{210990.2} & \num{53020.3} & \num{384645.7} \\
20 & 3600 & 207.330 & 0.006\% & 0.101 & -0.001 & \num{205438.0} & \num{48755.4} & \num{210279.1} & \num{60575.5} & \num{525048.0} \\
 \hline
    % 更多行...
  \end{tabular}
}
\end{table}

\clearpage  % 结束当前页并刷新所有浮动体
\begin{table}[H]
  \centering
{\footnotesize
  %\begin{tabularx}{\textwidth}{|*{11}{X|}}  % 15列自动调整宽度
  \begin{tabular}{|c|c|c c c c|c c c c|c|}
    \hline
    HistWinSize &	FcastHorzn & MSE&	$R^2$ &	Coef0&Intcpt&Trainpnl-l&Testpnl-l&Trainpnl-s&Testpnl-s&	Totalpnl \\
    \hline
30 & 1 & 0.064 & 1.099\% & 0.027 & 0.000 & \num{25630.2} & \num{5894.6} & \num{25342.7} & \num{6083.6} & \num{62951.1} \\
30 & 2 & 0.106 & 2.023\% & 0.046 & 0.000 & \num{70794.7} & \num{17390.1} & \num{70469.5} & \num{17540.6} & \num{176194.9} \\
30 & 5 & 0.270 & 1.080\% & 0.054 & 0.000 & \num{96766.8} & \num{24521.0} & \num{96974.6} & \num{23853.1} & \num{242115.5} \\
30 & 10 & 0.569 & 0.599\% & 0.060 & 0.000 & \num{109202.7} & \num{26370.3} & \num{107827.2} & \num{27070.1} & \num{270470.3} \\
30 & 20 & 1.174 & 0.313\% & 0.061 & 0.000 & \num{115197.4} & \num{28579.6} & \num{114535.8} & \num{28760.6} & \num{287073.4} \\
30 & 30 & 1.788 & 0.215\% & 0.062 & 0.000 & \num{118551.2} & \num{29646.7} & \num{116578.8} & \num{31519.2} & \num{296295.9} \\
30 & 60 & 3.665 & 0.117\% & 0.067 & 0.000 & \num{129432.6} & \num{31026.2} & \num{130623.1} & \num{31051.2} & \num{322133.1} \\
30 & 120 & 7.456 & 0.063\% & 0.066 & 0.000 & \num{136477.6} & \num{33359.1} & \num{138497.5} & \num{38310.2} & \num{346644.4} \\
30 & 240 & 14.973 & 0.025\% & 0.067 & -0.002 & \num{134445.5} & \num{30131.6} & \num{146999.6} & \num{39165.3} & \num{350742.0} \\
30 & 600 & 36.435 & 0.014\% & 0.064 & -0.017 & \num{101257.7} & \num{23231.9} & \num{189901.2} & \num{47545.2} & \num{361936.0} \\
30 & 1200 & 69.084 & 0.007\% & 0.073 & -0.022 & \num{95428.0} & \num{23254.4} & \num{207466.9} & \num{54486.5} & \num{380635.8} \\
30 & 3600 & 206.145 & 0.006\% & 0.099 & 0.001 & \num{88341.6} & \num{21040.7} & \num{84006.5} & \num{42334.5} & \num{235723.3} \\ 
    \hline
60 & 1 & 0.064 & 1.130\% & 0.027 & 0.000 & \num{25132.4} & \num{6194.4} & \num{24972.0} & \num{6413.2} & \num{62712.0} \\
60 & 2 & 0.104 & 1.952\% & 0.046 & 0.000 & \num{70438.6} & \num{17125.9} & \num{70068.4} & \num{17626.7} & \num{175259.6} \\
60 & 5 & 0.271 & 1.066\% & 0.053 & 0.000 & \num{96271.5} & \num{23831.5} & \num{95912.5} & \num{24430.7} & \num{240446.2} \\
60 & 10 & 0.573 & 0.660\% & 0.058 & 0.000 & \num{106348.5} & \num{27559.5} & \num{106231.6} & \num{28102.2} & \num{268241.8} \\
60 & 20 & 1.171 & 0.317\% & 0.060 & 0.000 & \num{113474.8} & \num{28269.3} & \num{113744.3} & \num{28794.8} & \num{284283.2} \\
60 & 30 & 1.762 & 0.201\% & 0.061 & 0.000 & \num{117401.7} & \num{28675.5} & \num{117454.8} & \num{29859.5} & \num{293391.5} \\
60 & 60 & 3.638 & 0.121\% & 0.066 & -0.001 & \num{131960.8} & \num{33033.3} & \num{136079.4} & \num{31921.5} & \num{332995.0} \\
60 & 120 & 7.417 & 0.057\% & 0.066 & -0.001 & \num{134444.9} & \num{34416.7} & \num{142060.5} & \num{33864.7} & \num{344786.8} \\
60 & 240 & 14.707 & 0.032\% & 0.065 & -0.002 & \num{129581.3} & \num{33475.7} & \num{142541.8} & \num{43627.4} & \num{349226.2} \\
60 & 600 & 36.490 & 0.010\% & 0.066 & -0.019 & \num{96620.3} & \num{27396.2} & \num{195226.5} & \num{41636.9} & \num{360879.9} \\
60 & 1200 & 68.952 & 0.008\% & 0.071 & -0.023 & \num{94535.9} & \num{25422.5} & \num{210497.9} & \num{49165.6} & \num{379621.9} \\
60 & 3600 & 207.066 & 0.003\% & 0.106 & -0.003 & \num{201098.6} & \num{53340.8} & \num{213044.9} & \num{55778.4} & \num{523262.7} \\
 \hline
120 & 1 & 0.063 & 1.117\% & 0.027 & 0.000 & \num{24800.0} & \num{6395.8} & \num{25238.0} & \num{5996.7} & \num{62430.5} \\
120 & 2 & 0.103 & 2.016\% & 0.045 & 0.000 & \num{69423.2} & \num{17592.4} & \num{69468.8} & \num{17606.1} & \num{174090.5} \\
120 & 5 & 0.268 & 1.020\% & 0.053 & 0.000 & \num{95864.0} & \num{23564.1} & \num{95455.4} & \num{23976.7} & \num{238860.2} \\
120 & 10 & 0.557 & 0.612\% & 0.059 & 0.000 & \num{106950.8} & \num{26479.2} & \num{106793.7} & \num{26556.4} & \num{266780.1} \\
120 & 20 & 1.142 & 0.314\% & 0.059 & 0.000 & \num{118102.9} & \num{30080.1} & \num{119835.3} & \num{28237.6} & \num{296255.9} \\
120 & 30 & 1.735 & 0.210\% & 0.060 & 0.000 & \num{116903.7} & \num{28607.5} & \num{114719.4} & \num{30563.2} & \num{290793.8} \\
120 & 60 & 3.554 & 0.123\% & 0.065 & 0.000 & \num{126575.6} & \num{31402.7} & \num{127096.8} & \num{31517.0} & \num{316592.1} \\
120 & 120 & 7.331 & 0.053\% & 0.066 & -0.001 & \num{134518.2} & \num{33364.0} & \num{142174.4} & \num{31116.5} & \num{341173.1} \\
120 & 240 & 14.725 & 0.030\% & 0.064 & -0.004 & \num{126744.3} & \num{34440.8} & \num{148422.3} & \num{35779.7} & \num{345387.1} \\
120 & 600 & 36.262 & 0.012\% & 0.064 & -0.016 & \num{98589.5} & \num{27840.5} & \num{185565.4} & \num{44341.5} & \num{356336.9} \\
120 & 1200 & 68.901 & 0.007\% & 0.069 & -0.022 & \num{93458.7} & \num{26527.5} & \num{205403.3} & \num{45697.0} & \num{371086.5} \\
120 & 3600 & 205.863 & 0.002\% & 0.104 & 0.000 & \num{90607.7} & \num{20286.4} & \num{89337.3} & \num{25754.7} & \num{225986.1} \\
 \hline
240 & 1 & 0.063 & 1.141\% & 0.027 & 0.000 & \num{24521.4} & \num{6232.5} & \num{24748.9} & \num{6118.8} & \num{61621.6} \\
240 & 2 & 0.101 & 1.971\% & 0.045 & 0.000 & \num{68687.9} & \num{17069.7} & \num{68732.8} & \num{17239.9} & \num{171730.3} \\
240 & 5 & 0.263 & 1.059\% & 0.053 & 0.000 & \num{93624.3} & \num{23936.7} & \num{94289.9} & \num{23604.5} & \num{235455.4} \\
240 & 10 & 0.554 & 0.577\% & 0.058 & 0.000 & \num{109290.3} & \num{27850.2} & \num{110588.2} & \num{26840.0} & \num{274568.7} \\
240 & 20 & 1.126 & 0.291\% & 0.059 & 0.000 & \num{112075.9} & \num{26873.4} & \num{112126.0} & \num{27216.5} & \num{278291.8} \\
240 & 30 & 1.715 & 0.219\% & 0.059 & 0.000 & \num{113528.5} & \num{29659.0} & \num{114411.2} & \num{29483.5} & \num{287082.2} \\
240 & 60 & 3.529 & 0.109\% & 0.065 & -0.001 & \num{128118.0} & \num{33101.6} & \num{131147.0} & \num{32998.9} & \num{325365.5} \\
240 & 120 & 7.349 & 0.051\% & 0.065 & 0.000 & \num{135784.6} & \num{27981.4} & \num{137299.4} & \num{35605.7} & \num{336671.1} \\
240 & 240 & 14.361 & 0.027\% & 0.065 & -0.007 & \num{117546.6} & \num{34743.7} & \num{155397.5} & \num{34685.1} & \num{342372.9} \\
240 & 600 & 35.996 & 0.007\% & 0.067 & -0.016 & \num{104355.6} & \num{33378.2} & \num{185707.3} & \num{37490.9} & \num{360932.0} \\
240 & 1200 & 68.423 & 0.007\% & 0.069 & -0.019 & \num{97473.0} & \num{28203.9} & \num{194109.7} & \num{44193.0} & \num{363979.6} \\
240 & 3600 & 205.929 & 0.003\% & 0.103 & 0.000 & \num{94804.6} & \num{23575.6} & \num{96004.8} & \num{10744.6} & \num{225129.6} \\
\hline    
 \end{tabular}
}
\end{table}

\end{document}